\documentclass[
reprint,
superscriptaddress,
nofootinbib,
amsmath,amssymb,
aps,
prd,
floatfix,
]{revtex4-2}

\usepackage{booktabs}
\usepackage{epic}
\usepackage{tikz}
\usepackage{xr-hyper} 
\usepackage{placeins}
\usepackage{graphicx}
\usepackage{dcolumn}
\usepackage{bm}
\usepackage{xcolor}
\usepackage{comment}
\usepackage{lipsum}
\usepackage{multirow}
\usepackage{soul}
\usepackage{orcidlink}
\usepackage[newcommands]{ragged2e}

\usepackage{hyperref}
\hypersetup{
    colorlinks=true,
    linkcolor=blue,
    filecolor=magenta,
    urlcolor=cyan,
    pdftitle={Methods},
    pdfpagemode=FullScreen,
}

\usepackage[colorinlistoftodos]{todonotes}
\usepackage{lineno}

\newenvironment{subfigure}[1]
  {\begin{minipage}[t]{#1}\centering}
  {\end{minipage}}

\newcommand{\paneltitle}[1]{%
  \par\smallskip
  \makebox[\linewidth][c]{\small #1}%
  \par
}

\newcommand{\paneltitlex}[2][0pt]{%
  \par\smallskip
  \makebox[\linewidth][c]{\hspace*{#1}\small #2}%
  \par
}

\begin{document}

\preprint{APS/123-QED}

\title{Probing Nuclear Effects with Transverse Kinematic Imbalance in Muon-neutrino Induced Charged-Current $\pi^0$ Production on Argon with the MicroBooNE Detector }%
\date{\today}

\newcommand{\ANL}{Argonne National Laboratory (ANL), Lemont, IL, 60439, USA}
\newcommand{\Bern}{Universit{\"a}t Bern, Bern CH-3012, Switzerland}
\newcommand{\BNL}{Brookhaven National Laboratory (BNL), Upton, NY, 11973, USA}
\newcommand{\UCSB}{University of California, Santa Barbara, CA, 93106, USA}
\newcommand{\Cambridge}{University of Cambridge, Cambridge CB3 0HE, United Kingdom}
\newcommand{\CIEMAT}{Centro de Investigaciones Energ\'{e}ticas, Medioambientales y Tecnol\'{o}gicas (CIEMAT), Madrid E-28040, Spain}
\newcommand{\Chicago}{University of Chicago, Chicago, IL, 60637, USA}
\newcommand{\Cincinnati}{University of Cincinnati, Cincinnati, OH, 45221, USA}
\newcommand{\CSU}{Colorado State University, Fort Collins, CO, 80523, USA}
\newcommand{\Columbia}{Columbia University, New York, NY, 10027, USA}
\newcommand{\Edinburgh}{University of Edinburgh, Edinburgh EH9 3FD, United Kingdom}
\newcommand{\FNAL}{Fermi National Accelerator Laboratory (FNAL), Batavia, IL 60510, USA}
\newcommand{\Granada}{Universidad de Granada, Granada E-18071, Spain}
\newcommand{\IIT}{Illinois Institute of Technology (IIT), Chicago, IL 60616, USA}
\newcommand{\ICL}{Imperial College London, London SW7 2AZ, United Kingdom}
\newcommand{\Indiana}{Indiana University, Bloomington, IN 47405, USA}
\newcommand{\Kansas}{The University of Kansas, Lawrence, KS, 66045, USA}
\newcommand{\KSU}{Kansas State University (KSU), Manhattan, KS, 66506, USA}
\newcommand{\Lancaster}{Lancaster University, Lancaster LA1 4YW, United Kingdom}
\newcommand{\LANL}{Los Alamos National Laboratory (LANL), Los Alamos, NM, 87545, USA}
\newcommand{\Louisiana}{Louisiana State University, Baton Rouge, LA, 70803, USA}
\newcommand{\Manchester}{The University of Manchester, Manchester M13 9PL, United Kingdom}
\newcommand{\MIT}{Massachusetts Institute of Technology (MIT), Cambridge, MA, 02139, USA}
\newcommand{\Michigan}{University of Michigan, Ann Arbor, MI, 48109, USA}
\newcommand{\MSU}{Michigan State University, East Lansing, MI 48824, USA}
\newcommand{\Minnesota}{University of Minnesota, Minneapolis, MN, 55455, USA}
\newcommand{\Nankai}{Nankai University, Nankai District, Tianjin 300071, China}
\newcommand{\NMSU}{New Mexico State University (NMSU), Las Cruces, NM, 88003, USA}
\newcommand{\NotreDame}{University of Notre Dame, Notre Dame, IN, 46556, USA}
\newcommand{\Oxford}{University of Oxford, Oxford OX1 3RH, United Kingdom}
\newcommand{\Pitt}{University of Pittsburgh, Pittsburgh, PA, 15260, USA}
\newcommand{\QMUL}{Queen Mary University of London, London E1 4NS, United Kingdom}
\newcommand{\Rutgers}{Rutgers University, Piscataway, NJ, 08854, USA}
\newcommand{\SLAC}{SLAC National Accelerator Laboratory, Menlo Park, CA, 94025, USA}
\newcommand{\SDSMT}{South Dakota School of Mines and Technology (SDSMT), Rapid City, SD, 57701, USA}
\newcommand{\Maine}{University of Southern Maine, Portland, ME, 04104, USA}
\newcommand{\TelAviv}{Tel Aviv University, Tel Aviv, Israel, 69978}
\newcommand{\UTA}{University of Texas, Arlington, TX, 76019, USA}
\newcommand{\Tufts}{Tufts University, Medford, MA, 02155, USA}
\newcommand{\VTech}{Center for Neutrino Physics, Virginia Tech, Blacksburg, VA, 24061, USA}
\newcommand{\Warwick}{University of Warwick, Coventry CV4 7AL, United Kingdom}


\affiliation{\ANL}
\affiliation{\Bern}
\affiliation{\BNL}
\affiliation{\UCSB}
\affiliation{\Cambridge}
\affiliation{\CIEMAT}
\affiliation{\Chicago}
\affiliation{\Cincinnati}
\affiliation{\CSU}
\affiliation{\Columbia}
\affiliation{\Edinburgh}
\affiliation{\FNAL}
\affiliation{\Granada}
\affiliation{\IIT}
\affiliation{\ICL}
\affiliation{\Indiana}
\affiliation{\Kansas}
\affiliation{\KSU}
\affiliation{\Lancaster}
\affiliation{\LANL}
\affiliation{\Louisiana}
\affiliation{\Manchester}
\affiliation{\MIT}
\affiliation{\Michigan}
\affiliation{\MSU}
\affiliation{\Minnesota}
\affiliation{\Nankai}
\affiliation{\NMSU}
\affiliation{\NotreDame}
\affiliation{\Oxford}
\affiliation{\Pitt}
\affiliation{\QMUL}
\affiliation{\Rutgers}
\affiliation{\SLAC}
\affiliation{\SDSMT}
\affiliation{\Maine}
\affiliation{\TelAviv}
\affiliation{\UTA}
\affiliation{\Tufts}
\affiliation{\VTech}
\affiliation{\Warwick}

\author{P.~Abratenko\,\orcidlink{0000-0001-6945-5941}}\affiliation{\Tufts} 
\author{D.~Andrade~Aldana\,\orcidlink{0009-0008-3143-3374}} \affiliation{\IIT}
\author{J.~Asaadi\,\orcidlink{0000-0001-6915-5279}}   \affiliation{\UTA}
\author{A.~Ashkenazi\,\orcidlink{0000-0002-1995-3851}}   \affiliation{\TelAviv}
\author{S.~Balasubramanian} \affiliation{\FNAL}
\author{B.~Baller\,\orcidlink{0000-0001-8731-9281}}  \affiliation{\FNAL}

\author{A.~Barnard\,\orcidlink{0000-0001-6117-1768}} \affiliation{\Oxford}

\author{G.~Barr\,\orcidlink{0000-0002-9763-1882}} \affiliation{\Oxford}
\author{D.~Barrow\,\orcidlink{0000-0001-5844-709X}} \affiliation{\Oxford}
\author{J.~Barrow\,\orcidlink{0000-0002-7319-3339}} \affiliation{\Minnesota} 
% ...

\collaboration{The MicroBooNE Collaboration}
\thanks{microboone\_info@fnal.gov}\noaffiliation

\begin{abstract}
Neutrino-nucleus cross-section measurements are needed to improve interaction modeling and to enable precision neutrino oscillation measurements in upcoming experiments such as the Deep Underground Neutrino Experiment (DUNE), Hyper-Kamiokande, and the Short-Baseline Neutrino program. Baryon-resonance neutrino interactions constitute a dominant contribution near the peak of the DUNE neutrino energy spectrum. We present the first measurement of muon neutrino charged-current resonance-like interactions on argon using transverse kinematic imbalance variables with the MicroBooNE detector. These observables are highly sensitive to the modeling of final-state interactions. This measurement probes kinematic imbalances using the reconstructed momenta of the muon, leading proton, and neutral pion. A comprehensive characterization of the $\pi^0$–proton final state is presented; however, none of the models considered are able to simultaneously reproduce all measured observables.

\end{abstract}

\maketitle
\section{Introduction}\label{sec:intro}
Neutrino oscillation measurements in accelerator-based experiments rely on reconstructing the incident neutrino energy from the particles observed in the final state. This reconstruction depends on accurate modeling of neutrino–nucleus interactions inside the detector medium ~\cite{Benhar2017,KatoriMartini2020}. As experiments move toward increasingly precise measurements, such as those planned at DUNE ~\cite{DUNECDR2020}, reliable interaction models are required to control systematic uncertainties in oscillation analyses~\cite{AlvarezRuso2018}.
Interactions on nuclear targets are complicated by nuclear effects that modify the momenta and multiplicities of particles emerging from the nucleus relative to the primary neutrino–nucleon interaction. In particular, final-state interactions (FSI) can alter hadron kinematics through processes such as pion absorption, charge exchange, and elastic or inelastic scattering within the nucleus. These effects distort the observable final-state kinematics used in neutrino energy reconstruction, making the accuracy of their modeling a significant source of systematic uncertainty in oscillation measurements, including those sensitive to the leptonic CP-violating phase~\cite{Coloma2014,Mosel2016}.
At neutrino energies in the range 1–4\,GeV, resonant pion production is a dominant interaction channel~\cite{Katori2012RMP}. This energy range overlaps with a substantial portion of the neutrino flux expected at DUNE. In this regime, pion production is largely mediated by baryon resonance excitation, most prominently the $\Delta(1232)$ resonance~\cite{ReinSehgal1981,Hernandez2013}. Measurements of neutrino–nucleus interactions with final-state pions therefore provide tests of models describing resonant pion production and the subsequent propagation of hadrons through the nuclear medium.
Neutral pion ($\pi^{0}$) production is also of experimental interest because $\pi^{0}$ mesons decay promptly into two photons, which can produce electromagnetic signatures that mimic electrons in detectors. As a result, $\pi^{0}$ production constitutes an important background to electron-neutrino appearance measurements~\cite{MiniBooNEpi02010}.

We report ten single-differential cross-section measurements of muon-neutrino charged-current (CC) interactions on argon in a resonance-like topology, characterized by a final state containing one muon, one $\pi^{0}$ meson, and at least one proton ($\nu_\mu$CC$Np1\pi^0$), using data collected with the MicroBooNE detector. The corresponding flux-integrated total cross section is measured to be $1.62 \times 10^{-38}$ cm$^{2}$/Ar. The signal includes events with muon kinetic energy ($T_\mu$) above 40\,MeV, at least one proton with $T_p$ above 60\,MeV, and no charged pions or heavier mesons with kinetic energy above 40\,MeV. The requirement of exactly 1$\pi^{0}$ and at least one proton provides sensitivity to nuclear effects through the kinematics of the hadronic system and final-state interactions  \cite{ReinSehgal1981}. Although this topology is largely associated with baryon-resonance excitation, the selected sample can also include contributions from nonresonant and deep-inelastic processes that produce the same observable final state.

This topology was previously investigated by the \mbox{MINERvA} collaboration~\cite{minerva_pi0}, which measured cross sections on carbon for three transverse kinematic imbalance (TKI) observables using a neutrino beam with a mean energy of approximately 3\,GeV. In contrast, the present measurement on argon includes single-differential cross sections as functions of the $T$ and scattering angle ($\cos\theta$) of the muon, the $\pi^{0}$, and the leading proton, as well as four TKI observables that characterize the momentum imbalance of the final-state system in the plane transverse to the neutrino direction. These observables provide enhanced sensitivity to both the primary interaction vertex and final-state interactions in neutrino--argon scattering~\cite{TKI_Ar,TKI}.

Several updates relative to the previous MicroBooNE CC $\pi^{0}$ inclusive measurement~\cite{uboone_ccpi0} are incorporated in this analysis. First, the analysis uses the full MicroBooNE Booster Neutrino Beam (BNB) dataset corresponding to $1.10 \times 10^{21}$ protons-on-target (POT)~\cite{MicroBooNEDetector2017}, providing significantly larger event statistics than the earlier measurement. Second, an updated event selection is employed, achieving an efficiency of 12.7\% compared to 8.6\% previously reported, and a purity of 75.6\% compared to the previous 69\%. Third, an updated final-state interaction reweighting procedure is applied to address the overprediction of the pion charge-exchange cross section in the \texttt{GENIE} intranuclear hadron transport model, which differs from the corresponding predictions of other neutrino event generators, as described in Sec.~\ref{sec:mc_simulation}. In addition, the signal definition is refined relative to Ref.~\cite{uboone_ccpi0} by requiring the presence of at least one proton in the final state~\cite{Hernandez2013}, thereby increasing the fraction of baryon-resonance interactions in the selected sample~\cite{Mosel2016}. Finally, this analysis employs an updated multiple Coulomb scattering (MCS) algorithm to estimate the momentum of exiting muons~\cite{mcs_paper}. Relative to the previous implementation, the improved muon energy reconstruction reduces biases and improves data--simulation agreement, resulting in better momentum resolution for partially contained muons.


\section{Transverse Kinematic Imbalance}

In neutrino--nucleon interactions, assuming the initial-state nucleon is at rest, momentum conservation requires that the vector sum of the outgoing lepton and hadronic system momenta equals the incoming neutrino momentum. Consequently, in the plane transverse to the neutrino direction, the net transverse momentum is expected to vanish when all final-state particles are accounted for. In neutrino--nucleus interactions, however, several nuclear effects can generate a nonzero transverse momentum imbalance. These include the initial momentum of the struck nucleon inside the nucleus (Fermi motion), final-state interactions that alter hadron kinematics through intranuclear scattering, and the presence of unobserved or experimentally inaccessible final-state particles. Observables that characterize this imbalance are referred to as TKI variables and provide sensitivity to nuclear effects in neutrino--nucleus scattering~\cite{TKI,T2KTKI2018,uboone_tki_ptxy}.

\begin{figure}[!htb]
    \centering
    \begin{subfigure}{0.7\linewidth}
        \centering
        \fbox{\includegraphics[width=\linewidth]{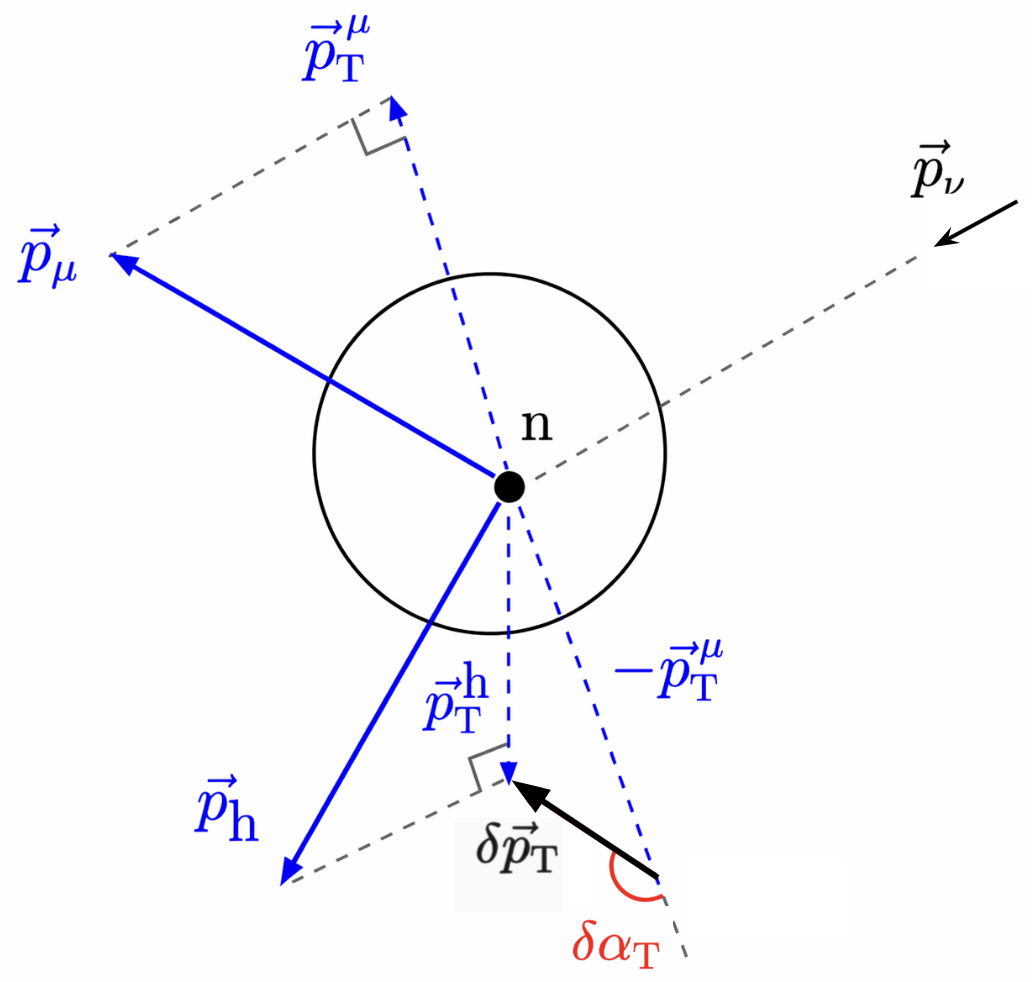}}
     \paneltitle{(a)}   
    \end{subfigure}
    \hfill
    \bigskip
    \begin{subfigure}{0.7\linewidth}
        \centering
        \fbox{\includegraphics[width=\linewidth]{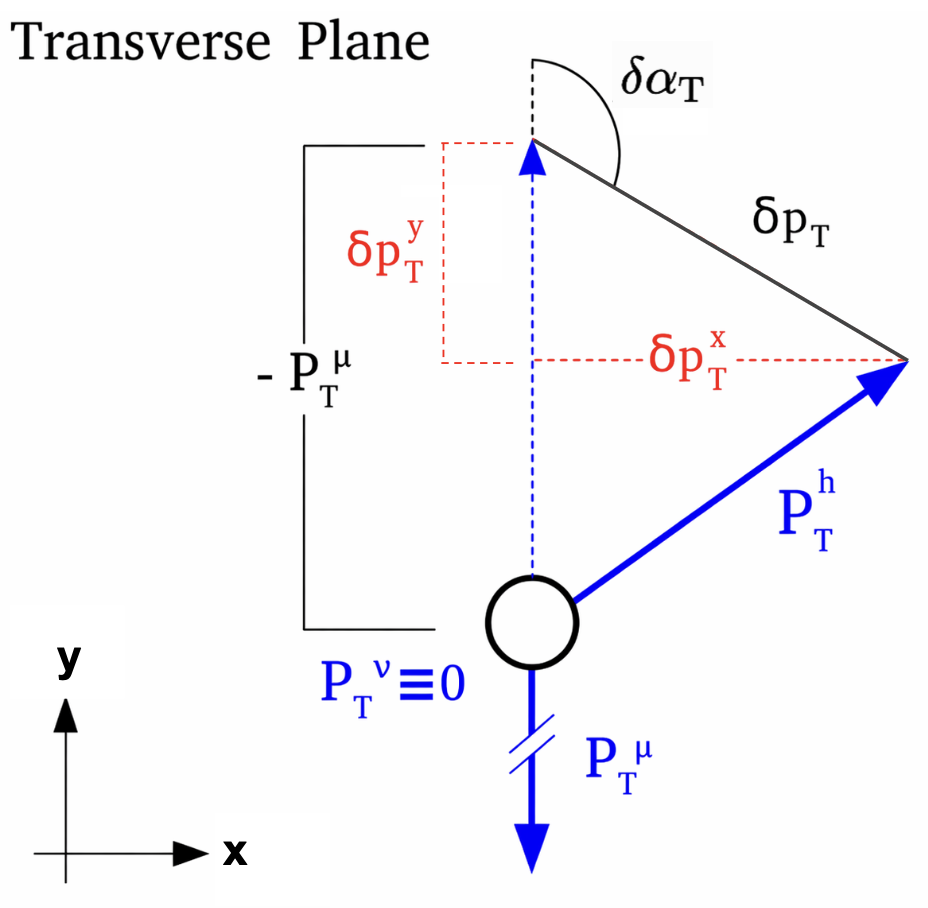}}
        \paneltitle{(b)}  
    \end{subfigure}
\caption{Illustration of the transverse kinematic imbalance variables used in this analysis~\cite{minerva_pi0}. Panel (a) defines $\delta \vec{p}_T$ and $\delta \alpha_T$, while panel (b) shows the decomposition into components $\delta p_T^x$ and $\delta p_T^y$.}
    \label{fig:tki_diagram}
\end{figure}

Figure~\ref{fig:tki_diagram} illustrates the TKI variables used in this work. The transverse plane is defined as perpendicular to the incoming neutrino beam direction. The leading hadronic momentum is given by
\begin{equation}
\vec{p}_h = \vec{p}_{\pi^0} + \vec{p}_p ,
\end{equation}
where $\vec{p}_{\pi^0}$ and $\vec{p}_p$ denote the three-momenta of the $\pi^{0}$ and the leading proton. Additional protons, when present, are not included in the TKI definitions. The muon momentum is denoted by $\vec{p}_\mu$.

The transverse momentum imbalance vector is defined as
\begin{equation}
\delta \vec{p}_T = \vec{p}_T^{\,\mu} + \vec{p}_T^{\,h},
\end{equation}
where $\vec{p}_T^{\,\mu}$ and $\vec{p}_T^{\,h}$ are the transverse components of the muon and hadronic momenta. Deviations of $\delta \vec{p}_T$ from zero indicate the presence of nuclear effects or undetected particles.

Two variables are commonly used to characterize this imbalance. The magnitude
\begin{equation}
\delta p_T = |\delta \vec{p}_T|,
\end{equation}
quantifies the overall size of the missing transverse momentum. The orientation of the imbalance is described by the transverse opening angle
\begin{equation}
\delta \alpha_T =
\arccos \left(
\frac{-\vec{p}_T^{\,\mu} \cdot \delta \vec{p}_T}
     {|\vec{p}_T^{\,\mu}|\,|\delta \vec{p}_T|}
\right),
\end{equation}
defined as the angle between $-\vec{p}_T^{\,\mu}$ and $\delta \vec{p}_T$. In the absence of nuclear effects, the hadronic system exactly balances the muon transverse momentum, resulting in a uniform distribution of $\delta \alpha_T$. Deviations from this configuration arise from nuclear dynamics such as Fermi motion and FSI which modify the magnitude and direction of the outgoing hadronic momentum~\cite{TKI}.

An alternative representation of the transverse momentum imbalance is obtained by decomposing $\delta \vec{p}_T$ into components defined with respect to the muon transverse momentum direction,
\begin{equation}
\delta p_T^x =
\frac{(\hat{z} \times \vec{p}_T^{\,\mu}) \cdot \delta \vec{p}_T}
{|\vec{p}_T^{\,\mu}|}, \qquad
\delta p_T^y =
-\frac{\vec{p}_T^{\,\mu} \cdot \delta \vec{p}_T}
{|\vec{p}_T^{\,\mu}|}.
\end{equation}
Here, $\delta p_T^y$ denotes the component parallel (or anti-parallel) to $\vec{p}_T^{\,\mu}$, while $\delta p_T^x$ is perpendicular to it, providing a convenient coordinate basis for $\delta \vec{p}_T$. In this representation, $\delta p_T^y$ probes momentum imbalance along the lepton direction in the transverse plane, whereas $\delta p_T^x$ characterizes transverse deflections orthogonal to it.
The distributions of $\delta p_T^x$ and $\delta p_T^y$ therefore carry complementary information about nuclear effects. The $\delta p_T^x$ distribution is expected to be symmetric about zero, since both the initial-state nucleon motion and the decay kinematics of the resonance are uncorrelated with the direction orthogonal to $\vec{p}_T^{\,\mu}$~\cite{ptx_sym}. In contrast, $\delta p_T^y$ is particularly sensitive to energy loss and momentum transfer within the nucleus, including modifications induced by final-state interactions and undetected final-state particles~\cite{minerva_pi0,T2K_pip_pTx,uboone_tki_ptxy}.

\section{THE MICROBOONE EXPERIMENT}
The MicroBooNE experiment uses a liquid argon time projection chamber (LArTPC) 
located on the BNB at Fermi National Accelerator Laboratory ~\cite{MicroBooNEDetector2017,BNB2009}. 
The detector contains an 85~tonne active liquid argon target with dimensions 
of 2.56\,m along the drift direction, 2.32\,m vertically, and 10.36\,m along the beam direction. 
An applied electric field of 273\,V/cm causes ionization electrons produced by charged particles 
to drift toward three wire readout planes located at the anode. The wire planes consist of two 
induction planes and one collection plane with 3\,mm wire spacing and orientations of 
0$^\circ$ and $\pm 60^\circ$ relative to the vertical axis. Signals induced on these wires are 
amplified and digitized by cryogenic electronics operating directly in the liquid argon, 
enabling high-resolution three-dimensional imaging and calorimetric reconstruction of 
charged-particle trajectories~\cite{MicroBooNEDetector2017}. 

The detector also includes 32 photomultiplier tubes that record prompt scintillation light produced 
during particle interactions. These light signals provide nanosecond-scale timing information, 
which is used to identify activity coincident with the neutrino beam spill and to reject 
cosmic-ray backgrounds in this surface detector~\cite{MicroBooNECosmic2018}. 
The fine spatial resolution and calorimetric capability of the LArTPC enable detailed studies 
of exclusive neutrino interaction final states on argon.

MicroBooNE is located approximately 470\,m downstream of the BNB target. 
The BNB is produced by 8\,GeV protons from the Fermilab Booster striking a beryllium target, 
followed by magnetic focusing of secondary hadrons and their decay in a 50\,m decay region~\cite{BNB2009}. The resulting neutrino beam has a mean energy of approximately 0.8\,GeV and 
is predominantly composed of $\nu_\mu$ (93.6\%), with smaller contributions from 
$\bar{\nu}_\mu$ (5.9\%) and $\nu_e+\bar{\nu}_e$ (0.6\%)~\cite{BNB2009}.

\section{MODEL DESCRIPTION} 
Monte Carlo (MC) simulation is used to determine signal selection efficiencies, 
background rates, the detector response, and systematic uncertainties. 
The simulation models the BNB flux, neutrino--argon interactions, 
and the detector response~\cite{BNB2009,MicroBooNEDetector2017}.

The neutrino flux is simulated using a \texttt{Geant4}-based beamline model 
that describes hadron production from 8\,GeV protons incident on a beryllium 
target, including magnetic horn focusing and hadron reinteractions prior 
to decay into neutrinos~\cite{BNB2009,GEANT4}.

Neutrino interactions with argon are simulated using the \texttt{GENIE} neutrino MC generator \texttt{(v3.0.6)}~\cite{GENIEv3}. The MicroBooNE model configuration \texttt{G18\_10a\_02\_11a} is used, with model parameters tuned to external data from T2K~\cite{uboone_tune}. The underlying interaction model includes a local Fermi gas description of the nuclear ground state together with models of quasielastic scattering, multinucleon interactions, resonance production, and coherent pion production.

Resonant pion production in the simulation is dominated by excitation of the $\Delta(1232)$ resonance, which provides the primary contribution to the selected signal in MicroBooNE. In \texttt{GENIE}, this process is modeled using the Berger--Sehgal implementation~\cite{BergerSehgal2007} of the Rein--Sehgal formalism, in which baryon resonances are produced and subsequently decayed into hadronic final states such as $\pi N$, $\pi\pi N$, and $\eta N$. The model includes lepton-mass effects, updated pion angular distributions, and parameters tuned to neutrino--deuterium data. However, several approximations remain relevant for $\Delta(1232)$-dominated channels: interference between resonances is neglected, the vector form factors used for $\Delta$ production are not fully consistent with modern electron-scattering constraints, and nuclear-medium modifications to the $N \to \Delta$ transition are not generally included~\cite{res_problems}. These limitations can affect both the normalization and kinematic distributions of the predicted pion-production events.

Final-state interactions are modeled using the \texttt{GENIE} \texttt{hA2018} model~\cite{GENIEv3,hA2018}. This semi-classical model assigns probabilities for hadronic re-interactions within the nucleus using parameterizations derived from pion--nucleus scattering data. In energy regions where experimental data are limited, the model incorporates results from intranuclear cascade calculations, primarily \texttt{CEM03}~\cite{cem_mashnik}, to extend the interaction probabilities across the relevant energy range, rather than performing an explicit microscopic transport simulation~\cite{fsi_models_genie}.

An updated treatment of pion charge-exchange (CEX) interactions is implemented in this analysis and will be included in the next release of \texttt{GENIE}. Previous studies indicate that the default \texttt{hA2018} model tends to overestimate the CEX rate~\cite{hA2018_issue}. This discrepancy originates from the way available pion--nucleus scattering data and model calculations were combined to construct the interaction probability parameterizations.  Recent developments incorporate an improved treatment of these data together with inputs from the \texttt{hN2018} intranuclear cascade model, which performs a full hadronic transport simulation~\cite{fsi_models_genie, GENIEGitHub}. To account for this effect in the present analysis, a dedicated reweighting procedure is applied to modify the CEX contribution in the simulation without regenerating the full MC sample. Details of this reweighting procedure are provided in Appendix~\ref{appendix:fsi_rw}.

Final-state particles are propagated through a detailed \texttt{Geant4}-based detector simulation~\cite{GEANT4}. The resulting ionization charge and scintillation light signals are modeled using the \texttt{LArSoft} framework~\cite{LArSoft}, which simulates charge drift, electronics response, optical photon transport, and detector noise~\cite{uboone_sim,uboone_recom}. 

The detector simulation is organized into distinct categories for signal and background events. The background event categories include EXT, which refers to cosmic-ray background events derived from the beam-off dataset and containing no BNB neutrino interactions; Cosmic, which corresponds to misidentified cosmic-ray background events from the BNB overlay MC simulation; and Dirt, which denotes neutrino interactions whose true interaction vertices occur outside the fiducial volume~\cite{Ben_0pNp}.\label{sec:mc_simulation}

\section{ANALYSIS OVERVIEW}

\subsection{Event Selection}

Candidate $\nu_\mu$ CC interactions are selected using the reconstruction and preselection framework described in Refs.~\cite{WCLee2018,WireCellImaging}. A brief summary of the reconstruction procedure and analysis-specific event selection is provided below.

Selected events are separated into fully contained (FC) and partially contained (PC) categories according to the containment of the final-state particles within the active detector volume. Fully contained events require all reconstructed final-state particles to remain inside the fiducial volume, whereas partially contained events permit only the reconstructed muon track to exit the detector while all other reconstructed particles remain contained. This categorization is motivated by the poorer momentum resolution associated with exiting muons. Since the transverse kinematic imbalance observables used in this analysis are particularly sensitive to final-state momentum reconstruction, events with additional exiting particles are excluded to avoid biases in the reconstructed momentum imbalance. The FC and PC samples are reconstructed independently and combined during the unfolding procedure used to extract the differential cross sections. Approximately 60\% of selected events are classified as FC, with the remaining 40\% belonging to the PC category.

For FC events, the muon and proton kinetic energies are reconstructed from their residual ranges. For PC events, the muon momentum is reconstructed using a MCS-estimator~\cite{mcs_paper}. Neutral pion energies are reconstructed from electromagnetic (EM) shower charge deposition using the Wire-Cell charge-to-energy conversion procedure. The collected ionization charge is first converted to deposited energy assuming an average ionization energy of 23.6\,eV per electron-ion pair~\cite{charge_to_ionization}. An additional multiplicative scale factor of 2.50 is then applied to EM showers to account for charge reconstruction biases and the average recombination factor for electromagnetic activity in liquid argon~\cite{WireCellImaging}.

\begin{figure}[!htbp]
  \centering
  \includegraphics[width=\linewidth]{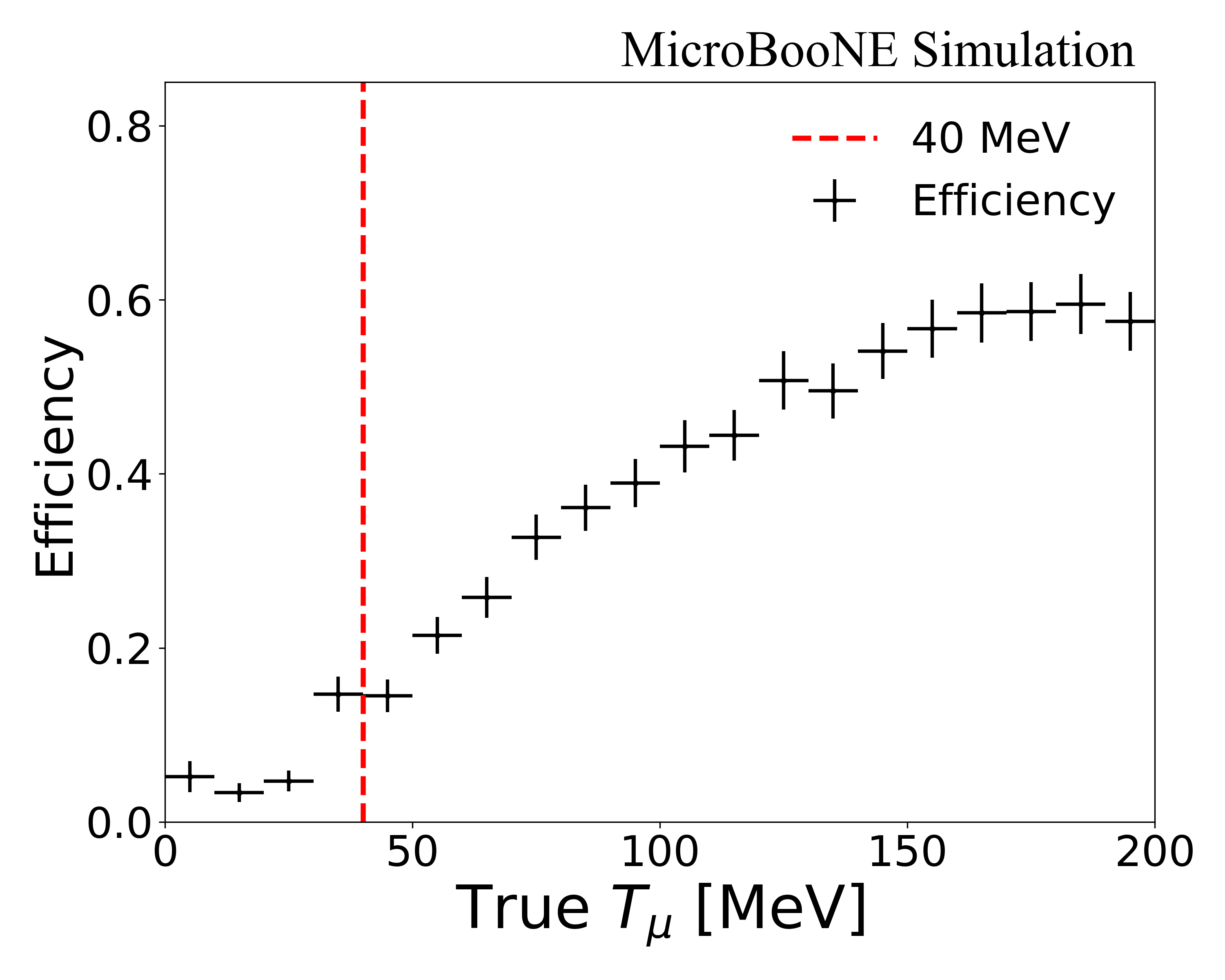}
\caption{Muon selection efficiency as a function of true muon kinetic energy for true $\nu_\mu$CC$Np1\pi^0$ events. The error bars represent statistical uncertainties.}
  \label{fig:muon_eff}
\end{figure}

The muon reconstruction efficiency for true $\nu_\mu$CC$Np1\pi^0$ interactions is shown as a function of true muon kinetic energy in Fig.~\ref{fig:muon_eff}. Reconstructed muon tracks are required to have a minimum track length of 5~cm to ensure reliable particle identification. This requirement significantly reduces the reconstruction efficiency below approximately 30\,MeV. To avoid regions of rapidly varying detector acceptance, selected events are required to contain a reconstructed muon with kinetic energy greater than 40\,MeV. For PC events, an additional minimum reconstructed track length requirement of 70~cm is imposed to ensure sufficient path length for the MCS-based momentum reconstruction~\cite{mcs_paper}.

Neutral-pions are reconstructed through the $\pi^{0}\rightarrow\gamma\gamma$ decay channel by identifying pairs of electromagnetic showers consistent with a common production point. Candidate photon showers are required to satisfy topology and calorimetric criteria optimized for EM shower identification. Additional selection requirements are imposed on the reconstructed photon energies, the photon conversion distances relative to the neutrino interaction vertex, the opening angle between the two photon candidates, the reconstructed invariant mass of the $\gamma\gamma$ system, and the spatial consistency between the reconstructed $\pi^0$ decay vertex and the neutrino interaction vertex~\cite{WireCellImaging,WCNCpi0}. These requirements are optimized to maximize the product of  signal efficiency times purity.

Proton reconstruction requirements follow the methodology developed in Ref.~\cite{Ben_0pNp}. Since this analysis additionally requires a reconstructed neutral pion in the final state, the impact of the proton reconstruction requirement is quantified using the ratio of selection efficiencies for events with and without reconstructed protons. The efficiency ratio is defined as

\begin{equation}
R(T_p) =
\frac{\varepsilon_{N p,1\pi^0}(T_p)}
     {\varepsilon_{X p,1\pi^0}(T_p)},
\end{equation}
where $\varepsilon_{Np,1\pi^0}$($T_p)$ is the efficiency for selecting events containing at least one reconstructed proton and one reconstructed $\pi^0$, and $\varepsilon_{Xp,1\pi^0}(T_p)$ is the corresponding efficiency for events containing a reconstructed $\pi^0$ independent of proton reconstruction. Here, $X\in\{0,N\}$ denotes events with either zero or at least one proton in the final state.

\begin{figure}[htbp]
  \centering
  \includegraphics[width=1\linewidth]{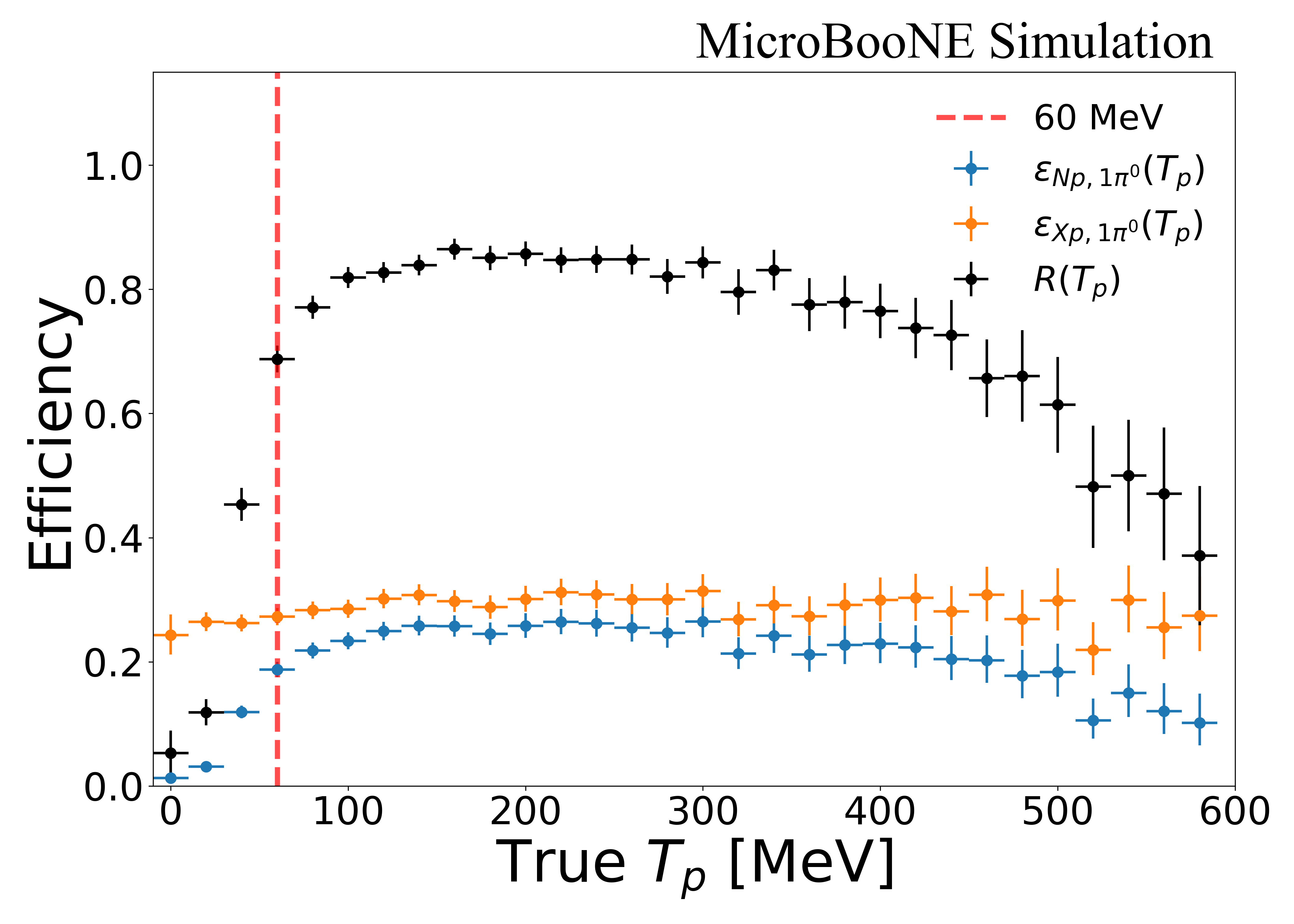}
\caption{%
\hyphenpenalty=10000
\exhyphenpenalty=10000
Ratio of proton selection efficiencies, $R(T_p)$, between $Np1\pi^0$ and $Xp1\pi^0$ samples as a function of the true leading-proton kinetic energy $T_p$. The error bars represent statistical uncertainties. A sharp decrease is observed below 60\,MeV, motivating the applied proton kinetic energy threshold.}
  \label{fig:proton-ratio}
\end{figure}

Figure~\ref{fig:proton-ratio} shows the resulting efficiency ratio as a function of the true leading-proton kinetic energy. A rapid decrease in efficiency is observed below approximately 60\,MeV due to the reduced reconstruction performance for short proton tracks. To ensure stable reconstruction efficiency, the leading proton is therefore required to have kinetic energy $T_p > $ 60\,MeV. No kinetic energy requirement is imposed on additional reconstructed protons. Simulation predicts that approximately 17\% of selected signal events contain additional protons in the final state.

After applying the muon, proton, and neutral-pion reconstruction requirements, the selected sample remains contaminated by CC deep-inelastic scattering (DIS) interactions, predominantly events containing additional mesons in the final state~\cite{GENIEv3}. To suppress these backgrounds and enhance the resonant single-$\pi^0$ topology, events containing reconstructed charged pion candidates with kinetic energy greater than 40\,MeV are rejected. In addition, residual electromagnetic activity not associated with the two photon showers assigned to the reconstructed $\pi^0$ candidate is required to have total reconstructed energy below 70\,MeV. The DIS rejection requirements reduce the residual DIS background contribution by approximately 4\%.

\begin{figure}[!htbp]
    \centering
        \includegraphics[width=\linewidth]{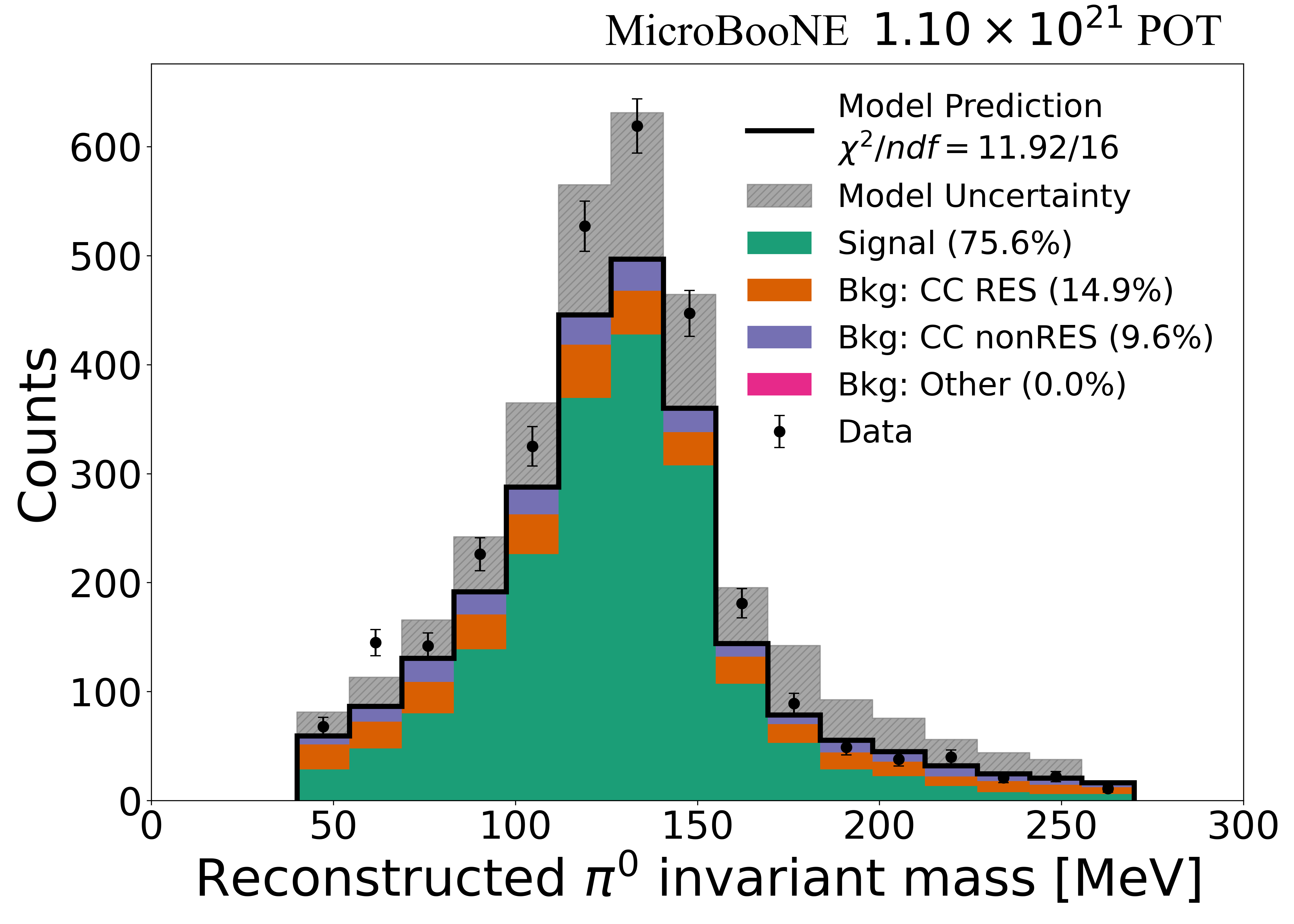}
    \caption{%
\hyphenpenalty=10000
\exhyphenpenalty=10000
Reconstructed $\pi^0$ invariant mass $m_{\pi^0}$ distribution after all selection criteria, separated into signal and background categories. The shaded band represents the total uncertainty, including both statistical and systematic contributions. The signal sample ($\nu_\mu$ CC$Np1\pi^0$) constitutes 75.6\% of the selected events. The dominant background arises from charged-current resonant (CC RES) interactions, primarily involving multi-pion production. Additional background contributions originate from charged-current nonresonant (CC nonRES) processes, including quasielastic and deep-inelastic scattering. A selection requirement of $40 < m_{\pi^0} < 270$ MeV is applied, resulting in the sharp edges observed in the distribution.
}

    \label{fig:pi0_mass}
\end{figure}

Figure~\ref{fig:pi0_mass} shows the reconstructed $\pi^0$ invariant mass distribution after all selection requirements are applied. The selected sample is dominated by $\nu_\mu$ CC$Np1\pi^0$ signal interactions, which constitute 75.6\% of the selected events. The dominant residual backgrounds arise from CC resonant interactions with multi-pion final states, followed by nonresonant CC processes including  quasielastic and DIS interactions. Smaller contributions originate from  Dirt events, and Cosmic-ray backgrounds.

\subsection{Reconstruction Performance}

After applying all selection criteria, the signal selection efficiency is estimated to be 12.7\% in simulation, with a corresponding sample purity of 75.6\%. In the full detector exposure, a total of 2988 candidate events are selected in data, consistent with the prediction, as shown in Fig.~\ref{fig:pi0_mass}.

The reconstruction performance is evaluated separately for the FC and PC samples for all observables entering the TKI calculation. Since the TKI variables depend on the reconstructed momentum magnitude and direction of the final-state particles ($\mu$, $p$, and $\pi^0$), both kinetic-energy and angular resolutions are quantified for each species. The angular resolution is evaluated using the polar ($\theta$) and azimuthal ($\phi$) angles of the reconstructed particle direction relative to the incoming neutrino direction. For particle kinetic energies and $\delta p_T$, the resolution is defined as the fractional difference $(X_\mathrm{reco}-X_\mathrm{true})/X_\mathrm{true}$. For angular observables, as well as $\delta p_T^x$ and $\delta p_T^y$, the resolution is defined as the absolute difference $X_\mathrm{reco}-X_\mathrm{true}$. Given that the central regions of the distributions are approximately Gaussian, the bias and resolution are extracted from a Gaussian fit performed within $\pm1$ RMS of the most probable value. Restricting the fit to this range reduces the impact of non-Gaussian tails. No significant dependence of the resolution on particle kinetic energy is observed for any of the species considered; accordingly, no binning in kinetic energy is applied.

The kinetic energy resolution is 3.3\% for contained muons and 16\% for exiting muons. The difference arises from the reconstruction method: contained-muon energy is determined from range, whereas exiting-muon energy is estimated using MCS, which has intrinsically lower precision. Contained and exiting muons have similar angular resolutions, with absolute-difference resolutions less than 0.07\,rad for $\theta$ and 0.105\,rad for $\phi$, and negligible bias.

For the leading proton and $\pi^0$, the kinetic-energy and angular resolutions are evaluated separately. In the FC sample, the $\pi^0$ kinetic-energy resolution is 23\%, while the proton kinetic-energy resolution is 2.3\%. Both particles have angular resolutions less than 0.07~rad for $\theta$ and 0.111~rad for $\phi$. The $\pi^0$ angular resolution benefits from the reconstruction strategy, in which the momentum direction is estimated using the vector from the reconstructed neutrino interaction vertex to the shower start position~\cite{WireCellImaging}. 

Although the proton and $\pi^0$ are required to be fully contained, their resolutions are found to be consistent between the FC and PC event samples. This agreement provides a cross-check that the event categorization does not introduce significant biases in the reconstruction of hadronic final-state kinematics.

The fractional resolution of the missing transverse momentum magnitude, $\delta p_T$, is 21\% for the FC sample and 28\% for the PC sample. For the transverse momentum components, the $\delta p_T^x$ absolute resolution is similar in the two samples, with values of 67\,MeV (FC) and 71\,MeV (PC). In contrast, the $\delta p_T^y$ resolution is larger in the PC sample, 139\,MeV compared to 88\,MeV for FC, consistent with its stronger dependence on the muon momentum reconstruction. The transverse opening angle, $\delta \alpha_T$, has an absolute-difference resolution less than 0.22\,rad for both samples.

Table~\ref{table:resolution_summary_boxed} summarizes the fitted bias and resolution values for all reconstructed particles and derived kinematic observables discussed in this section.

\begin{table*}[htbp]
\centering
\renewcommand{\arraystretch}{1.1}
\caption{
Summary of the reconstruction resolution (Res) and bias for fully contained (FC) 
and partially contained (PC) event samples. Absolute resolutions (with units) 
are reported for angular observables and transverse momentum components, 
while fractional resolutions are quoted for kinetic-energy observables and 
$\delta p_T$. Bias values correspond to the mean reconstructed minus true value 
for each observable.
}
\label{table:resolution_summary_boxed}
\begin{tabular}{|c|c|c|c|c|c|}
\hline\hline
\text{Category} & \text{Observable} 
& \text{FC Res} & \text{PC Res} 
& \text{FC Bias} & \text{PC Bias} \\
\hline

$\mu$ 
& $T_\mu$ (frac.) & 0.033 & 0.160 & 0.009 & -0.014 \\
& $\theta_\mu$ & 0.069 & 0.056 & 0.012 & 0.008 \\
& $\phi_\mu$   & 0.094 & 0.105 & -0.001 & -0.001 \\
\hline

$p$   
& $T_{p}$ (frac.) & 0.023 & 0.022 & -0.005 & -0.006 \\
& $\theta_{p}$ & 0.068 & 0.063 & 0.015 & 0.013 \\
& $\phi_{p}$   & 0.111 & 0.101 & 0.001 & 0.001 \\
\hline

$\pi^0$ 
& $T_{\pi^0}$ (frac.) & 0.226 & 0.252 & -0.015 & -0.019 \\
& $\theta_{\pi^0}$  & 0.047 & 0.050 & 0.0002 & 0.001 \\
& $\phi_{\pi^0}$    & 0.064 & 0.068 & 0.0006 & 0.005 \\
\hline

TKI 
& $\delta p_T$ (frac.) & 0.211 & 0.282 & -0.014 & 0.012 \\
& $\delta p_T^x$ (MeV) & 66.61 & 70.79 & 1.09 & -0.439 \\
& $\delta p_T^y$ (MeV) & 88.33 & 139.08 & -3.39 & -22.14 \\
& $\delta \alpha_T$    & 0.210 & 0.219 & 0.004 & 0.041 \\
\hline\hline
\end{tabular}
\end{table*}

\subsection{Efficiencies}

The signal selection efficiency is evaluated in bins of the true kinematic variables used in the differential cross-section measurement. In each true bin, the efficiency is defined as the ratio of the number of signal events that satisfy all reconstruction and selection requirements to the total number of generated signal events in that bin, as defined in the Supplemental Material~\cite{supp}.
 
The efficiency plots provided in the Supplemental Material~\cite{supp} show the combined FC and PC selection efficiency as functions of the ten observables considered in this analysis. The efficiency is generally uniform across the transverse kinematic imbalance variables. For the muon, the efficiency decreases at low kinetic energy, primarily due to the phase-space requirement applied to the PC sample. A reduction is also observed at low values of $\cos\theta_\mu$, corresponding to backward-going muons, which have shorter visible track lengths and are therefore more difficult to reconstruct accurately~\cite{WCLee2018}.

For the leading proton and the $\pi^0$, the efficiency decreases at high kinetic energies. In the proton case, the reduction is more pronounced, as high-energy protons are more likely to undergo secondary hadronic interactions in the detector material, which complicates track reconstruction and particle identification~\cite{MicroBooNEDetector2017}. The $\pi^0$ efficiency also decreases at high energies and in the forward region. In both cases, the opening angle between the decay photons becomes small, leading to overlapping electromagnetic showers that are more difficult to separate and accurately reconstruct~\cite{WCpi02023}.

\subsection{Systematic Uncertainties} \label{main_sys}
The systematic uncertainty evaluation includes contributions from the neutrino flux prediction, neutrino--argon interaction modeling, hadron--argon reinteractions, and detector response effects~\cite{BNB2009,uboone_tune,EPJCDetectorSys}. These uncertainties are propagated to the extracted cross sections through covariance matrices constructed for the ten measured observables while preserving the correlations among them.

Uncertainties in the neutrino flux prediction arise primarily from hadron production in interactions of the 8\,GeV proton beam with the beryllium target, as well as from beamline-related effects, including variations in the magnetic horn current and target alignment ~\cite{BNB2009}. The impact of these effects is evaluated through a reweighting procedure applied to the simulated flux prediction.

Uncertainties associated with neutrino--argon interaction modeling are evaluated by varying the parameters of the \texttt{GENIE} neutrino event generator within their associated uncertainties, following the procedure described in Ref.~\cite{uboone_tune}. These variations account for uncertainties in the modeling of quasielastic scattering, multinucleon processes, resonance production, DIS, coherent pion production, and other interaction channels relevant to the signal and background predictions. The resulting variations affect both the signal selection efficiency and the estimated background contributions. Additional uncertainties associated with intranuclear hadronic reinteractions of final-state particles within the argon nucleus are also included.

Systematic uncertainties associated with hadron--argon reinteractions in the detector medium are evaluated using the \texttt{Geant4Reweight} package~\cite{geant4_rw}. The inelastic cross sections for proton- and pion-induced interactions are varied by approximately 20\%, consistent with uncertainties derived from available external hadron--argon scattering data. These variations account for uncertainties in secondary hadronic interactions occurring as particles propagate through the detector material and are included in the overall flux uncertainty evaluation.

Uncertainties in the detector response are evaluated using dedicated simulation samples in which detector model parameters are varied~\cite{EPJCDetectorSys,light_yeild}. These variations account for effects such as the wire response, space charge distortions, electron--ion recombination, and scintillation light yield. Owing to the limited statistical precision of the simulated samples, the small number of events per bin introduces significant statistical fluctuations, resulting in an overestimation of the detector covariance matrix that propagates across bins through their correlations. To mitigate this effect, Gaussian process regression (GPR) smoothing is applied. GPR provides a Bayesian framework to model the distribution, suppressing statistical fluctuations by introducing correlations between nearby bins via a radial basis kernel. The kernel length scales are chosen to correspond to the reconstruction resolutions of the relevant observables, as summarized in Table~\ref{table:resolution_summary_boxed}. Additional details on the implementation and parameter choices are provided in~\cite{london_3d}.

Additional subdominant contributions to the systematic uncertainties arise from Dirt backgrounds and the finite size of the MC sample. Additional normalization uncertainties arise from POT counting and the determination of the number of argon nuclei in the fiducial volume~\cite{BNB2009}.

The fractional systematic uncertainties, detailed in the Supplemental Material~\cite{supp}, are presented for the single-differential cross sections of each observable, including both transverse kinematic imbalance variables and final-state particle kinematics. The corresponding figures also decompose the total uncertainty into contributions from individual sources.
For the integrated cross sections, the neutrino flux uncertainty is the dominant contribution, at 7.3\%. This is followed by cross section model uncertainties at the 5.1\% level, and detector-related and data statistical uncertainties, each contributing at the 3.5\% level. Subdominant contributions arise from Dirt backgrounds and MC statistical uncertainties, both at the 1.5\% level.
Notably, cross section model uncertainties remain subdominant relative to the total uncertainty in all bins, indicating a limited dependence of the measurement on the underlying interaction model. This feature enhances the sensitivity of the analysis to the physics encoded in the measured observables.

\section{CROSS SECTION EXTRACTION}
The differential cross sections are extracted using a blockwise unfolding approach~\cite{blockwise}. In this method, each measured distribution is unfolded separately, while groups of related observables are treated as blocks in the construction of the covariance matrix. This procedure accounts for detector acceptance, efficiency, and resolution effects for each observable and preserves the correlations among different distributions by propagating them from reconstruction to truth space. The resulting covariance matrix therefore captures both the statistical and systematic correlations across the full set of measured observables.

The nominal-flux-averaged differential cross section in true bin $\mu$ is given by
\begin{equation}
\left\langle \frac{d \sigma}{d\mathbf{x}} \right\rangle_\mu
=
\frac{\sum_i R_{\mu i} (M_i - B_i)}{\Phi_{\mathrm{nom}}\, T\, \Delta \mathbf{x}_\mu},
\end{equation}
where $R_{\mu i}$ denotes the unfolding matrix element that maps reconstructed bin $i$ to true bin $\mu$. Here, ``nominal flux-averaged'' indicates that the cross section is defined with respect to the assumed central neutrino flux prediction, rather than the unknown true flux realized in the data, as discussed in Refs.~\cite{real_flux, wc_methods}. 
The unfolding matrices for all ten observables are provided in the Supplemental Material~\cite{supp}. 
The quantities $M_i$ and $B_i$ represent the measured and predicted background event counts in reconstructed bin $i$, respectively. 
The normalization factors $\Phi_{\mathrm{nom}}$ and $T$ correspond to the integrated nominal neutrino flux prediction and the number of target nuclei in the fiducial volume, respectively. 
The bin width of observable $\mathbf{x}$ is denoted by $\Delta \mathbf{x}_\mu$.

The covariance between two unfolded measurements in true bins $\mu$ and $\lambda$, corresponding to observables $\mathbf{x}$ and $\mathbf{y}$, is given by
\begin{equation}
\operatorname{Cov}\!\left(
\left\langle \frac{d \sigma}{d\mathbf{x}} \right\rangle_\mu,
\left\langle \frac{d \sigma}{d\mathbf{y}} \right\rangle_\lambda
\right)
=
\frac{
\sum_{i,j}
R_{\mu i}\,
\operatorname{Cov}\!\left(M_i, M_j\right)\,
R_{\lambda j}
}{
\Phi^2\, T^2\, \Delta \mathbf{x}_\mu\, \Delta \mathbf{y}_\lambda
}.
\end{equation}
Here, $\operatorname{Cov}(M_i, M_j)$ represents the total covariance between reconstructed bins $i$ and $j$, including both statistical and systematic contributions. This expression propagates all uncertainties through the unfolding procedure and retains the full bin-to-bin and inter-observable correlations in the final result.

The unfolding is performed using the Wiener-SVD technique with a second-derivative regularization condition~\cite{wiener_svd}. 
The impact of regularization on the unfolded results is captured by the regularization matrix $A_C$, which maps theoretical predictions to the regularized unfolded space. 
This matrix is applied to model predictions prior to comparison with the measured cross sections and is provided in the Supplemental Material~\cite{supp}.

\section{Model Validation}

The model-validation procedure assesses the ability of the simulation to reproduce the reconstructed distributions observed in data. The simulation includes the neutrino interaction model, flux prediction, detector response, and the associated systematic uncertainties~\cite{Ben_0pNp,wc_methods}. The validation is performed through goodness-of-fit (GoF) tests in reconstructed space, where the observed data are compared directly with the nominal MC prediction, defined as the central prediction obtained without systematic variations. To enhance the sensitivity of these tests, conditional constraints derived from well-measured observables are applied. This method exploits correlations among reconstructed variables to obtain conditionally constrained model predictions, thereby increasing sensitivity to potential mismodeling. The conditional-constraint formalism follows the approach described in Ref.~\cite{wc_methods}. In this framework, correlations among reconstructed observables are explicitly incorporated to produce constrained predictions for a given observable using information from other measured kinematic quantities. These constraints are applied exclusively in reconstructed space and are used only for validation purposes. They are not applied during the unfolding procedure itself, which is performed using the nominal model prediction without additional conditioning.

The performance of the validation framework is evaluated through dedicated fake-data studies presented in the Supplemental Material~\cite{supp}. In these tests, simulated data samples generated with alternative model variations are treated as pseudo-data and propagated through the full analysis chain. The comparison between reconstructed pseudo-data and the nominal prediction provides a detector-level GoF test. The pseudo-data are then unfolded to true space, and the residual tension between the unfolded result and the generator-level prediction is quantified using the $\chi^2/\mathrm{ndf}$ statistic, where ($\mathrm{ndf}$) is the number of degrees of freedom~\cite{wc_methods}. These studies demonstrate that reconstructed-level constrained GoF tests provide substantially greater sensitivity to mismodeling than comparisons performed only after unfolding to true space. In all tested variations, the unfolded spectra are consistent with the corresponding generator-level truth within the estimated uncertainties, with residual biases smaller than $2\sigma$.

\begin{figure*}[htbp]
\centering
\setlength{\tabcolsep}{2pt}

\begin{subfigure}{0.45\textwidth}
    \centering
    \includegraphics[clip,trim={1.1cm 0.9cm 2.6cm 1.0cm},width=\linewidth]{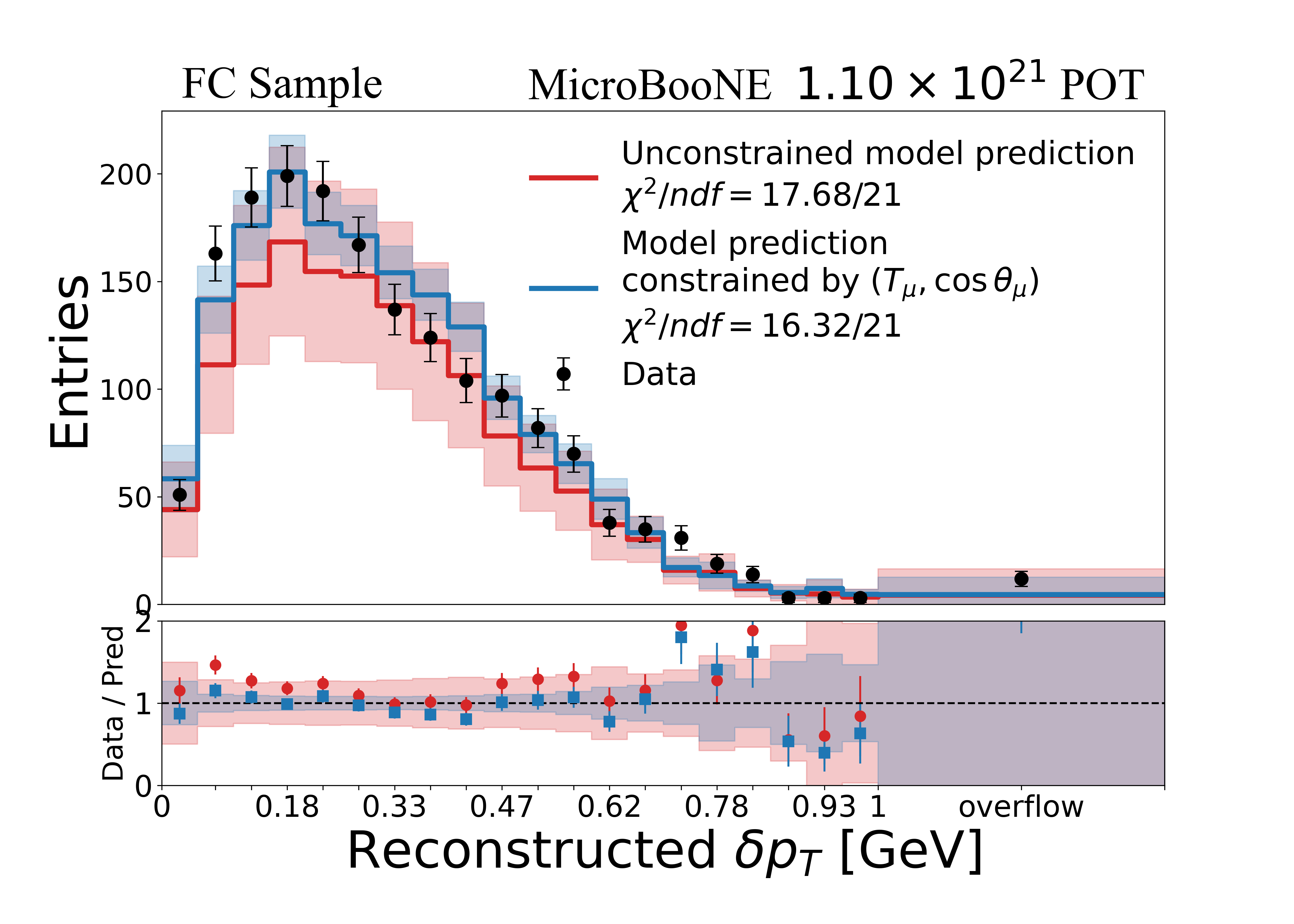}
    \paneltitle{(a) $\delta p_T$}
\end{subfigure}
\begin{subfigure}{0.45\textwidth}
    \centering
    \includegraphics[clip,trim={1.1cm 0.9cm 2.6cm 1.0cm},width=\linewidth]{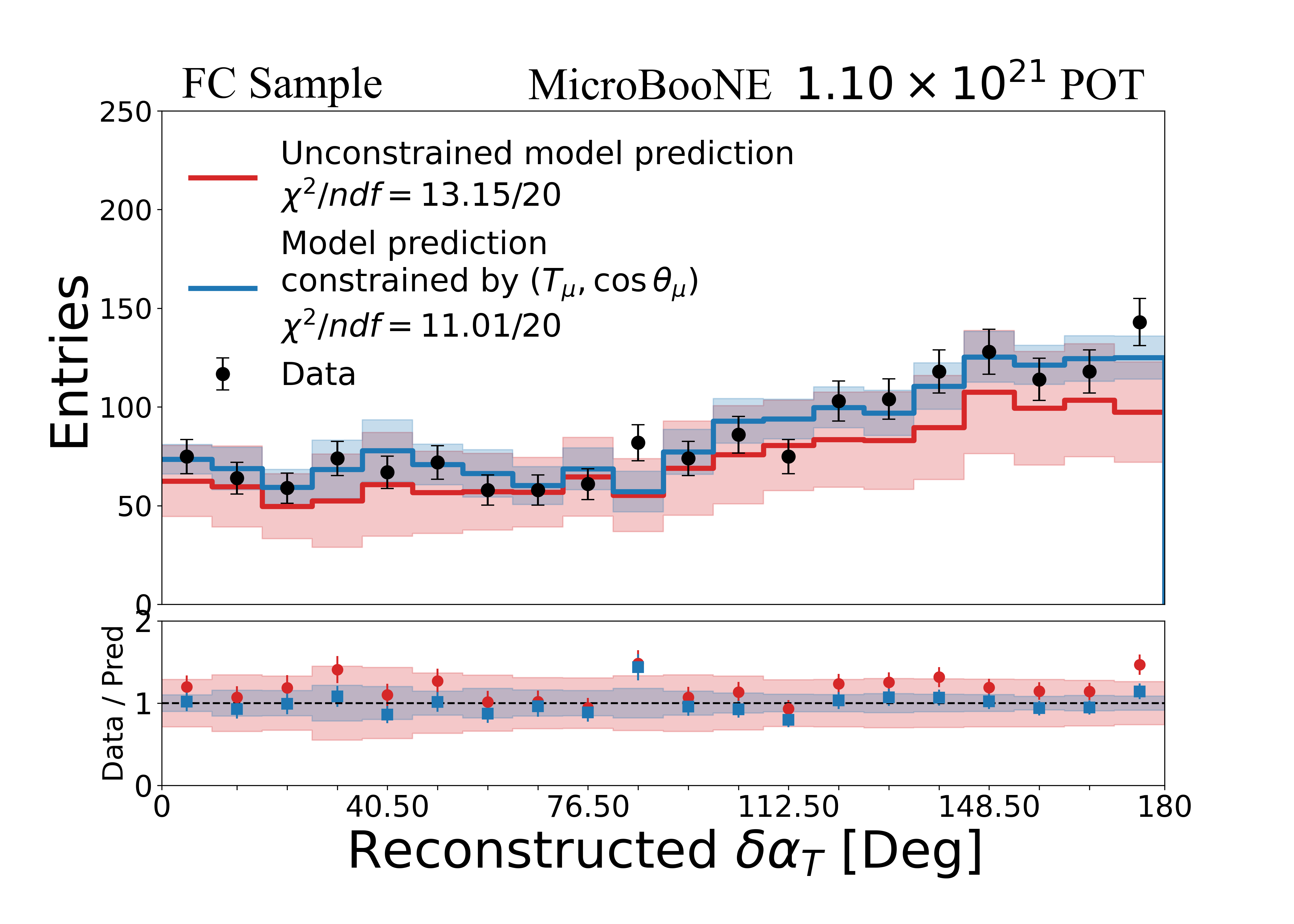}
    \paneltitle{(b) $\delta \alpha_T$}
\end{subfigure}

\begin{subfigure}{0.45\textwidth}
    \centering
    \includegraphics[clip,trim={1.1cm 0.9cm 2.6cm 1.0cm},width=\linewidth]{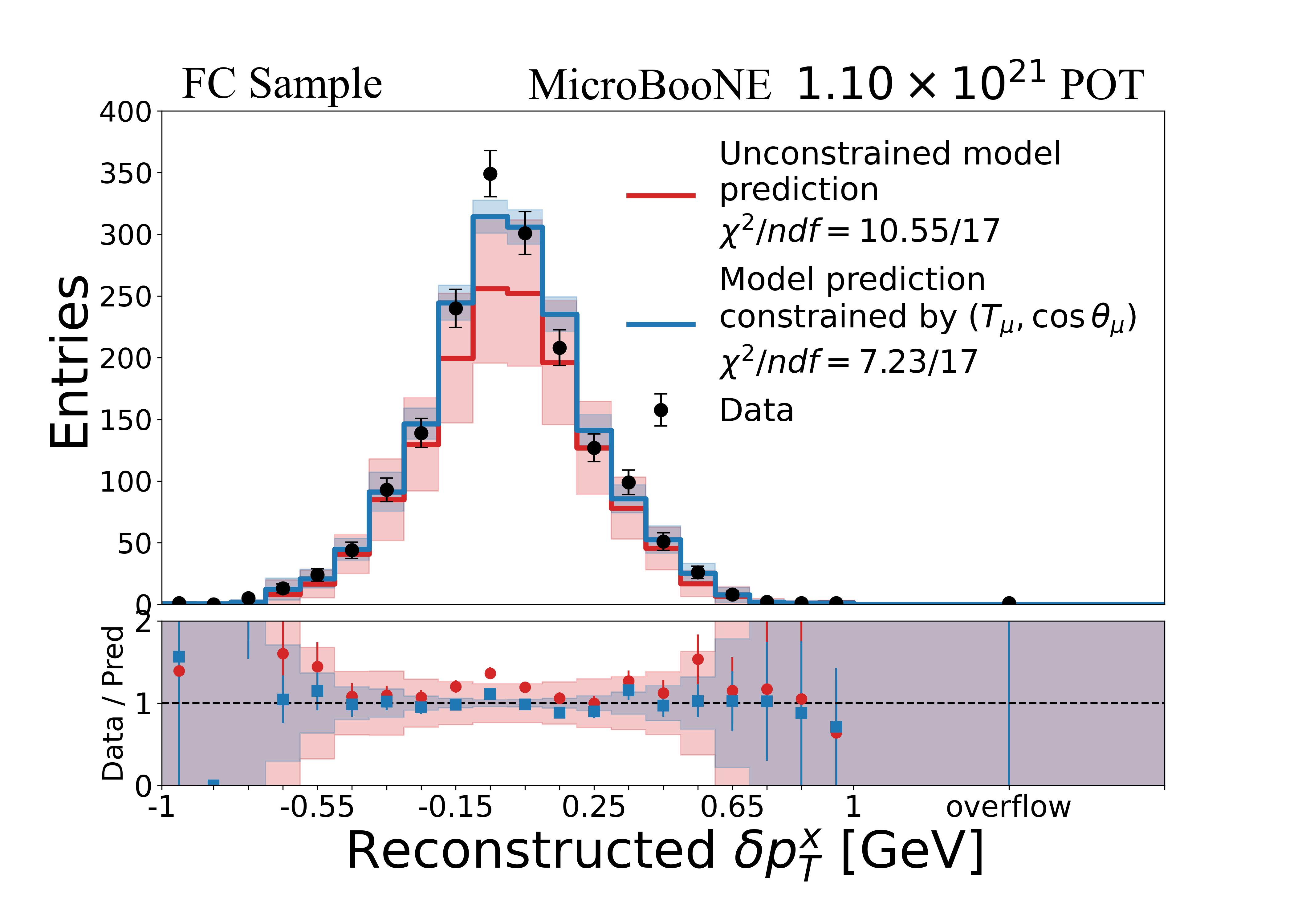}
    \paneltitle{(c) $\delta p_T^{x}$}
\end{subfigure}
\begin{subfigure}{0.45\textwidth}
    \centering
    \includegraphics[clip,trim={1.1cm 0.9cm 2.6cm 1.0cm},width=\linewidth]{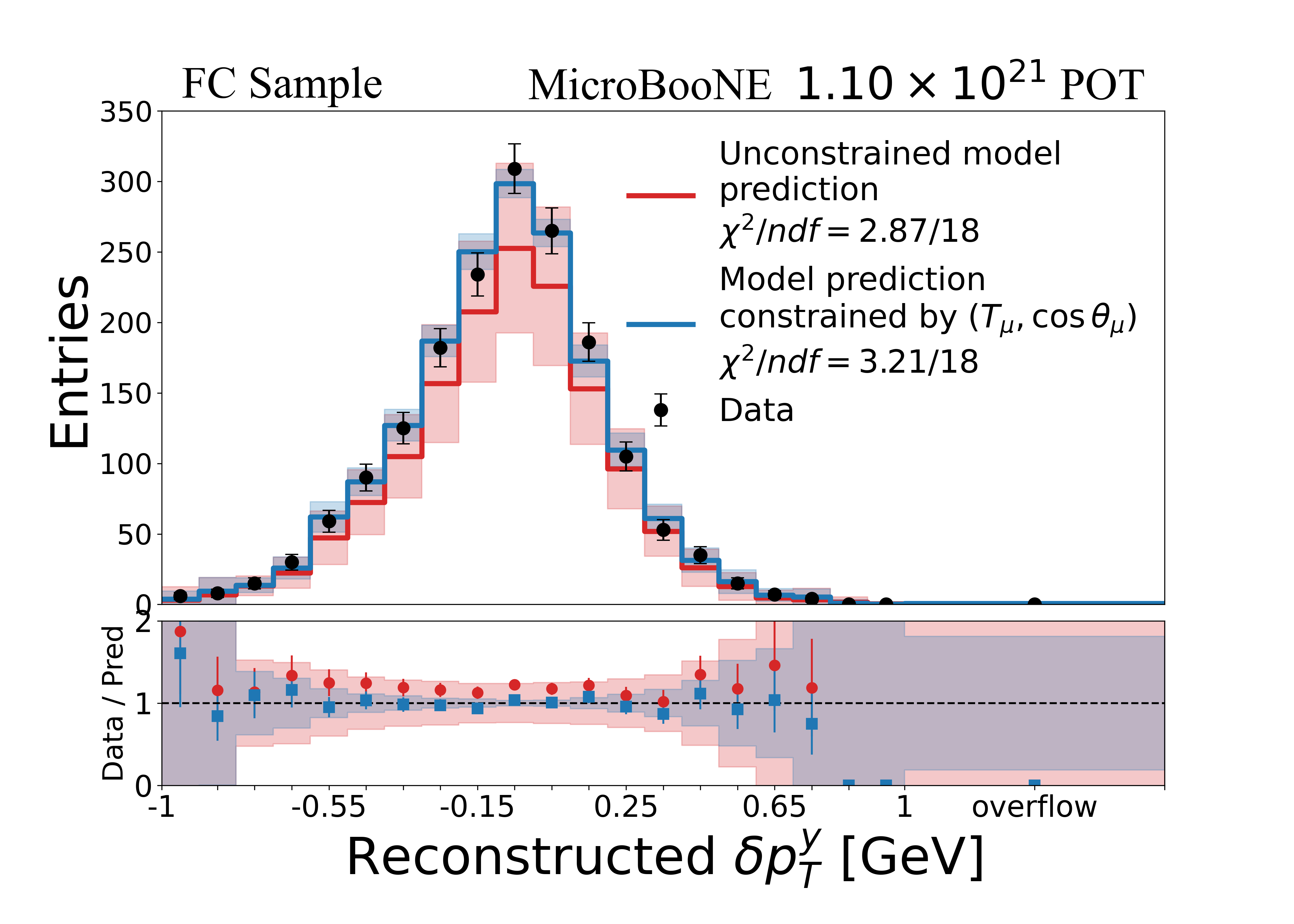}
    \paneltitle{(d) $\delta p_T^{y}$}
\end{subfigure}

\caption{
\hyphenpenalty=10000
\exhyphenpenalty=10000
Reconstructed distributions of transverse kinematic imbalance observables for FC events: 
(a) transverse momentum imbalance $\delta p_T$, 
(b) transverse opening angle $\delta \alpha_T$, 
(c) transverse momentum component $\delta p_T^{x}$, and 
(d) transverse momentum component $\delta p_T^{y}$. 
The nominal model prediction is shown in red and the constrained model prediction in blue, with the constraint derived from the reconstructed data distributions as described in Section~\ref{sec:model_val}. 
The shaded bands indicate the total model systematic and MC statistical uncertainties, while the data error bars represent statistical uncertainties. Bins with zero or one predicted event are excluded from the $\chi^2/\mathrm{ndf}$ calculation.}
\label{fig:fc_tki_rd}
\end{figure*}

Having established the sensitivity of the validation procedure through these fake-data studies, we apply the reconstructed-level GoF tests to the data to assess the level of agreement between the model and the data.
As the fully contained sample comprises the majority of selected events, we present here the reconstructed FC distributions for all ten observables considered in this analysis. The corresponding partially contained distributions are provided in the Supplemental Material~\cite{supp}. Figures ~\ref{fig:fc_basic_rd}--\ref{fig:fc_tki_rd} show the reconstructed FC distributions for the full set of observables used in this measurement. Table~\ref{tab:gof_rd_sum_prd} lists the $\chi^2/\mathrm{ndf}$ values for the FC and PC samples, where the effective number of degrees of freedom accounts for the removal of bins with negligible predicted event yields.

An excess of data is observed in several bins of the muon kinetic energy distribution at low $T_\mu$. This excess is, however, consistent with the assigned systematic uncertainties, as reflected by a goodness-of-fit of $\chi^2/\mathrm{ndf} = 12.33/21$. For the muon angular distribution, $\cos\theta_\mu$, good agreement between data and simulation is observed, with $\chi^2/\mathrm{ndf} = 18.02/20 $. The $T_\mu$ and $\cos\theta_\mu$ distributions provide correlated constraints on the model, with each observable contributing to the overall consistency of the description.

For the hadronic observables, the $\pi^0$ kinetic energy and $\cos\theta_{\pi^0}$ distributions are constrained using the muon and leading-proton kinematics, while the proton observables are constrained using the muon and $\pi^0$ kinematics. In all cases, the reconstructed-level tests yield $\chi^2/\mathrm{ndf} < 1$, indicating good agreement between data and simulation after applying the conditional constraints. The transverse kinematic imbalance observables are constrained using the muon kinematics ($T_\mu,\cos\theta_\mu)$. The reconstructed-level GoF tests show good agreement between data and simulation for all four TKI variables, with $\chi^2/\mathrm{ndf} \le 1$.

\begin{figure*}[htbp]
\centering
\setlength{\tabcolsep}{4pt}

\begin{subfigure}{0.48\textwidth}
    \centering
    \includegraphics[clip,trim={1.1cm 0.9cm 2.6cm 1.0cm},width=\linewidth]{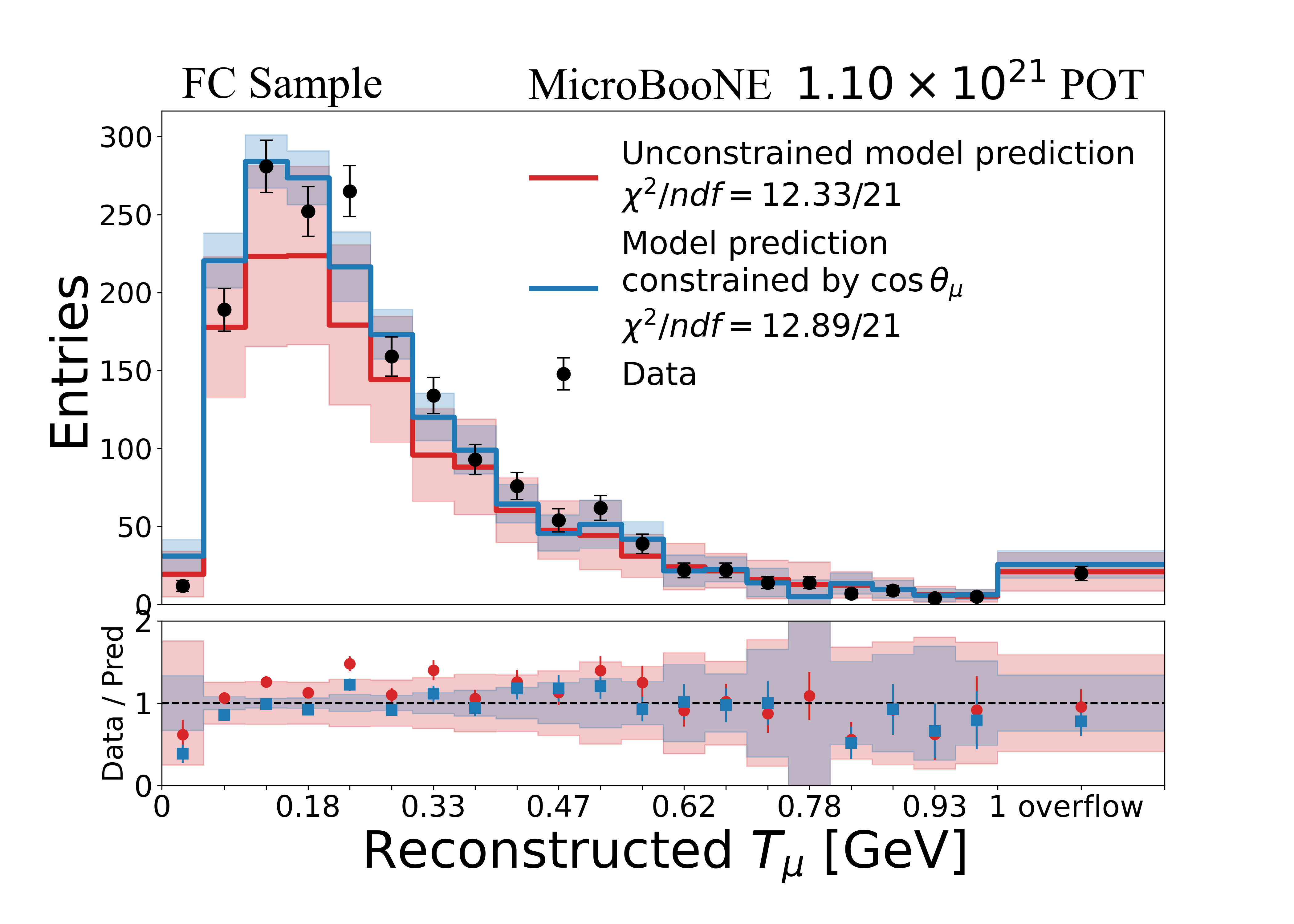}
    \paneltitle{(a) $T_{\mu}$}
\end{subfigure}
\begin{subfigure}{0.48\textwidth}
    \centering
    \includegraphics[clip,trim={1.1cm 0.9cm 2.6cm 1.0cm},width=\linewidth]{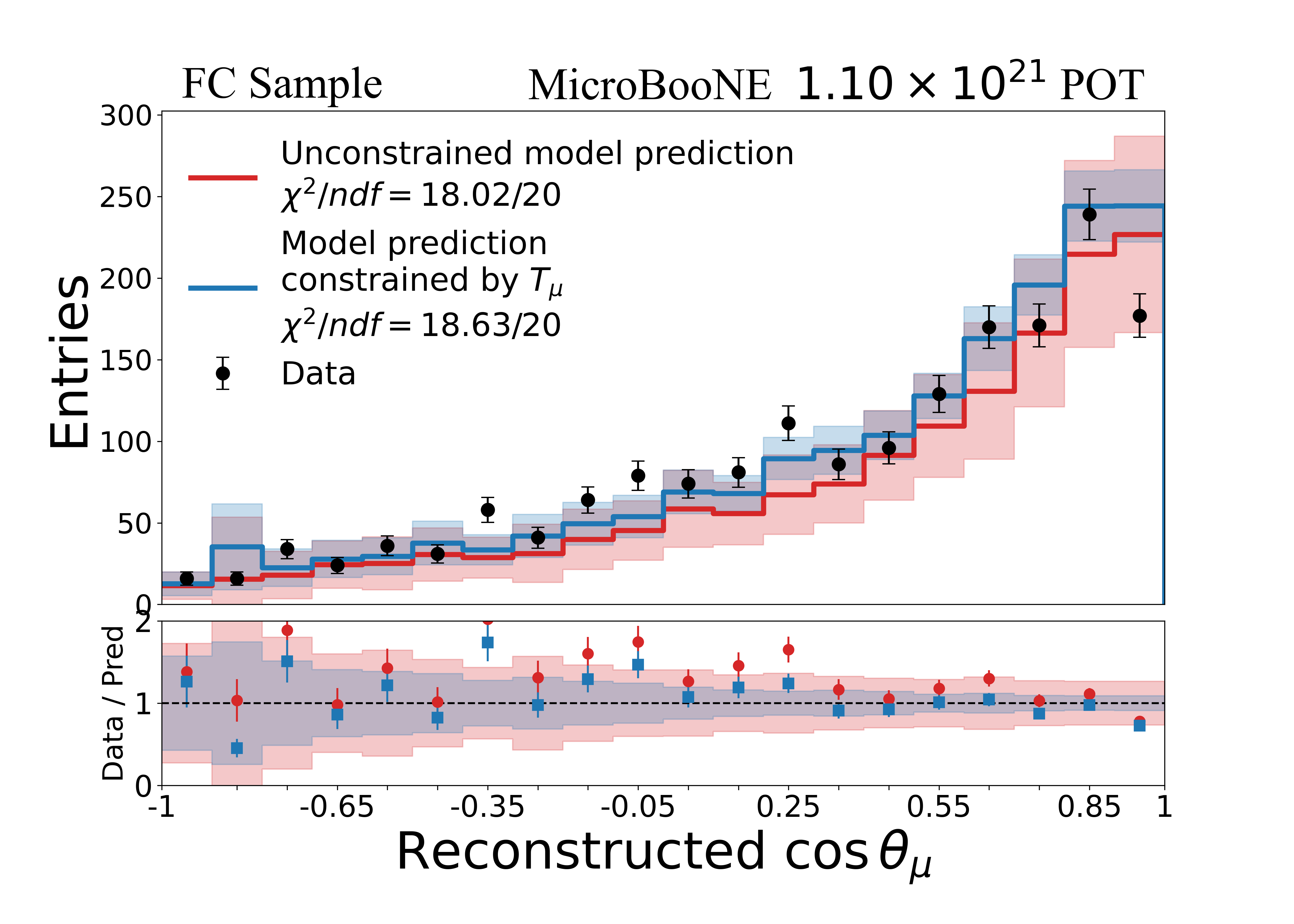}
    \paneltitle{(b) $\cos\theta_{\mu}$}
\end{subfigure}

\begin{subfigure}{0.48\textwidth}
    \centering
    \includegraphics[clip,trim={1.1cm 0.9cm 2.6cm 1.0cm},width=\linewidth]{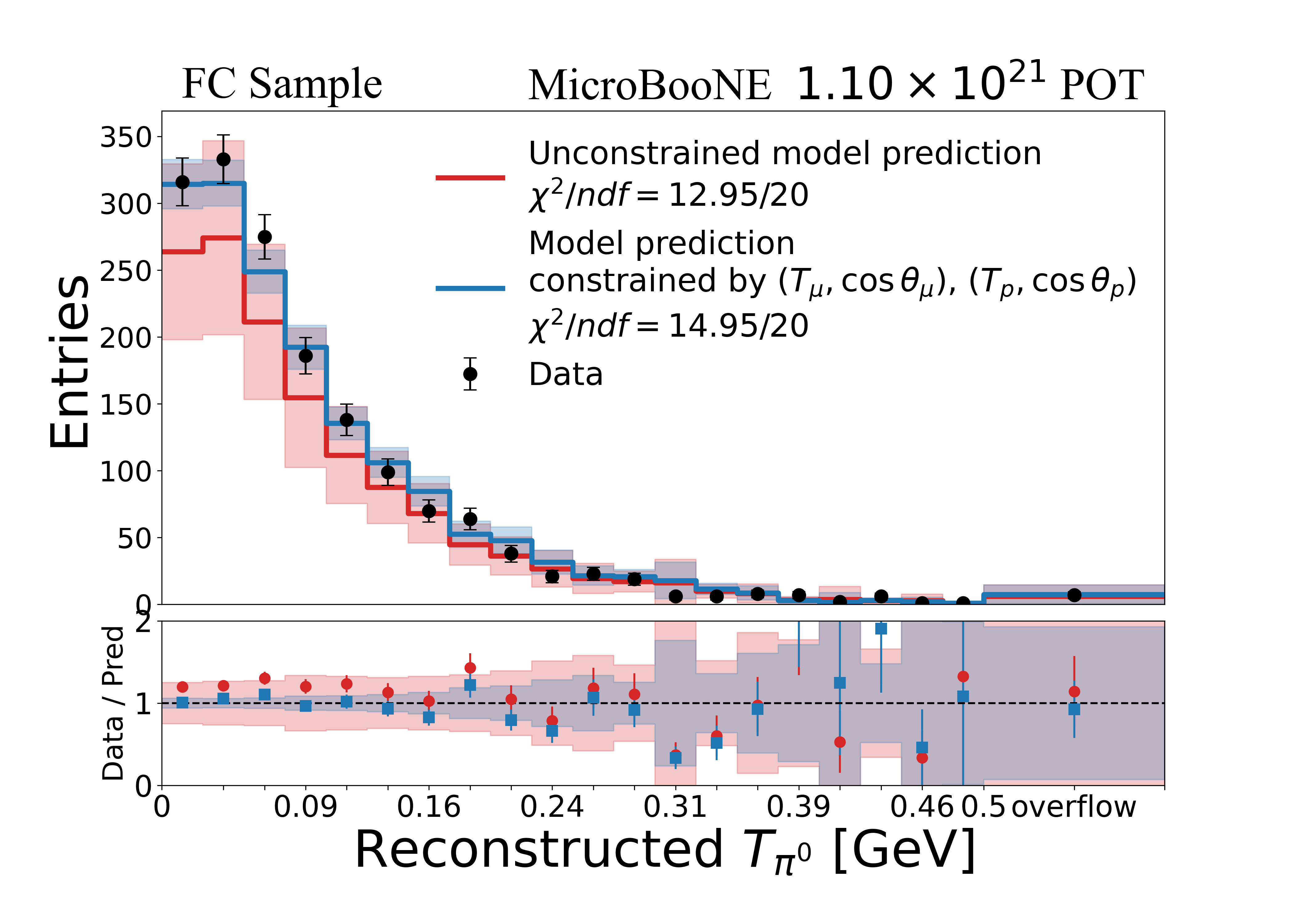}
    \paneltitle{(c) $T_{\pi^0}$}
\end{subfigure}
\begin{subfigure}{0.48\textwidth}
    \centering
    \includegraphics[clip,trim={1.1cm 0.9cm 2.6cm 1.0cm},width=\linewidth]{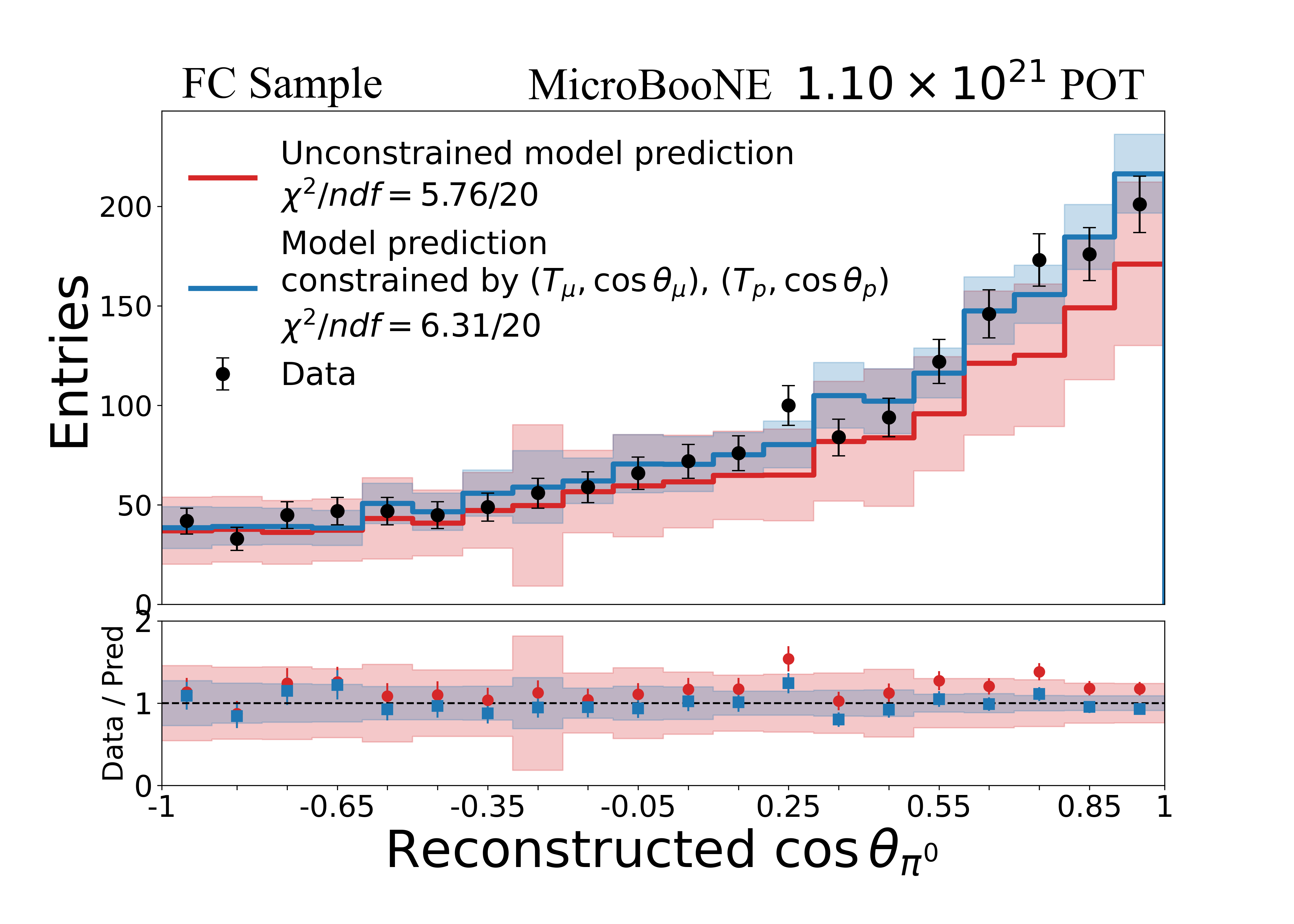}
    \paneltitle{(d) $\cos\theta_{\pi^0}$}
\end{subfigure}

\begin{subfigure}{0.48\textwidth}
    \centering
    \includegraphics[clip,trim={1.1cm 0.9cm 2.6cm 1.0cm},width=\linewidth]{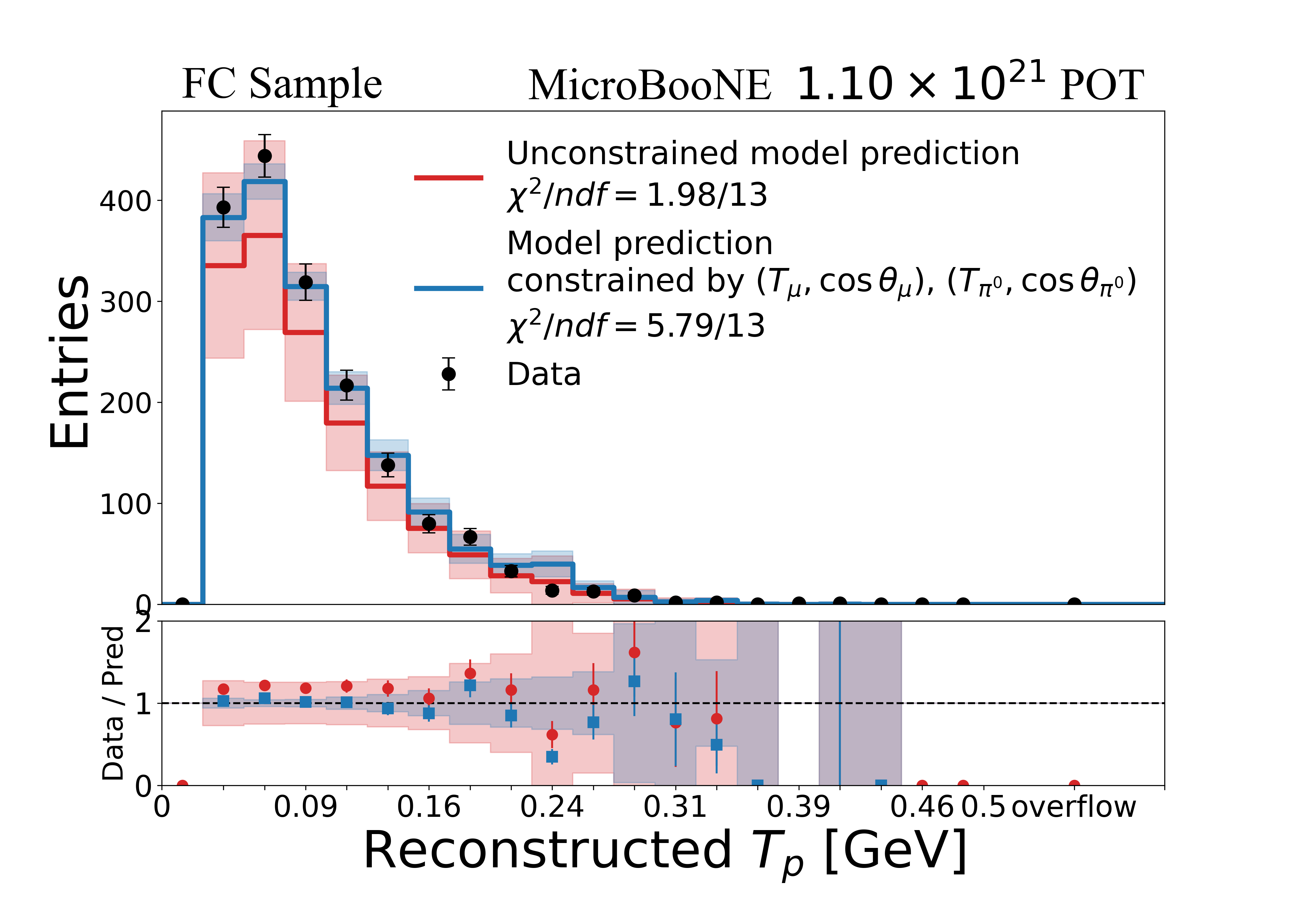}
    \paneltitle{(e) $T_{p}$}
\end{subfigure}
\begin{subfigure}{0.48\textwidth}
    \centering
    \includegraphics[clip,trim={1.1cm 0.9cm 2.6cm 1.0cm},width=\linewidth]{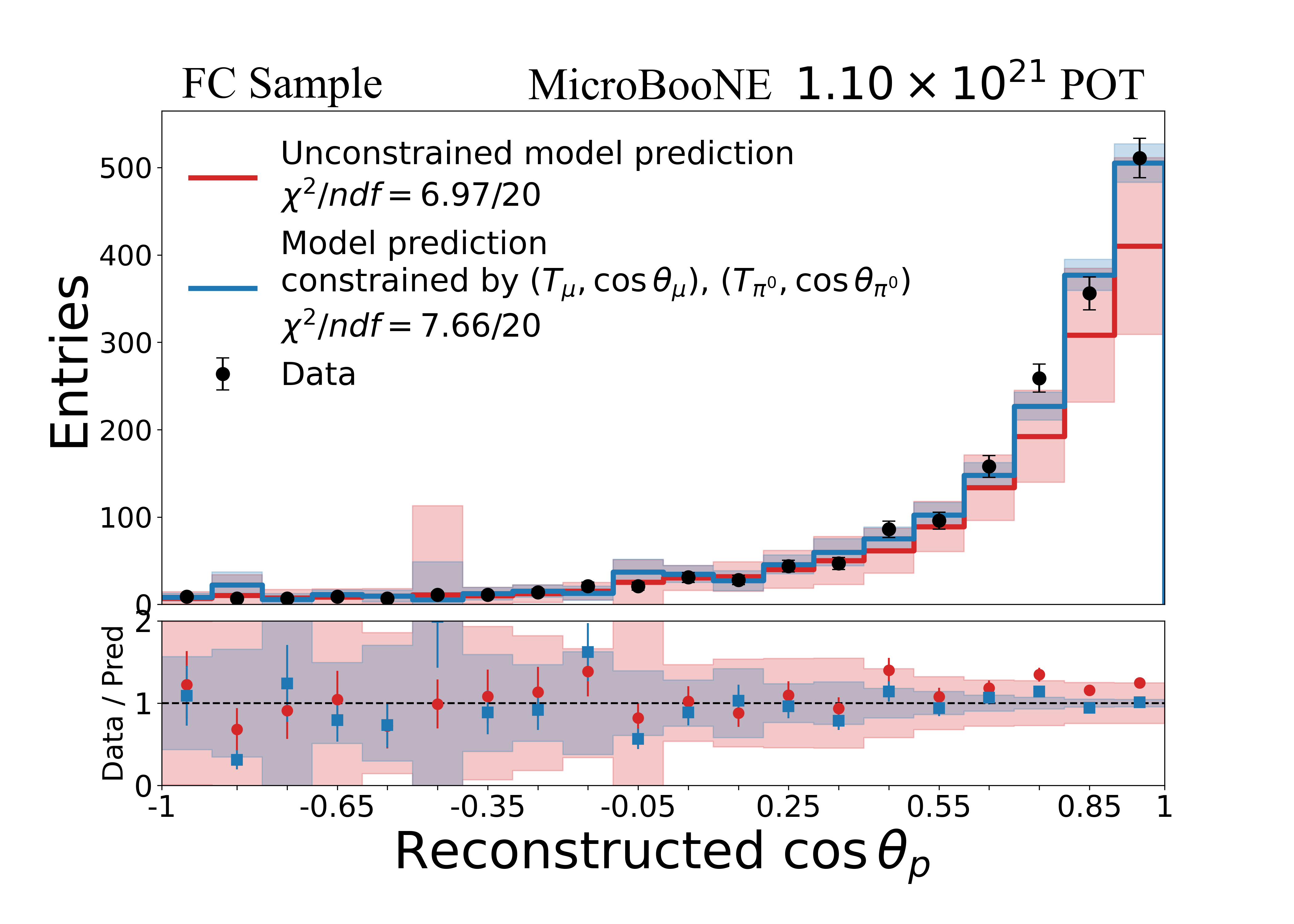}
    \paneltitle{(f) $\cos\theta_{p}$}
\end{subfigure}

\caption{
Reconstructed distributions of basic kinematic observables for FC events: 
(a) muon kinetic energy, (b) muon angle, 
(c) $\pi^0$ kinetic energy, (d) $\pi^0$ angle, 
(e) proton kinetic energy, and (f) proton angle. 
The nominal model prediction is shown in red and the constrained model prediction in blue. 
The constraint is applied using the reconstructed distributions measured in data for each observable, as described in Section~\ref{sec:model_val}. 
The shaded bands represent the combined model systematic and MC statistical uncertainties, while the data error bars indicate statistical uncertainties. Bins with zero or one predicted event are excluded from the $\chi^2/\mathrm{ndf}$ calculation.
}
\label{fig:fc_basic_rd}
\end{figure*}

\begin{table*}[!htbp]
\centering
\caption{Summary of the $\chi^2$ values for the FC and PC distributions for unconstrained (Unconstr.) and constrained (Constr.) model predictions. 
Bins with zero or one predicted event are excluded from the $\chi^2/\mathrm{ndf}$ calculation. 
$P$-values corresponding to each $\chi^2$ and $\mathrm{ndf}$ are shown in parentheses.}
\renewcommand{\arraystretch}{1.02}
\resizebox{\textwidth}{!}{
\begin{tabular}{|c|c|c|cc|c|cc|}
\hline\hline
& & \multicolumn{3}{c|}{\text{FC}} & \multicolumn{3}{c|}{\text{PC}} \\
\cline{3-8}
\text{Observable} & \text{Constraint}
& $\mathrm{ndf}$ & \text{Unconstr.} & \text{Constr.}
& \text{ndf} & \text{Unconstr.} & \text{Constr.} \\
\hline

$T_\mu$  & $\cos\theta_\mu$
& 21 & 12.33 (0.930) & 12.89 (0.912)
& 21 & 5.35 (1.000) & 6.06 (0.999) \\
\hline

$\cos\theta_\mu$ & $T_\mu$
& 20 & 18.02 (0.586) & 18.63 (0.546)
& 20 & 8.05 (0.992) & 8.71 (0.986) \\
\hline

$T_{\pi^0}$  & ($T_\mu$,$\cos\theta_\mu)$,($T_p$,$\cos\theta_p)$
& 20 & 12.95 (0.880) & 14.95 (0.779)
& 20 & 11.17 (0.942) & 15.06 (0.773) \\
\hline

$\cos\theta_{\pi^0}$ & ($T_\mu$,$\cos\theta_\mu)$,($T_p$,$\cos\theta_p)$
& 20 & 5.76 (0.999) & 6.31 (0.998)
& 20 & 8.83 (0.985) & 11.59 (0.930) \\
\hline

$T_p$  & ($T_{\mu},\cos\theta_\mu),$($T_{\pi^0},\cos\theta_{\pi^0})$
& 13 & 1.98 (1.000) & 5.79 (0.953)
& 13 & 3.37 (0.996) & 3.75 (0.994) \\
\hline

$\cos\theta_p$ & ($T_{\mu},\cos\theta_\mu),$($T_{\pi^0},\cos\theta_{\pi^0})$
& 20 & 6.97 (0.997) & 7.66 (0.994)
& 20 & 3.43 (1.000) & 3.48 (1.000) \\
\hline

$\delta p_T$  & ($T_{\mu},\cos\theta_\mu)$
& 21 & 17.68 (0.669) & 16.32 (0.752)
& 21 & 21.75 (0.414) & 20.88 (0.466) \\
\hline

$\delta \alpha_T$  & ($T_{\mu},\cos\theta_\mu)$
& 20 & 13.15 (0.871) & 11.01 (0.946)
& 20 & 12.73 (0.889) & 14.30 (0.815) \\
\hline

$\delta p_T^x$  & ($T_{\mu},\cos\theta_\mu)$
& 17 & 10.55 (0.879) & 7.23 (0.980)
& 16 & 5.83 (0.990) & 5.22 (0.995) \\
\hline

$\delta p_T^y$  & ($T_{\mu},\cos\theta_\mu)$
& 18 & 2.87 (1.000) & 3.21 (1.000)
& 17 & 8.04 (0.966) & 8.28 (0.960) \\
\hline\hline
\end{tabular}
}
\label{tab:gof_rd_sum_prd}
\end{table*}

\label{sec:model_val}

\section{Results}
The unfolded differential cross sections for the ten observables are presented in this section. Each distribution is unfolded in parallel, while correlations between observables are propagated from reconstruction to truth space through the blockwise framework described previously~\cite{blockwise}. The unfolded results in the regularized phase space are compared to predictions from several neutrino interaction generators. These include \texttt{GENIE v3.0.6} with the MicroBooNE tune (hereafter the \texttt{G18T} model), the \texttt{GENIE AR23\_20i\_00\_0000} DUNE baseline configuration (hereafter the \texttt{AR23} model), \texttt{NuWro 25.11}, \texttt{NEUT v5.4.0.1}, and \texttt{GiBUU 2025}. All generator predictions are processed within the \texttt{NUISANCE} framework~\cite{NUISANCE}.

The \texttt{G18T} model corresponds to \texttt{GENIE v3.0.6} with the MicroBooNE generator tune described in Ref.~\cite{uboone_tune}. This configuration employs a local Fermi gas nuclear model with quasielastic scattering based on the Nieves formalism, including 2p2h contributions and random-phase approximation (RPA) corrections~\cite{nieves,nieves_2p2h}. Resonant pion production in the $\Delta(1232)$ region is modeled using the Berger--Sehgal formalism~\cite{BergerSehgal2007}, while nonresonant background processes follow the Bodek--Yang model~\cite{Yang_2009}. Final-state interactions are modeled using the empirical \texttt{hA} cascade model~\cite{hA2018}. The corresponding FSI reweighting procedure, described in Appendix~\ref{appendix:fsi_rw}, is applied throughout this analysis.

The \texttt{AR23} model, corresponding to the DUNE baseline configuration of \texttt{GENIE}, is included in the comparison to the data. This configuration employs a local Fermi gas nuclear model with a spectral-function–like treatment of the initial nuclear state~\cite{dune_model_SP} and uses the SuSAv2 model for the 2p2h contribution~\cite{SuSAv2}. The nominal \texttt{AR23} configuration uses the \texttt{hA2018} model for final-state interactions. In this work, it is updated to use the corrected \texttt{hA2025} model described in the Supplemental Material~\cite{supp}, without applying any reweighting.

The \texttt{NuWro 25.11} generator~\cite{nuwro25_11} employs a local Fermi gas nuclear model with quasielastic scattering based on the Nieves formalism, including 2p2h contributions and RPA corrections~\cite{nieves,nieves_2p2h}. In the $\Delta(1232)$ region, \texttt{NuWro} replaces its previous $\Delta$-dominance model with the Ghent hybrid model, which provides a unified description of single-pion production across resonance and higher invariant-mass regions~\cite{nuwro_updates}. Nonresonant contributions are modeled using the Bodek--Yang approach~\cite{Yang_2009}, while the hybrid framework ensures a more consistent transition between resonance and DIS regimes. Final-state interactions are simulated using an intranuclear cascade model based on the Salcedo--Oset framework~\cite{Salcedo}.

The \texttt{NEUT v5.4.0.1} generator~\cite{Neut,Hayato2021NEUT} also employs a local Fermi gas nuclear model with the Nieves quasielastic formalism, including 2p2h processes and RPA corrections~\cite{nieves,nieves_2p2h}. Resonant pion production is modeled using the Berger--Sehgal approach~\cite{BergerSehgal2007}, and nonresonant background contributions follow the Bodek--Yang parameterization~\cite{Yang_2009}. The transition between resonance and DIS regimes is handled through a dedicated invariant-mass transition region. Final-state interactions are simulated using an intranuclear cascade model based on the Salcedo--Oset framework~\cite{Salcedo}.

\begin{figure*}[!htbp]
\centering
\setlength{\tabcolsep}{4pt}

\begin{subfigure}{0.46\textwidth}
  \centering
  \includegraphics[width=\linewidth]{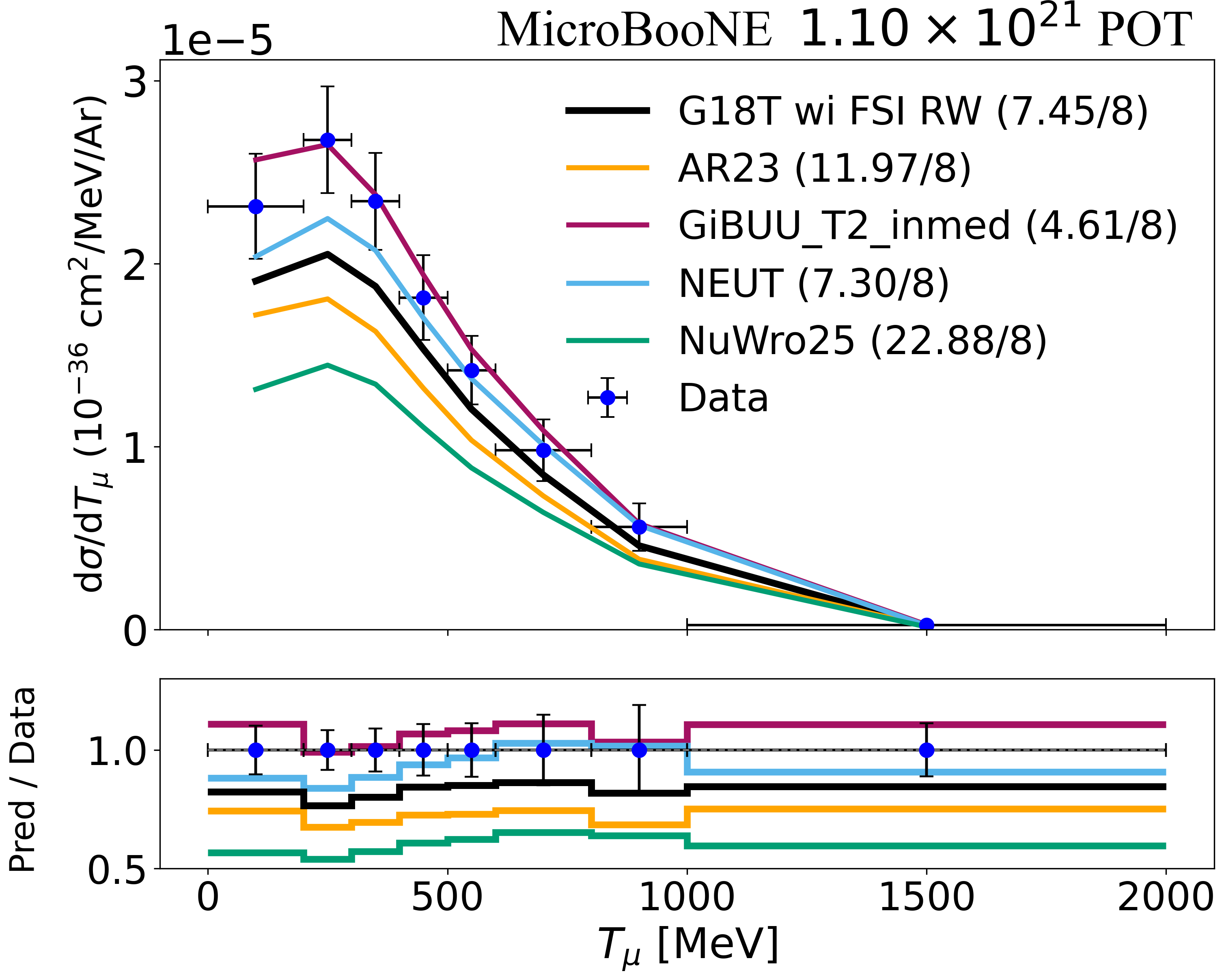}
  \paneltitlex[0.4cm]{(a) $T_\mu$}
\end{subfigure}
\begin{subfigure}{0.46\textwidth}
  \centering
  \includegraphics[width=\linewidth]{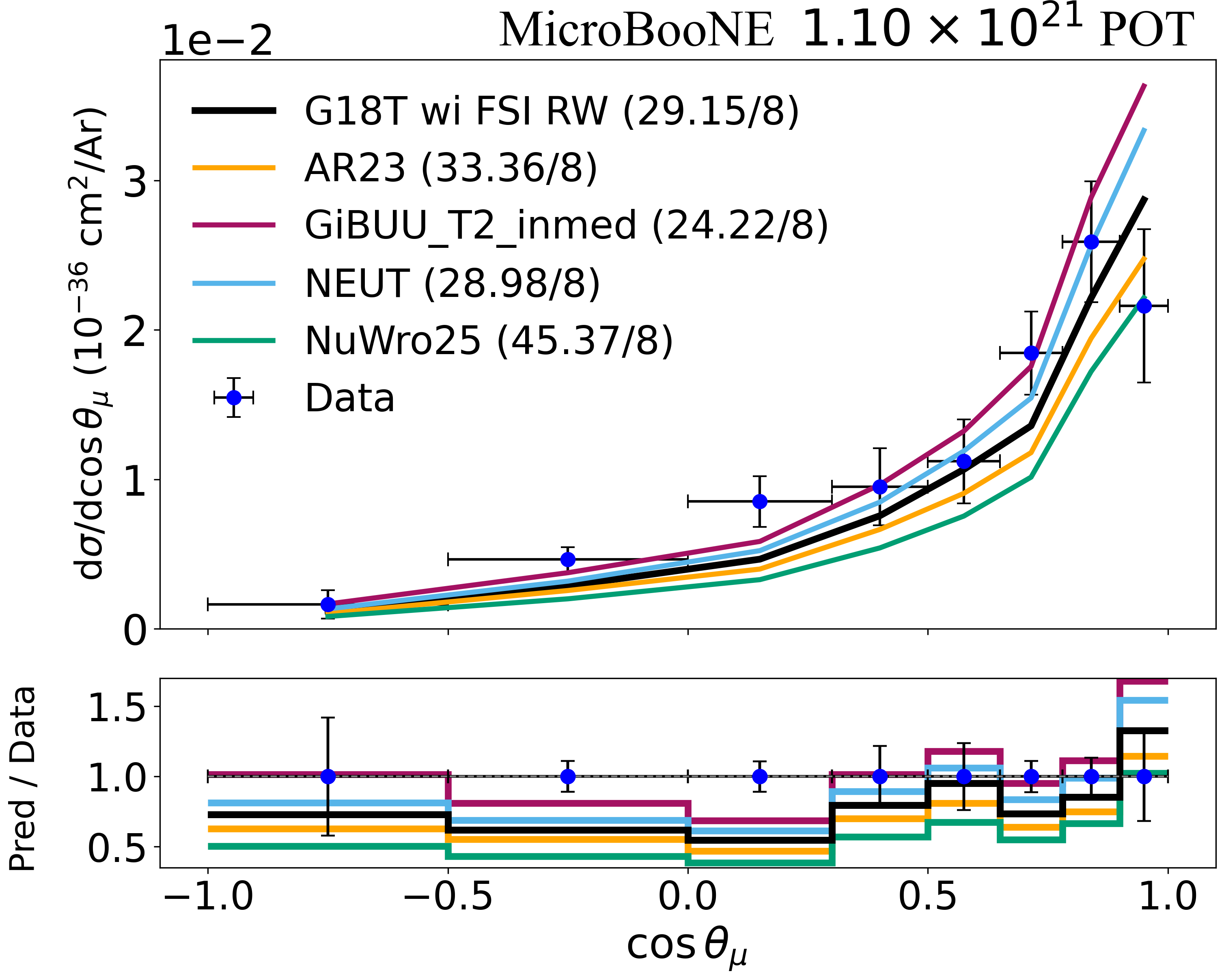}
  \paneltitlex[1.1cm]{(b) $\cos\theta_\mu$}
\end{subfigure}

\begin{subfigure}{0.46\textwidth}
  \centering
  \includegraphics[width=\linewidth]{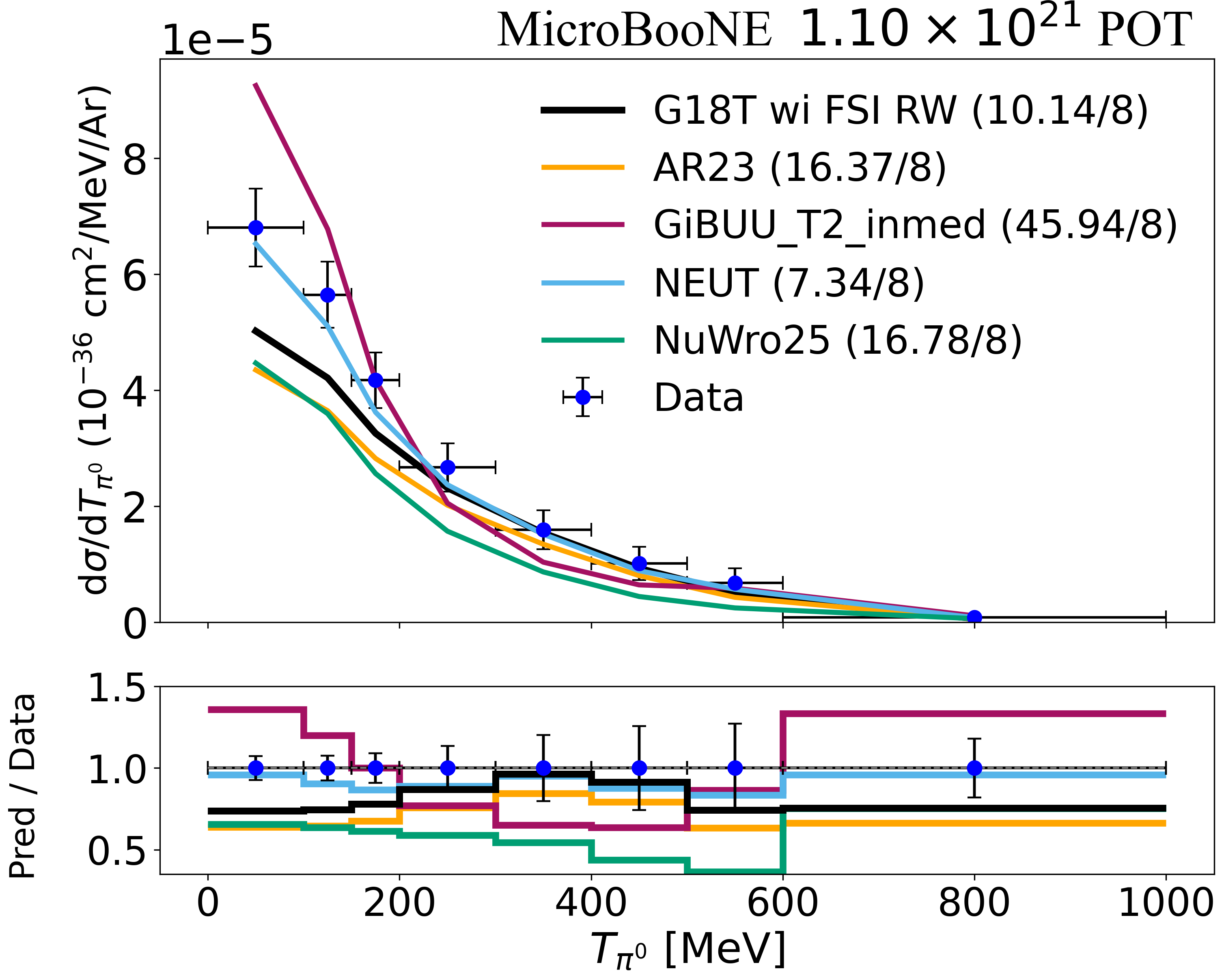}
  \paneltitlex[0.4cm]{(c) $T_{\pi^0}$}
\end{subfigure}
\begin{subfigure}{0.46\textwidth}
  \centering
  \includegraphics[width=\linewidth]{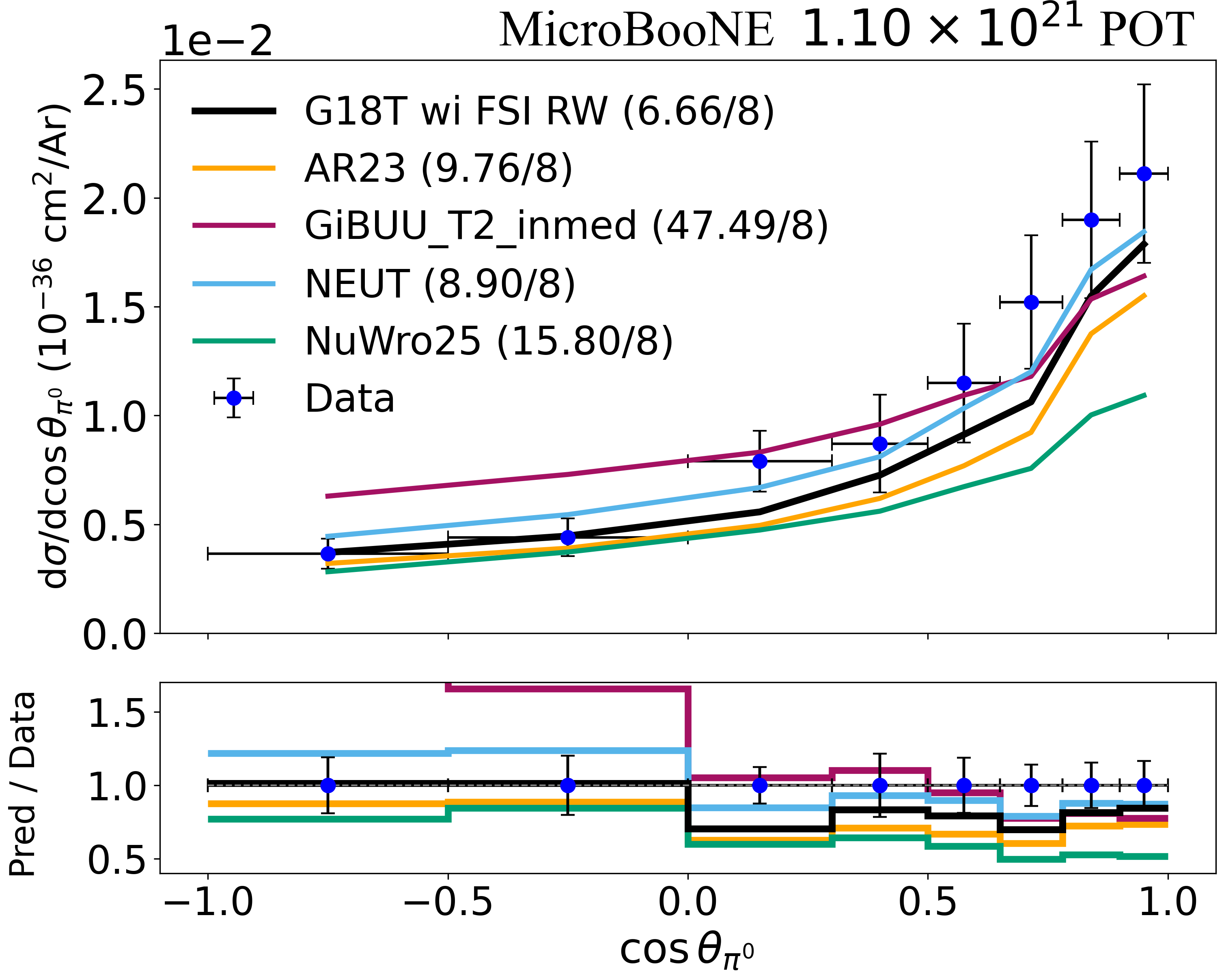}
  \paneltitlex[1.1cm]{(d) $\cos\theta_{\pi^0}$}
\end{subfigure}

\begin{subfigure}{0.46\textwidth}
  \centering
  \includegraphics[width=\linewidth]{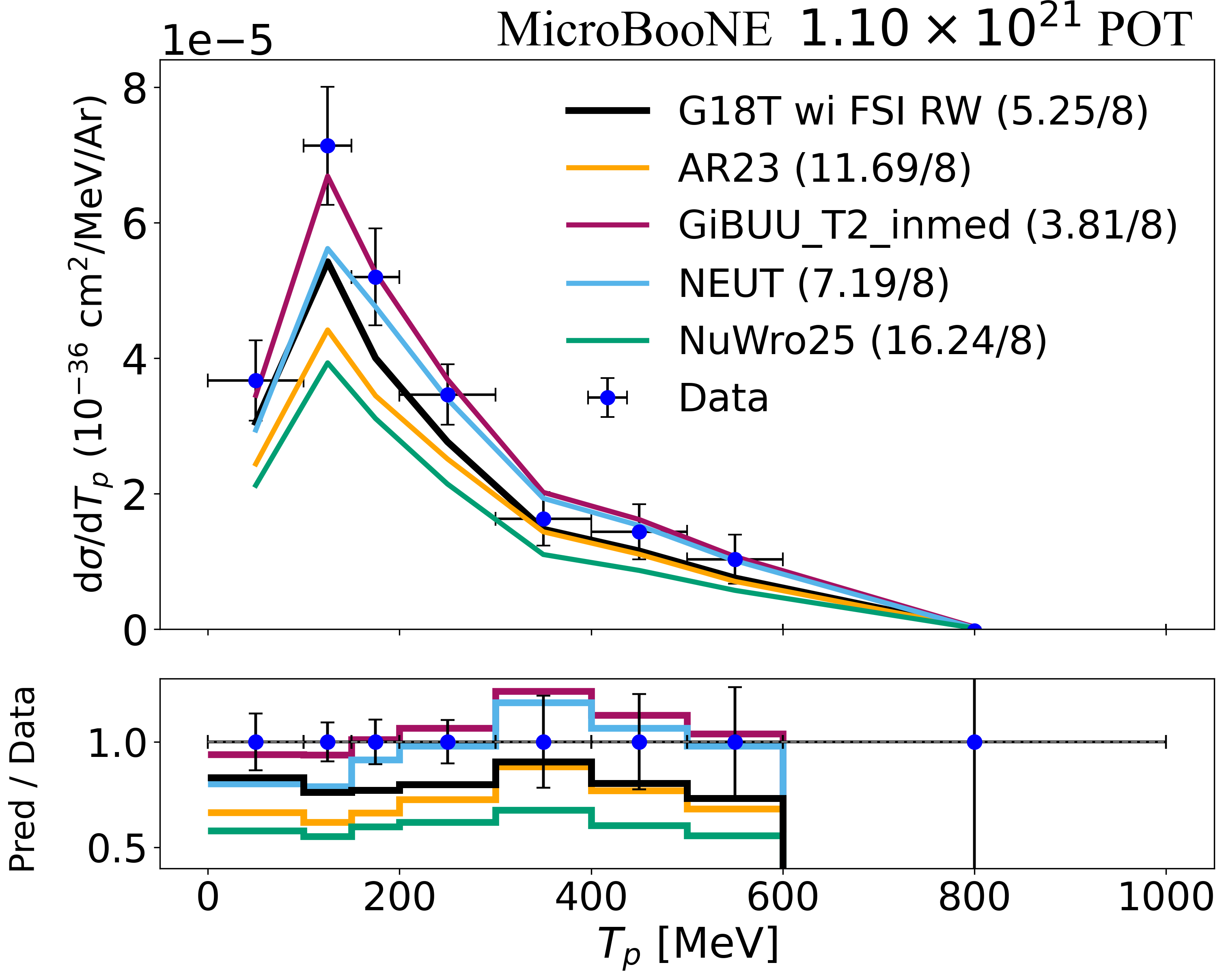}
  \paneltitlex[0.4cm]{(e) $T_p$}
\end{subfigure}
\begin{subfigure}{0.46\textwidth}
  \centering
  \includegraphics[width=\linewidth]{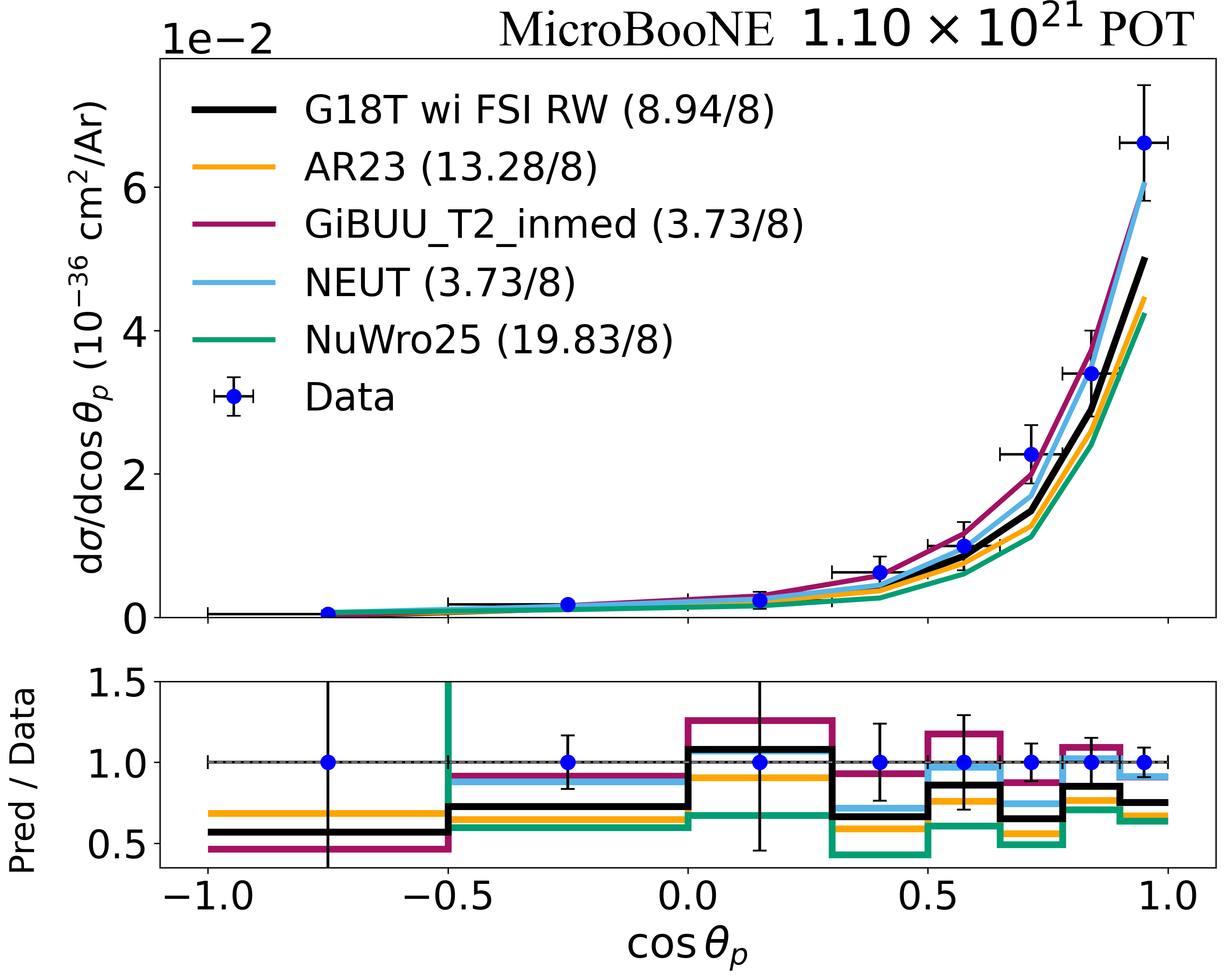}
  \paneltitlex[1.1cm]{(f) $\cos\theta_p$}
\end{subfigure}
\caption{
\hyphenpenalty=10000
\exhyphenpenalty=10000
Final unfolded differential cross sections for the $T$ and $\cos\theta$ observables: muon ((a)--(b)), $\pi^0$ ((c)--(d)), and proton ((e)--(f)). 
The error bars represent the total post-unfolding uncertainty, including statistical and systematic contributions propagated through the full covariance matrix. 
The displayed generator predictions are smeared with the regularization matrix obtained during the unfolding procedure. 
The corresponding $\chi^2$/ndf values, computed using the full covariance matrix, are reported in the legend of each panel.
}
\label{fig:final_unfolded_basic}
\end{figure*}

\begin{figure*}[!htbp]
\centering
\setlength{\tabcolsep}{4pt}

\begin{subfigure}{0.46\textwidth}
  \centering
  \includegraphics[width=\linewidth]{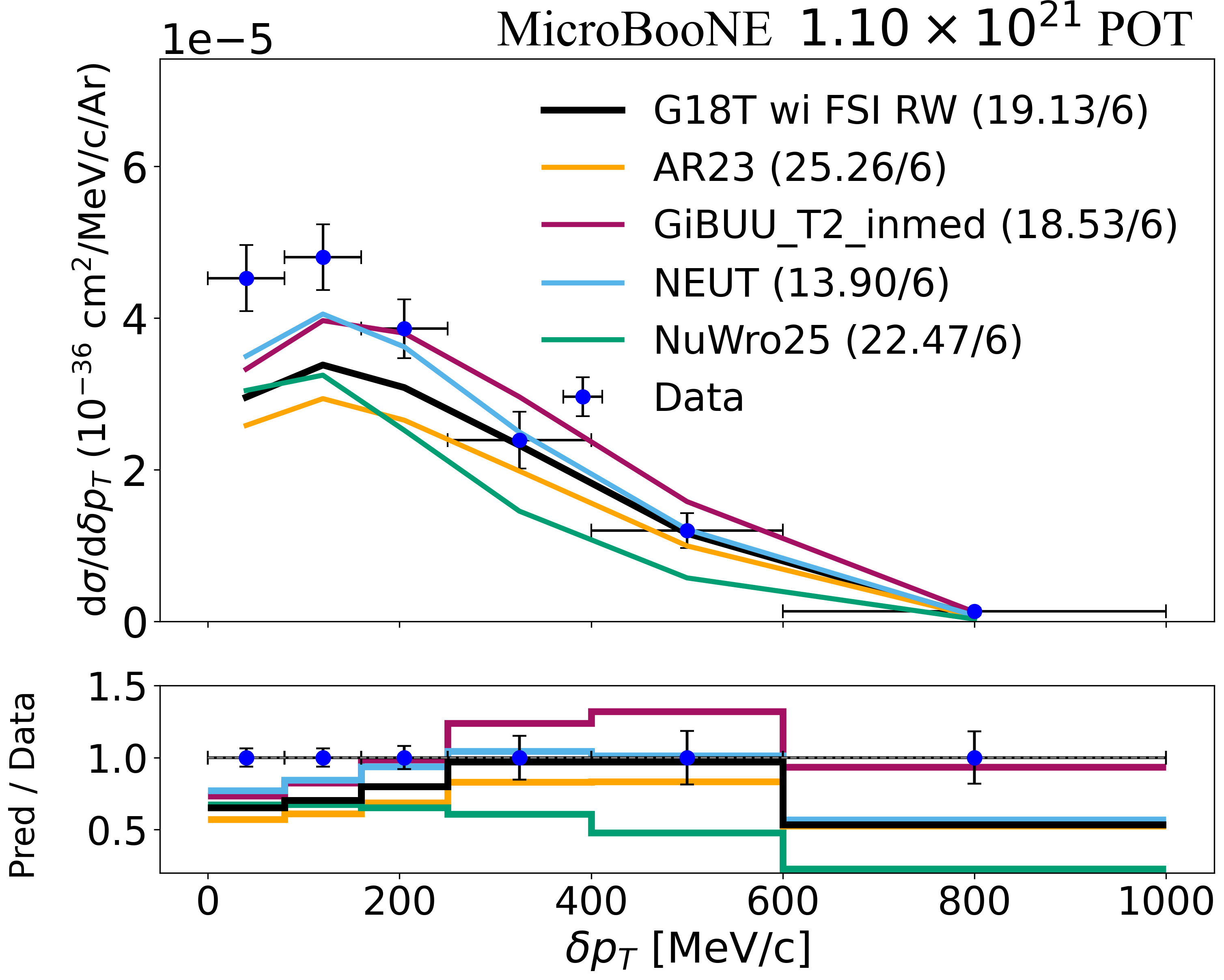}
  \paneltitle{(a) $\delta p_T$}
\end{subfigure}
\begin{subfigure}{0.46\textwidth}
  \centering
  \includegraphics[width=\linewidth]{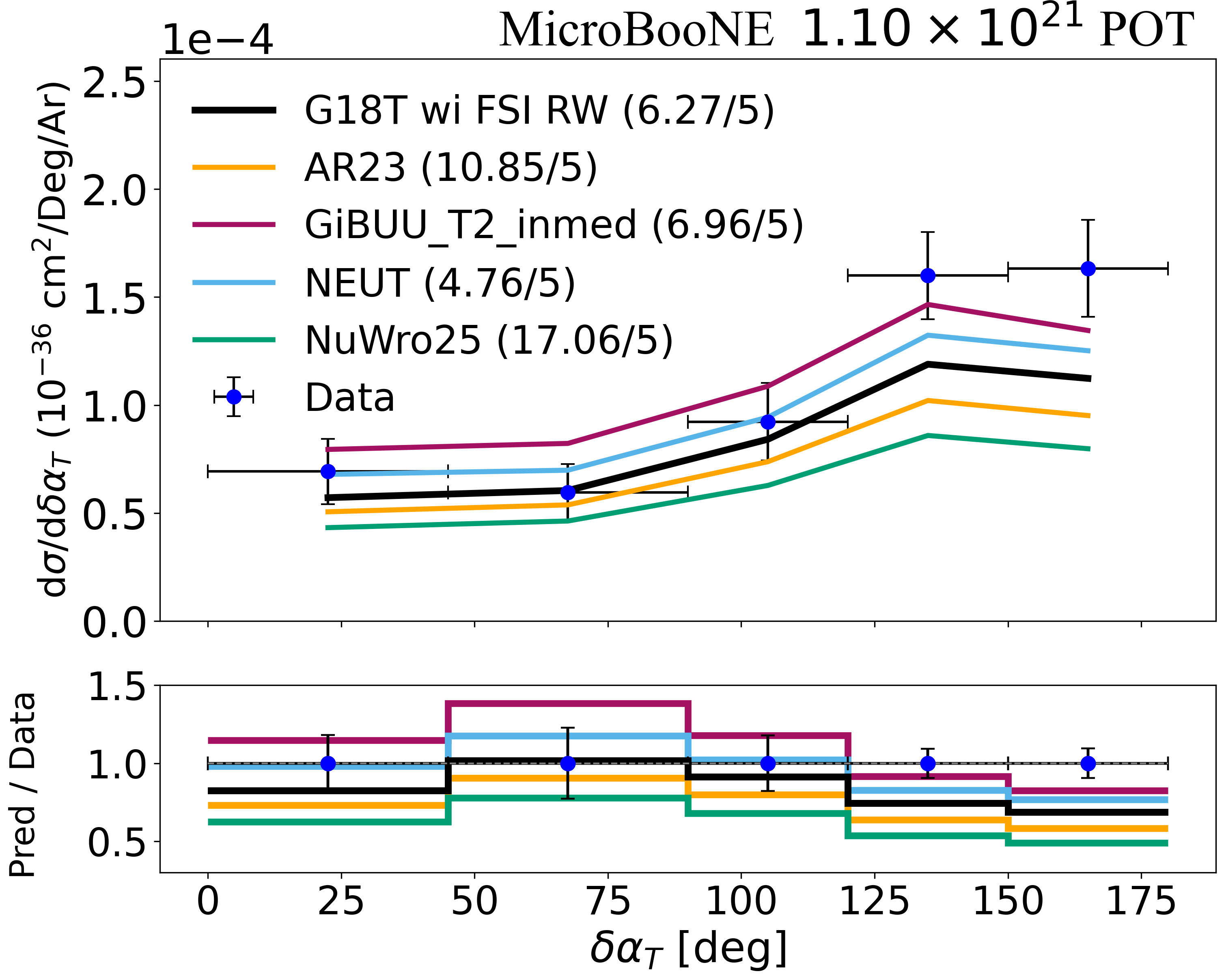}
  \paneltitlex[0.9cm]{(b) $\delta \alpha_T$}
\end{subfigure}

\begin{subfigure}{0.46\textwidth}
  \centering
  \includegraphics[width=\linewidth]{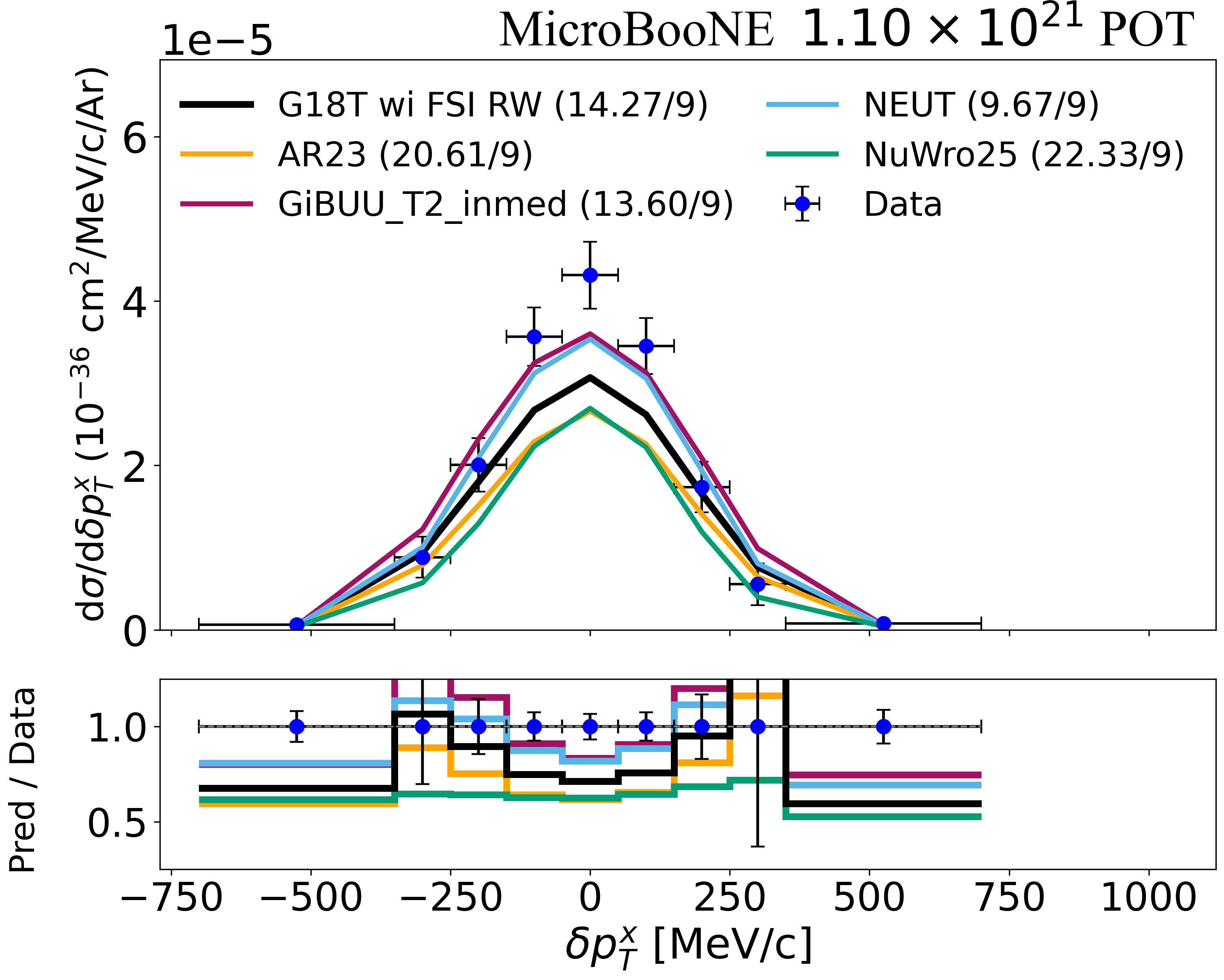}
  \paneltitle{(c) $\delta p_T^{x}$}
\end{subfigure}
\begin{subfigure}{0.46\textwidth}
  \centering
  \includegraphics[width=\linewidth]{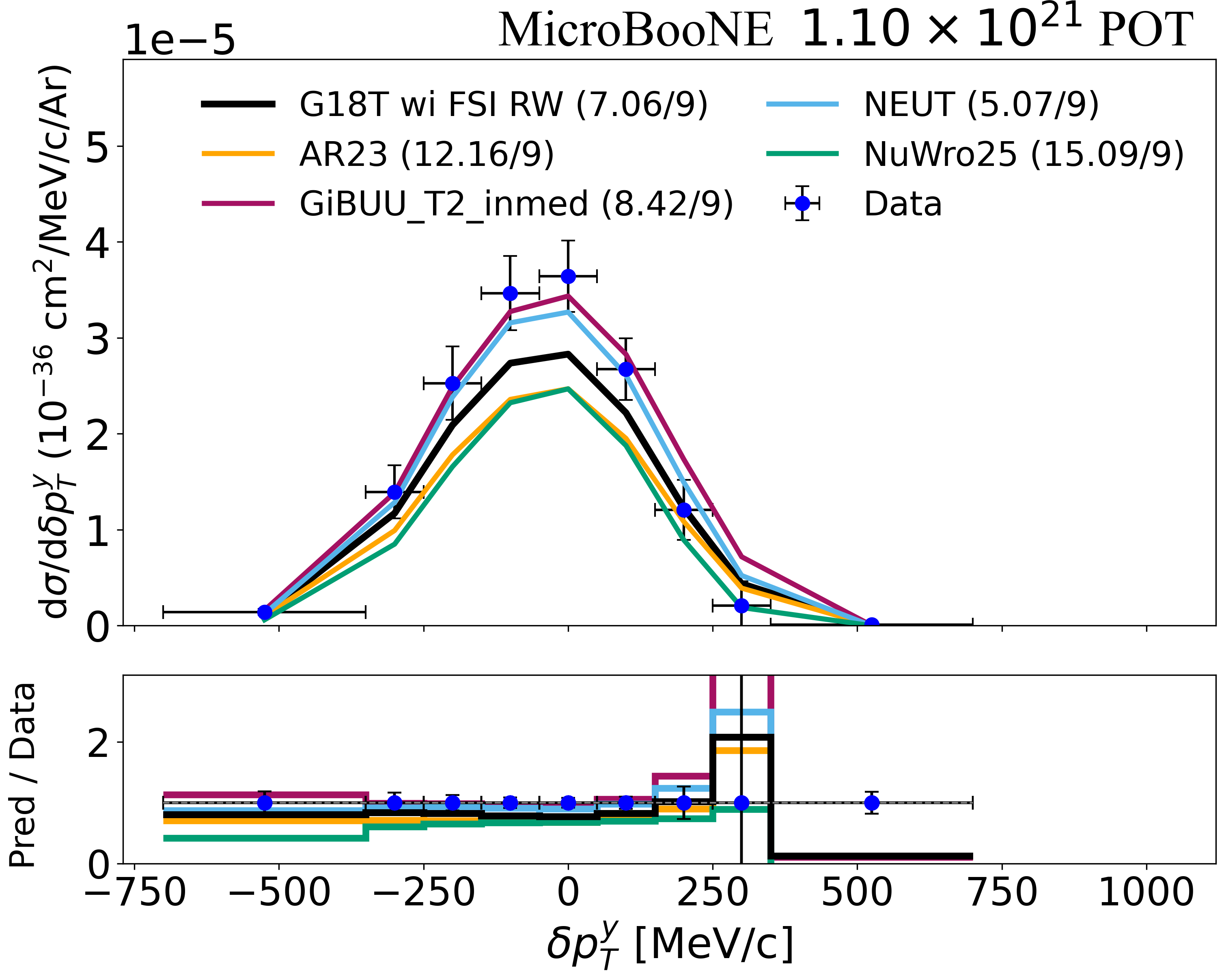}
  \paneltitlex[0.9cm]{(d) $\delta p_T^{y}$}
\end{subfigure}

\caption{%
\hyphenpenalty=10000
\exhyphenpenalty=10000
Final unfolded differential cross sections for transverse kinematic imbalance observables. 
The error bars represent the total post-unfolding uncertainty, including statistical and systematic contributions propagated through the full covariance matrix. 
The generator predictions are smeared with the regularization matrix obtained during the unfolding procedure, and $\chi^2$/ndf values are reported in each panel legend. 
For the $\delta p_T^y$ observable, the \texttt{NuWro} prediction in the final bin becomes negative after applying the regularization matrix, reflecting negative entries in that matrix.
}
\label{fig:final_unfolded_tki}
\end{figure*}

The \texttt{GiBUU 2025} framework~\cite{gibuu_fw,gibuu_pi0} employs a unified transport model that treats the primary neutrino–nucleon interaction and subsequent final-state interactions consistently within a common nuclear potential. The model includes quasielastic scattering, resonance production, 2p2h processes, and DIS, with nonresonant contributions constrained by electron–nucleus scattering data. The \texttt{GiBUU\_T2\_inmed} configuration is used in this analysis. This configuration applies an isospin scaling factor of two to nonresonant pion production and meson-exchange current 2p2h processes, and additionally includes in-medium modifications to the nucleon--nucleon cross sections. This model configuration is adopted because it yields improved agreement with the measured distributions relative to the default \texttt{GiBUU} prediction. Similar behavior was previously observed in comparisons to earlier pion production MicroBooNE measurements, as discussed in Ref.~\cite{gibuu_pi0}.
Figures~\ref{fig:final_unfolded_basic}--\ref{fig:final_unfolded_tki} present the unfolded differential cross sections for all measured observables, while Table~\ref{tab:chi2_summary_unfolded_xs} summarizes the corresponding $\chi^2/\mathrm{ndf}$ values quantifying the agreement between the measurements and generator predictions. The pion production models implemented in \texttt{G18T} and \texttt{AR23} are similar, leading to correspondingly similar levels of agreement with the data.

\begin{table*}[!htbp]
\centering
\renewcommand{\arraystretch}{1.}
\caption{
Summary of the $\chi^2$ values obtained from comparisons of the unfolded 
differential cross-section distributions with predictions from different 
neutrino interaction generators: \texttt{G18T} (with FSI reweighting), \texttt{AR23}, \texttt{NEUT}, 
\texttt{NuWro25.11}, and \texttt{GiBUU\_T2\_inmed}. 
The $\chi^2$ values are calculated using the block covariance matrix for each observable, incorporating both statistical and systematic uncertainties. The number of degrees of freedom is shown explicitly in the second column. 
}
\renewcommand{\arraystretch}{1.02}
\begin{tabular}{|c|c|c|c|c|c|c|}
\hline\hline

\text{Observable} 
& \text{ndf}
& \texttt{G18T (FSI RW)} 
& \texttt{AR23}
& \texttt{NEUT} 
& \texttt{NuWro25.11} 
& \texttt{GiBUU\_T2\_inmed} \\
\hline

$T_\mu$ 
& 8 & 7.45 & 11.97 & 7.30 & 22.88 & 4.61 \\
\hline

$\cos\theta_\mu$
& 8 & 29.15 & 33.36 & 28.98 & 45.37 & 24.22 \\
\hline

$T_{\pi^0}$ 
& 8 & 10.14 & 16.37 & 7.34 & 16.78 & 45.94 \\
\hline

$\cos\theta_{\pi^0}$
& 8 & 6.66 & 9.76 & 8.90 & 15.80 & 47.49 \\
\hline

$T_p$ 
& 8 & 5.25 & 11.69 & 7.19 & 16.24 & 3.81 \\
\hline

$\cos\theta_p$
& 8 & 8.94 & 13.28 & 3.73 & 19.83 & 3.73 \\
\hline

$\delta p_T$ 
& 6 & 19.13 & 25.26 & 13.90 & 22.47 & 18.53 \\
\hline

$\delta p_T^x$ 
& 9 & 14.27 & 20.61 & 9.67 & 22.33 & 13.60 \\
\hline

$\delta p_T^y$ 
& 9 & 7.06 & 12.16 & 5.07 & 15.09 & 8.42 \\
\hline

$\delta \alpha_T$ 
& 5 & 6.27 & 10.85 & 4.76 & 17.06 & 6.96 \\
\hline\hline
\end{tabular}
\label{tab:chi2_summary_unfolded_xs}
\end{table*}

The kinetic energy distributions of the final-state particles, shown in panels (a), (c), and (e) of Fig.~\ref{fig:final_unfolded_basic}, are generally well reproduced by the generators in terms of their overall shapes. Among the models considered, \texttt{NEUT} provides the best overall agreement with the data, yielding $\chi^2/\mathrm{ndf} = 7.30/8$ for $T_\mu$, $7.34/8$ for $T_{\pi^0}$, and $7.19/8$ for $T_p$. In contrast, the \texttt{GiBUU\_T2\_inmed} prediction exhibits substantial disagreement in the $\pi^0$ observables, with $\chi^2/\mathrm{ndf} = 45.94/8$ for $T_{\pi^0}$ and $47.49/8$ for $\cos\theta_{\pi^0}$.

The differing levels of agreement among the final-state particles reflect the distinct physical processes governing their kinematics. The muon kinetic energy is primarily sensitive to the dynamics at the neutrino interaction vertex, including the modeling of the underlying neutrino--nucleon interaction and parameters such as the resonant axial mass, $M_A^{\mathrm{res}}$. In contrast, the kinematics of the outgoing hadrons are additionally modified by intranuclear effects as the particles propagate through the nuclear medium. These observables are influenced by a combination of factors, including the production and decay of intermediate baryon resonances, nuclear-medium effects, and the treatment of final-state interactions, whose interplay can differ significantly among generators. The disagreement observed for the \texttt{GiBUU\_T2\_inmed} prediction in the $\pi^0$ observables may therefore indicate differences in the modeling of hadronic transport and reinteraction effects, while also reflecting the broader uncertainties associated with neutrino--nucleus interaction modeling in the resonance region.

This behavior is particularly noteworthy in the context of the discussion presented in Ref.~\cite{gibuu_pi0}, which highlighted the importance of a \mbox{MicroBooNE} pion TKI measurement for resolving the differing in-medium correction preferences inferred from \mbox{MINERvA} and previous \mbox{MicroBooNE} pion production measurements. While the \mbox{MINERvA} CC pion data favor relatively weak in-medium modifications, earlier \mbox{MicroBooNE} CC $\pi^0$ and NC $\pi^0$ measurements showed improved agreement with configurations incorporating stronger in-medium effects. In the present analysis, most measured cross sections similarly favor the inclusion of in-medium corrections, consistent with the earlier MicroBooNE results. However, significant shape discrepancies remain in the reconstructed $\pi^0$ kinematic distributions, where the \texttt{GiBUU\_T2\_inmed} prediction fails to accurately reproduce the data. The persistence of these discrepancies, also observed in previous MicroBooNE pion measurements, points to unresolved tensions in the current description of pion propagation through the nuclear medium.

The angular observables $\cos\theta_\mu$, $\cos\theta_{\pi^0}$, and $\cos\theta_p$, shown in panels (b), (d), and (f) of Fig.~\ref{fig:final_unfolded_basic}, exhibit more pronounced shape discrepancies between the data and generator predictions. Model predictions over $\cos\theta_{\mu}$ exhibit model excess in the forward-angle region. All generators show a significant shape discrepancy with the data in this region of phase space, resulting in relatively large $(\chi^2/\mathrm{ndf})$ values. This disagreement may be associated with the modeling of Pauli blocking effects, which suppress interactions involving low-momentum nucleons and thereby influence the predicted rate of strongly forward-going muons.

In contrast, the $\pi^0$ and proton angular distributions display discrepancies in the opposite direction, with all generators underestimating the data at forward angles. The differing behavior between the leptonic and hadronic observables further illustrates their sensitivity to distinct components of the interaction model. While the muon angle primarily reflects the kinematics of the initial neutrino--nucleon scattering process, the hadron angular distributions are strongly influenced by FSI within the nuclear medium. The observed discrepancies therefore point to potential shortcomings in the modeling of nuclear transport processes, including nucleon rescattering, pion absorption, and CEX interactions, which can significantly alter the directions and energies of the outgoing hadrons before they exit the nucleus.

Additional insight into nuclear effects is provided by the TKI observables shown in Fig.~\ref{fig:final_unfolded_tki}. These observables are designed to isolate deviations from simple quasifree neutrino--nucleon scattering arising from nuclear motion and final-state interactions. In the absence of nuclear effects, the transverse momenta of the final-state particles would balance exactly, resulting in $\delta p_T = 0$. Nuclear effects such as Fermi motion, multinucleon interactions, and hadronic reinteractions within the nucleus broaden the transverse momentum imbalance distributions.

The measured $\delta p_T$ distribution indicates that the data favor a smaller transverse momentum imbalance than predicted by most generators. Consistent behavior is observed in the $\delta p_T^x$ and $\delta p_T^y$ distributions, which represent the components of the transverse momentum imbalance parallel and perpendicular to the outgoing muon transverse direction, respectively. In both observables, the data exhibit a narrower peak near zero than predicted by the generators considered. Together, these results suggest that the overall level of transverse momentum broadening in the nuclear medium may be overestimated in the generators considered. This discrepancy could originate from an overly strong treatment of FSI, differences in the modeling of the initial-state nucleon momentum distribution, or inaccuracies in the production and propagation of secondary hadrons within the nucleus.

The transverse opening angle $\delta\alpha_T$ provides complementary information by probing the direction of the transverse imbalance vector relative to the outgoing lepton transverse momentum. Values near $\delta\alpha_T \approx180^\circ$ correspond to configurations in which the residual transverse momentum is oriented opposite to the outgoing muon direction, consistent with a reduction of the hadronic transverse momentum relative to the lepton system. In this region, all generators underestimate the data. Such configurations are commonly associated with a ``deceleration'' of the hadronic system arising from FSI processes including nucleon rescattering, pion absorption, and CEX interactions.

The combined behavior of the $\delta p_T$-based observables and the $\delta\alpha_T$ distribution suggests that the disagreement between the data and generators cannot be interpreted solely as an overall under- or overestimation of FSI strength. While the data favor reduced broadening in the magnitude of the transverse momentum imbalance, they simultaneously exhibit a stronger preference for the imbalance vector to be oriented opposite to the outgoing muon direction. This indicates that the generators may inadequately model the correlation between the magnitude and direction of the transverse momentum imbalance, pointing to deficiencies in the current description of hadron transport and reinteraction dynamics within the nuclear medium.

\texttt{NEUT} provides the most consistent description of the individual measurements, exhibiting relatively low ($\chi^2/\mathrm{ndf}$) values across a range of kinematic distributions. When the full covariance matrix is used to account for correlations among the measurements, \texttt{GENIE} yields the best overall agreement with the data. The difference between \texttt{GENIE} and \texttt{NEUT} is modest, amounting to only six units in ($\chi^2$). In contrast, both generators provide a substantially better description of the data than \texttt{AR23} and \texttt{GiBUU}. The resulting ($\chi^2/\mathrm{ndf}$) values are listed in Table~\ref{tab:chi2_overall_ndf}.

None of the generators achieves uniformly good agreement across all measured observables, indicating that significant modeling uncertainties remain in the description of neutrino--nucleus interactions in this channel.

\begin{table}[!htbp]
\centering
\begin{tabular}{l c}
\hline
Model & $\chi^2$ \\
\hline
\texttt{G18T (FSI RW)}  & 84.14    \\
\texttt{NEUT}    & 89.99  \\
\texttt{NuWro25.11} & 92.32  \\
\texttt{\mbox{GiBUU\_T2\_inmed}}    & 163.71  \\
\texttt{AR23}    & 97.03  \\
\hline
\end{tabular}
\caption{Total $\chi^2$ values for the 10 observables using the full covariance matrix for each generator model ($\mathrm{ndf}=77$).}
\label{tab:chi2_overall_ndf}
\end{table}

Overall, the combined behavior of the angular observables and the TKI variables highlights the sensitivity of this measurement to both the primary neutrino interaction mechanism and subsequent nuclear transport processes. These results indicate that improved modeling of hadronic re-scattering and nuclear effects is required to achieve a consistent description of all observables.

\section{Conclusion}\label{sec:Conclusion}
We report a measurement of muon-neutrino CC resonance-like interactions on argon using the MicroBooNE detector. The analyzed final state comprises a muon, a single $\pi^0$, and one or more protons with a kinetic energy threshold of 60\,MeV. The analysis reports ten single-differential cross sections in final-state kinematic and transverse kinematic imbalance observables, extracted from the full MicroBooNE exposure of $1.10 \times 10^{21}$ POT. Pion production in this channel is inherently complex due to the interplay of intermediate baryon resonances and the multi-particle final state. This work presents a detailed characterization of the $\mu + \pi^0 +$ leading-proton final state.

This analysis improves upon previous measurements through several key developments, including the use of the full dataset, enhanced event selection with improved efficiency and purity, and a refined signal definition requiring at least one proton in the final state. It also incorporates an MCS–based reconstruction for exiting muon momentum and an improved treatment of pion final-state interactions via reweighting of the \texttt{GENIE} \texttt{hA} model. The unfolding procedure employs a blockwise approach that preserves correlations among observables. A dedicated model-validation study, based on reconstructed-level conditional goodness-of-fit tests and validated with fake-data studies, finds no evidence of mismodeling beyond the assigned uncertainties.

The dominant physics processes in this channel are CC excitation of the $\Delta(1232)$ resonance, its decay to a $\pi^{0}p$ final state, and the subsequent hadronic FSI within the nucleus. None of the evaluated generators provides a satisfactory description of all ten measured observables simultaneously. The muon kinematic distributions, which are primarily sensitive to the modeling of resonance production at the interaction vertex, exhibit variations among generators comparable to those observed for the hadronic observables. The hadronic distributions, which are additionally shaped by FSI in the nuclear medium, generally display a similar level of model spread and agreement with the data, with the notable exception of the \texttt{GiBUU} predictions for the $\pi^{0}$ kinematic observables. The TKI observables, which combine sensitivity to both initial-state nuclear dynamics and FSI effects, reinforce these conclusions while providing additional constraints on the treatment of nuclear effects in the generators.

The unfolded cross sections reveal shape discrepancies with respect to predictions from \texttt{GENIE}, \texttt{NuWro}, \texttt{NEUT}, and \texttt{GiBUU}, most prominently in the muon angular distribution. The TKI observables probe transverse momentum imbalance arising from nuclear effects and hadronic re-scattering, and in several regions of phase space the data favor smaller imbalance than predicted by the models. Among the generators considered, \texttt{NEUT} provides the most consistent overall agreement with the data, although no model successfully reproduces all measured distributions simultaneously.

Compared to the MicroBooNE CC $\pi^{0}$ inclusive measurement~\cite{uboone_ccpi0}, which reported only muon and $\pi^{0}$ kinematic distributions, our results show similar features in the muon angular distribution ($\cos\theta_\mu$). In both measurements, the data exhibit a suppression of forward-going muons relative to model predictions, a behavior that may be associated with nuclear effects such as Pauli blocking. For the muon kinetic energy distribution, however, our measurement shows a persistent model underprediction across the full measured range. This behavior is consistent with the previous result below $T_\mu \approx 0.8$ GeV, although we do not observe the model overprediction at higher energies reported in that analysis.

For the $\pi^{0}$ observables, we observe broader discrepancies between data and simulation than previously reported. In particular, the $\pi^{0}$ angular distribution exhibits significant shape disagreement, while the $\pi^{0}$ energy and momentum distributions display trends similar to those seen in the muon kinematics, with the models generally underpredicting the measured cross sections.

Comparison with the MINERvA CC1$\pi^{0}N p$ measurement on a carbon target~\cite{minerva_pi0} reveals both similarities and important differences in the shared observables. The transverse imbalance variable $\delta p_T^x$ (denoted $\delta p_{TT}$ in the MINERvA analysis) is predicted to be broader than observed in data in both measurements, although the discrepancy is substantially larger in the MINERvA result. The $\delta \alpha_T$ distribution highlights a more significant difference between the two analyses. In our measurement, the models underpredict the data at large $\delta \alpha_T$, a region associated with deceleration of the hadronic system in the final state. By contrast, the MINERvA measurement reports generally good agreement between data and simulation in both shape and normalization, with only a mild overall overprediction. Notably, that analysis used an earlier \texttt{GENIE} configuration~\cite{genie_v2}, which may contribute to the observed differences between the two measurements.

This work constitutes the first measurement of resonance-dominated $\nu_{\mu}$ interactions on argon using TKI observables. The detailed characterization of the multi-particle final state and its kinematic correlations provides stringent constraints on models of resonance production and hadronic transport in nuclei. These results will inform ongoing model development and are essential for reducing systematic uncertainties in long-baseline neutrino oscillation experiments. In particular, they provide valuable input for future experiments such as DUNE, where resonance production is a dominant interaction channel.

\section{ACKNOWLEDGEMENTS}
This document was prepared by the MicroBooNE collaboration using the resources of the Fermi National Accelerator Laboratory (Fermilab), a U.S. Department of Energy, Office of Science, Office of High Energy Physics HEP User Facility. Fermilab is managed by Fermi Forward Discovery Group, LLC, acting under Contract No. 89243024CSC000002. MicroBooNE is supported by the
following: 
the U.S. Department of Energy, Office of Science, Offices of High Energy Physics and Nuclear Physics; 
the U.S. National Science Foundation; 
the Swiss National Science Foundation; 
the Science and Technology Facilities Council (STFC), part of United Kingdom Research and Innovation (UKRI);
the Royal Society (United Kingdom);
the UKRI Future Leaders Fellowship;
the NSF AI Institute for Artificial Intelligence and Fundamental Interactions;
and the European Union’s Horizon 2020 research and innovation programme under the Marie Sk\l{}odowska-Curie grant agreement No. 101003460 (PROBES). Additional support for the laser calibration system and cosmic ray tagger was provided by the Albert Einstein Center for Fundamental Physics, Bern, Switzerland. We also acknowledge the contributions of technical and scientific staff to the design, construction, and operation of the MicroBooNE detector as well as the contributions of past collaborators to the development of MicroBooNE analyses, without whom this work would not have been possible. For the purpose of open access, the authors have applied a Creative Commons Attribution (CC BY) public copyright license to any Author Accepted Manuscript version arising from this submission.
\FloatBarrier

\makeatletter
\section*{Appendix A: Genie FSI Reweighting}

The MC sample used in this analysis was generated with the \texttt{GENIE} \texttt{hA2018} FSI model. Subsequent studies identified a deficiency in the treatment of pion--nucleus interactions, arising from inaccuracies in the pion--nucleus cross sections used to determine interaction fate fractions as a function of pion kinetic energy. This results in an overestimation of the pion CEX probability and, consequently, an enhanced production of $\pi^0$ after FSI in \texttt{hA2018}. 

\begin{figure}[!htbp]
    \centering
    \includegraphics[width=1\linewidth]{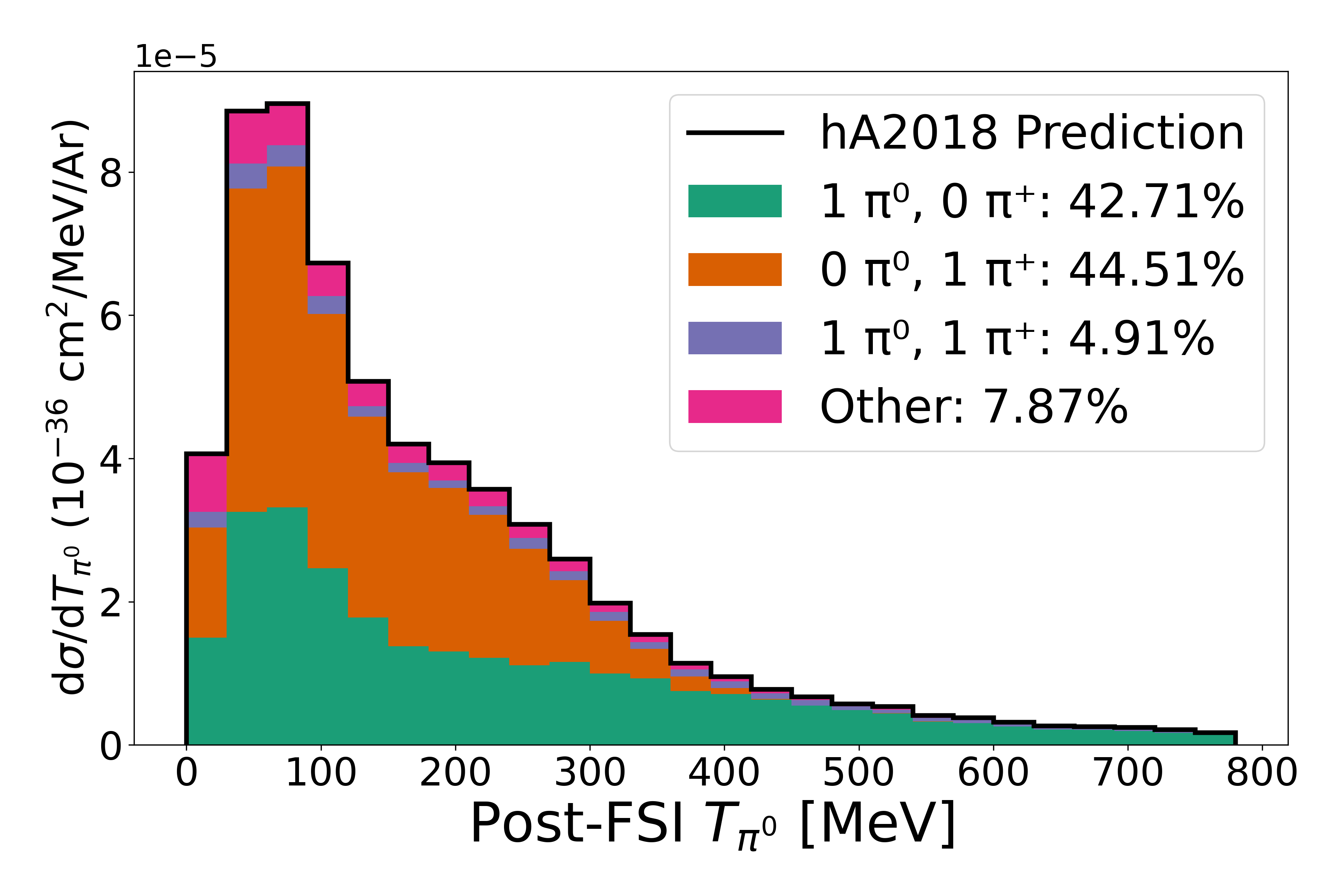}
    \caption{Neutral-pion kinetic-energy cross section predicted by the \texttt{GENIE} \texttt{hA2018} model. Stacked contributions indicate the pre-FSI pion origin of the $\nu_\mu$CC~$1\pi^0Np$ final state. The green component corresponds to charged pions undergoing CEX; the red component denotes primary $\pi^0$ mesons that survive FSI.}
    \label{fig:hA2018_pi0_sources}
\end{figure}

\begin{figure}[!htbp]
    \centering

    \begin{subfigure}{\linewidth}
        \centering
        \includegraphics[width=\linewidth]{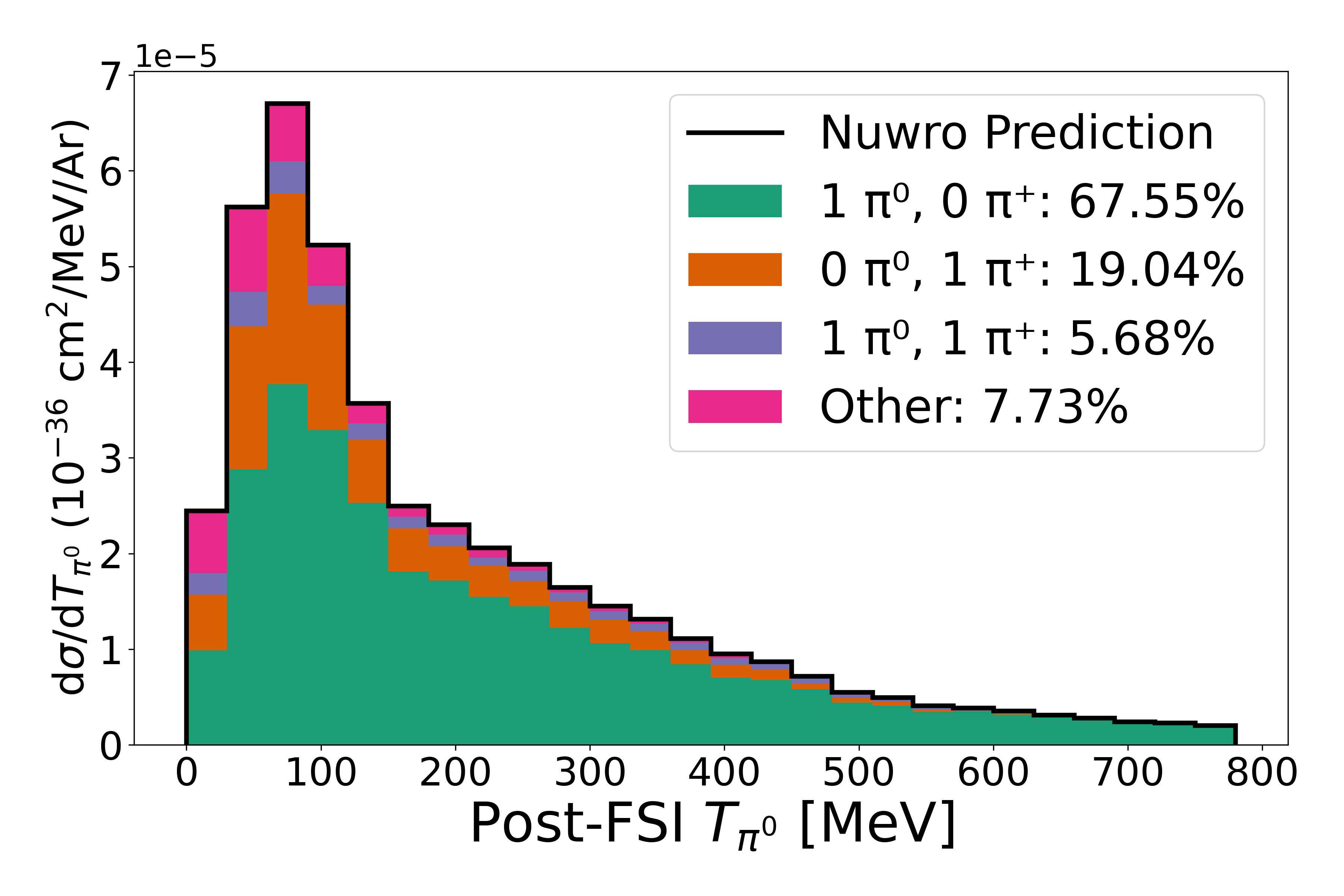}
        \paneltitle{(a) \texttt{NuWro}}
        \label{fig:nuwro_pi0_sources}
    \end{subfigure}

    \vspace{0.5cm}

    \begin{subfigure}{\linewidth}
        \centering
        \includegraphics[width=\linewidth]{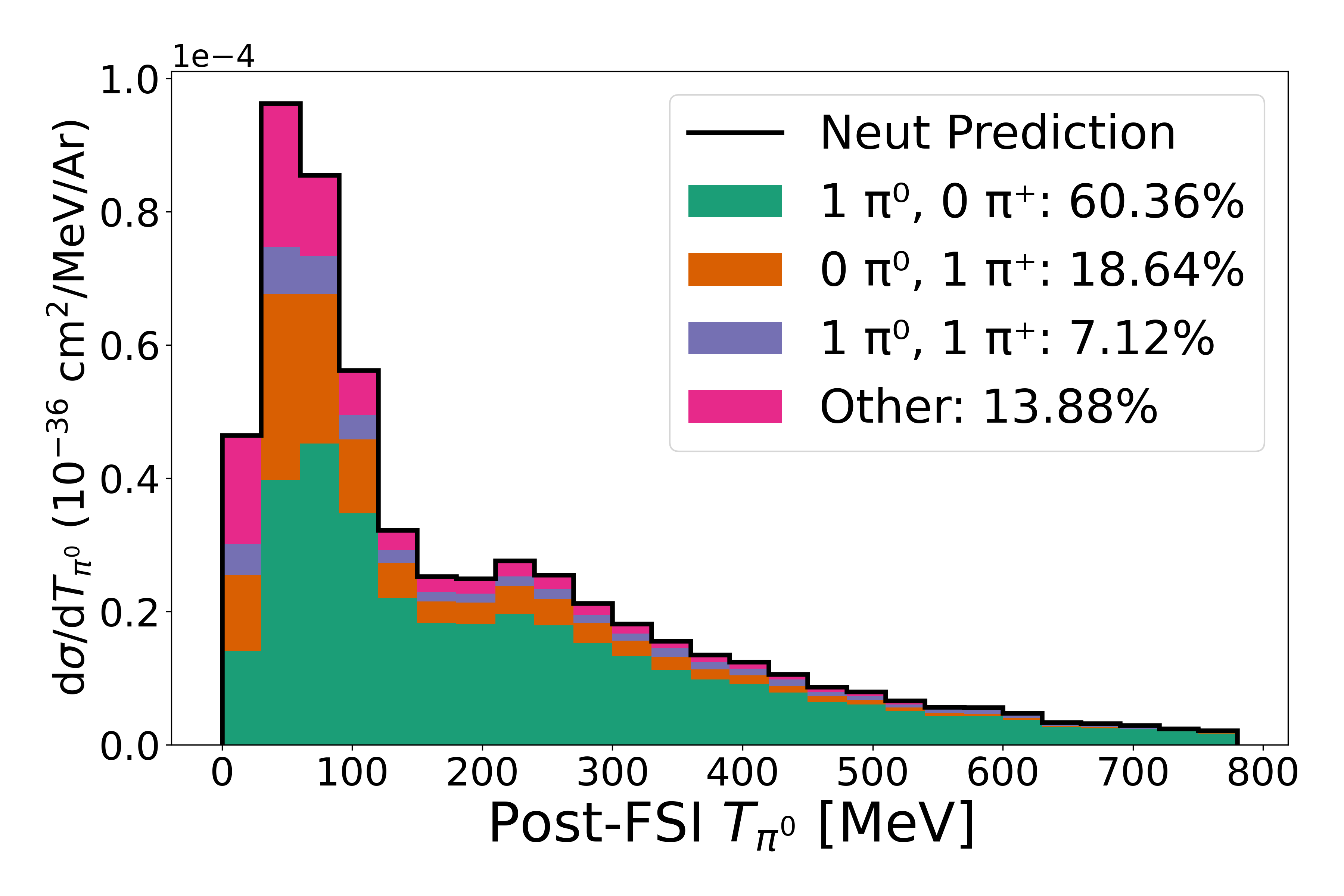}
        \paneltitle{(b) \texttt{NEUT}}
        \label{fig:neut_pi0_sources}
    \end{subfigure}

    \caption{Neutral-pion kinetic-energy cross sections predicted by (a) \texttt{NuWro} and (b) \texttt{NEUT}. Color conventions are as in Fig.~\ref{fig:hA2018_pi0_sources}.}
    \label{fig:nuwro_neut_pi0_sources}
\end{figure}

Figure~\ref{fig:hA2018_pi0_sources} shows the predicted composition of the post-FSI $\pi^0$ sample in CC resonance production. In \texttt{hA2018}, approximately 44\% of $\pi^0$ mesons originate from $\pi^+ \rightarrow \pi^0$ charge exchange. This fraction is significantly larger than that predicted by other generators. As shown in Fig.~\ref{fig:nuwro_neut_pi0_sources}, both \texttt{NuWro} and \texttt{NEUT} predict a CEX contribution of $\approx$19\%, indicating that \texttt{hA2018} overestimates this channel. 

\begin{figure}[!htbp]
    \centering
    \includegraphics[width=1\linewidth]{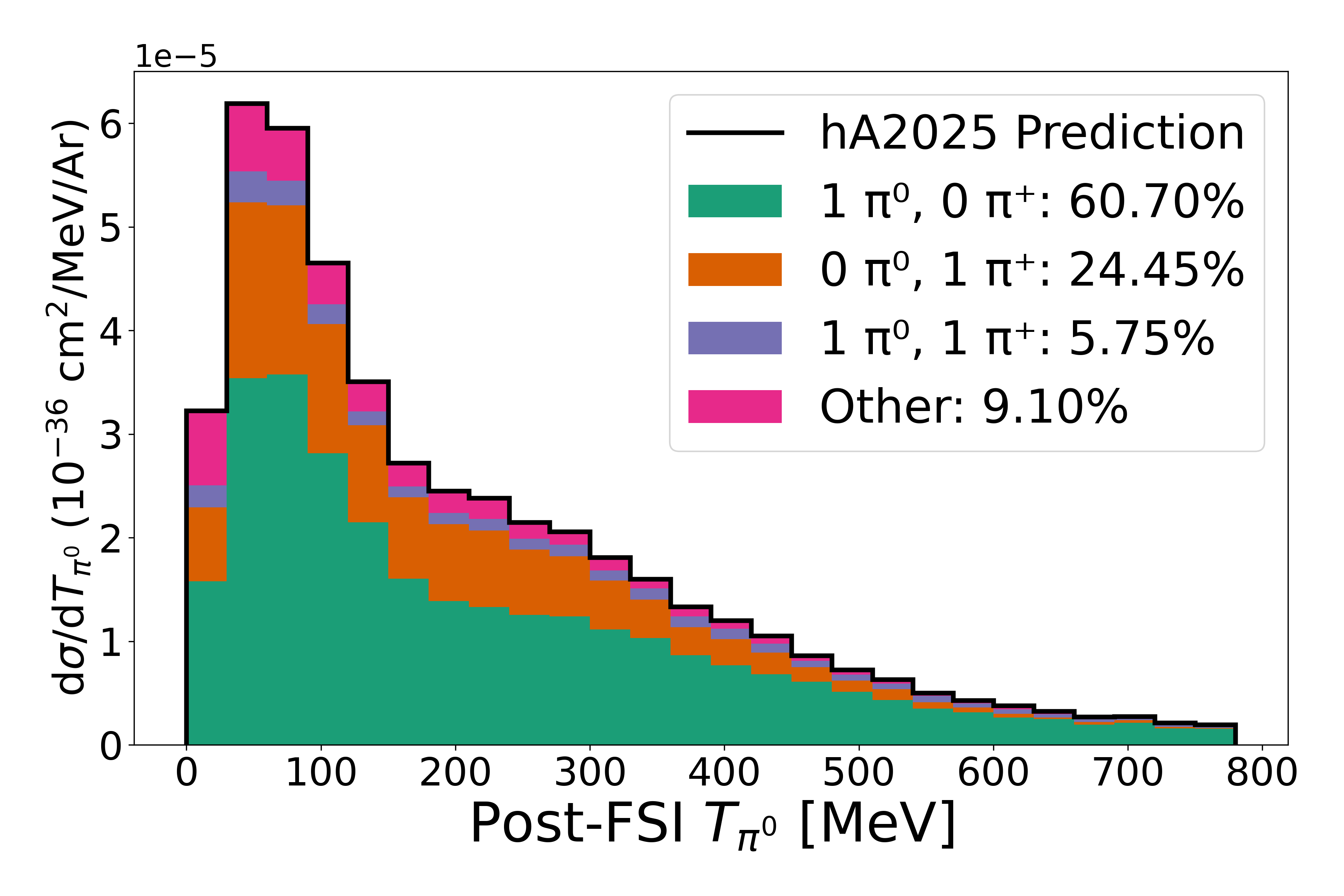}
    \caption{Neutral-pion kinetic-energy cross section predicted by the \texttt{GENIE} \texttt{hA2025} model. Color conventions are as in Fig.~\ref{fig:hA2018_pi0_sources}.}
    \label{fig:hA2025_pi0_sources}
\end{figure}

An updated model, \texttt{hA2025}, corrects this discrepancy. As shown in Fig.~\ref{fig:hA2025_pi0_sources}, the predicted CEX contribution is reduced and consistent with \texttt{NuWro} and \texttt{NEUT}. 

To incorporate the updated FSI model without regenerating the full sample, a reweighting procedure is applied to events generated with \texttt{hA2018}. Event weights are constructed as the bin-by-bin ratio of the two-dimensional post-FSI $\pi^0$ distributions in ($T_{\pi^0}, \cos\theta_{\pi^0})$ predicted by \texttt{hA2025} and \texttt{hA2018}. 

\begin{figure}[!htbp]
    \centering
    \includegraphics[width=1\linewidth]{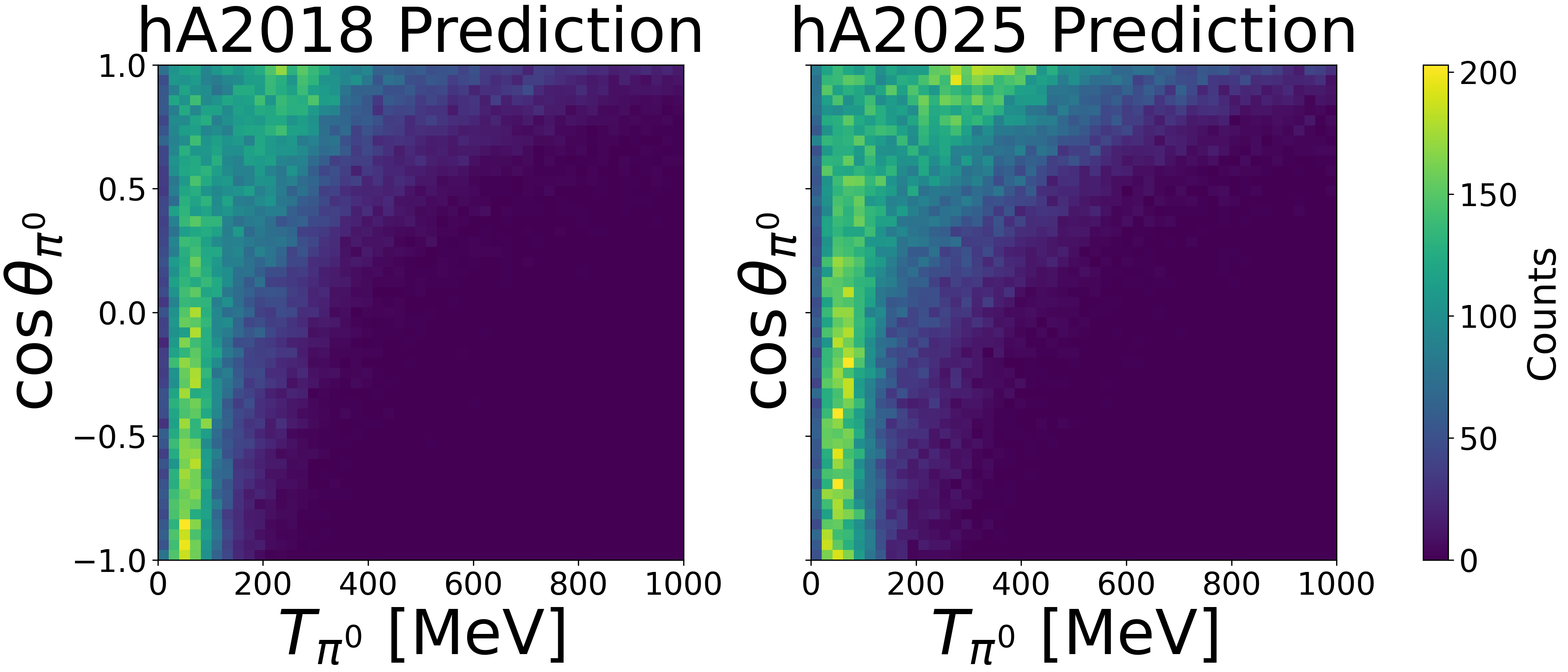}
    \caption{Two-dimensional distributions of $\pi^0$ kinetic energy and $\cos\theta$ for the \texttt{hA2018} (left) and \texttt{hA2025} (right) models.}
    \label{fig:2d_pi0_ke_cos_hA}
\end{figure}

\begin{figure}[!htbp]
    \centering
    \includegraphics[width=1\linewidth]{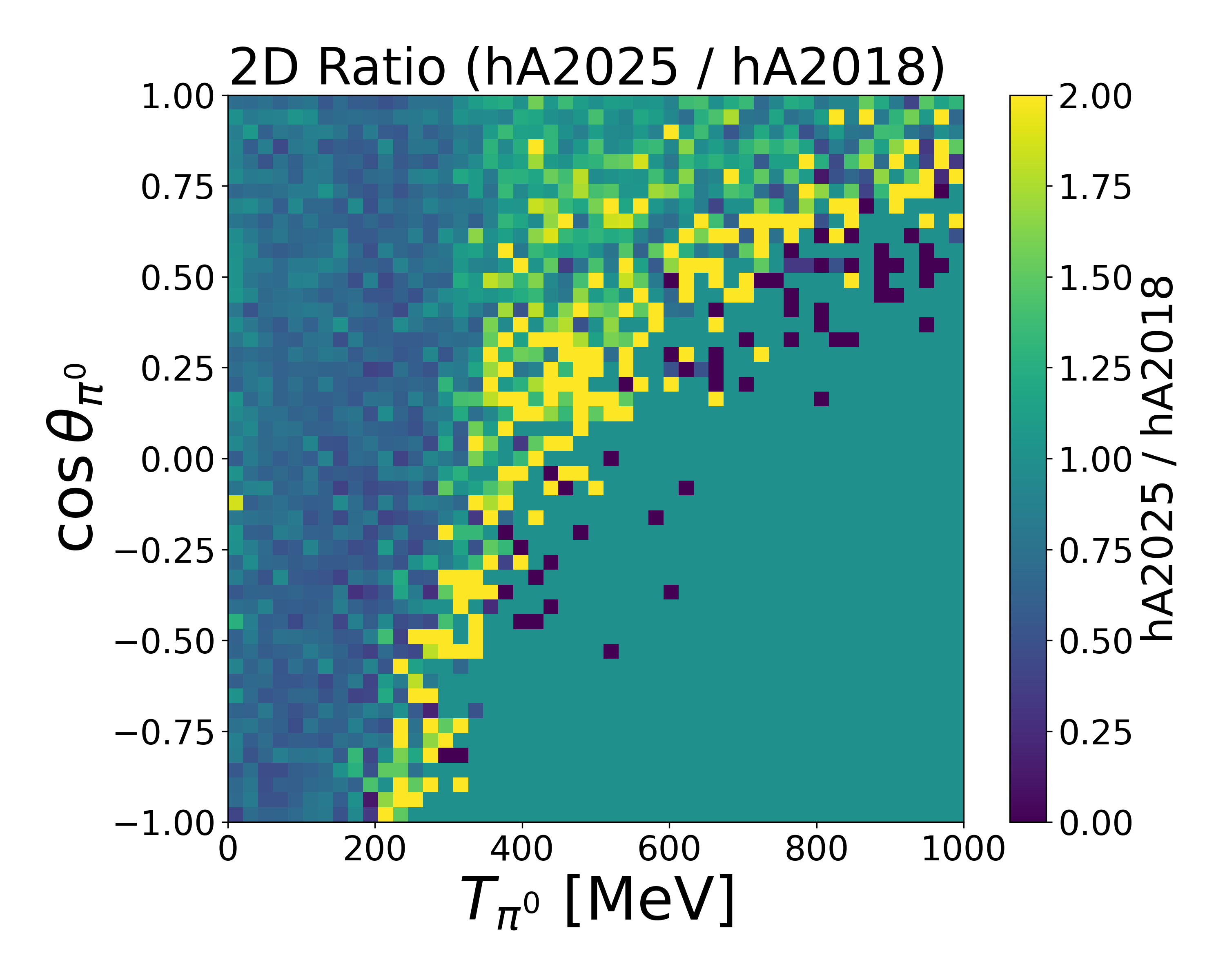}
    \caption{Ratio of the $\pi^0$ $(T,   \cos\theta)$ distributions between \texttt{hA2025} and \texttt{hA2018}, used to construct the reweighting factors.}
    \label{fig:hA_ratio}
\end{figure}

The underlying phase-space distributions are shown in Fig.~\ref{fig:2d_pi0_ke_cos_hA}, and their ratio, used to define the weights, is shown in Fig.~\ref{fig:hA_ratio}. 

\begin{figure}[!tbhp]
    \centering
    \includegraphics[width=1\linewidth]{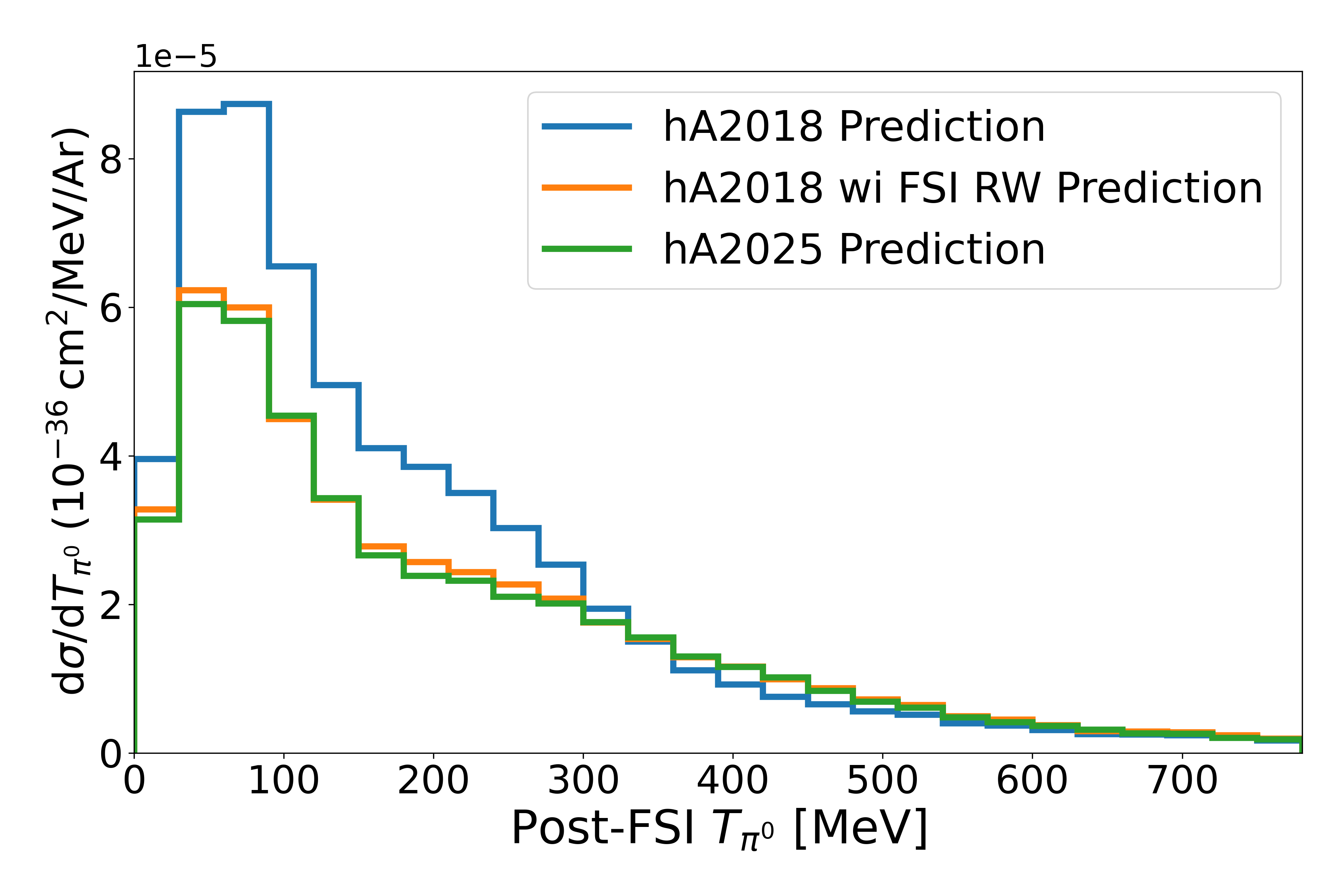}
    \caption{$T_{\pi^0}$ distributions for \texttt{hA2018} (with and without reweighting) and \texttt{hA2025}. The reweighted \texttt{hA2018} distribution is consistent with \texttt{hA2025}.}
    \label{fig:hA2018_2025_wi_rw}
\end{figure}

The effect of the reweighting is shown in Fig.~\ref{fig:hA2018_2025_wi_rw}, which compares the $T_{\pi^0}$ distributions from the nominal \texttt{hA2018}, reweighted \texttt{hA2018}, and \texttt{hA2025} samples. The reweighted distribution reproduces the \texttt{hA2025} prediction within statistical precision, validating the procedure. 

\phantomsection
\def\@currentlabel{A}
\label{appendix:fsi_rw}
\makeatother

\bibliography{bib.bib}

@article{KatoriMartini2020,
  author = {Katori, Teppei and Martini, Marco},
  title = {Neutrino--nucleus cross sections for oscillation experiments},
  journal = {Prog. Part. Nucl. Phys.},
  volume = {112},
  pages = {103773},
  year = {2020},
  doi = {10.1016/j.ppnp.2020.103773}
}

@article{Benhar2017,
  author = {Benhar, Omar and Huber, Patrick and Mariani, Camillo and Meloni, Davide},
  title = {Neutrino--nucleus interactions and the determination of oscillation parameters},
  journal = {Phys. Rep.},
  volume = {700},
  pages = {1--47},
  year = {2017},
  doi = {10.1016/j.physrep.2017.07.004}
}

@article{AlvarezRuso2018,
  author = {Alvarez-Ruso, L. and others},
  collaboration = {NuSTEC Collaboration},
  title = {{NuSTEC} White Paper: Status and challenges of neutrino--nucleus scattering},
  journal = {Prog. Part. Nucl. Phys.},
  volume = {100},
  pages = {1--68},
  year = {2018},
  doi = {10.1016/j.ppnp.2018.01.006}
}

@article{Coloma2014,
  author = {Coloma, Pilar and Huber, Patrick and Jen, Cheng-Ming and Mariani, Camillo},
  title = {Neutrino--nucleus cross section uncertainties and their role in long-baseline oscillation experiments},
  journal = {Phys. Rev. D},
  volume = {89},
  pages = {073015},
  year = {2014},
  doi = {10.1103/PhysRevD.89.073015}
}

@article{Mosel2016,
  author = {Mosel, Ulrich},
  title = {Neutrino interactions with nucleons and nuclei: Importance for long-baseline experiments},
  journal = {Phys. Rep.},
  volume = {645},
  pages = {1--60},
  year = {2016},
  doi = {10.1016/j.physrep.2016.06.002}
}

@article{ReinSehgal1981,
  author = {Rein, Dieter and Sehgal, Lalit M.},
  title = {Neutrino-excitation of baryon resonances and single pion production},
  journal = {Ann. Phys.},
  volume = {133},
  pages = {79--153},
  year = {1981},
  doi = {10.1016/0003-4916(81)90242-6}
}

@article{Hernandez2013,
  author = {Hern{\'a}ndez, E. and Nieves, J. and Vicente Vacas, M. J.},
  title = {Single pion production in neutrino--nucleon interactions},
  journal = {Phys. Rev. D},
  volume = {87},
  pages = {113009},
  year = {2013},
  doi = {10.1103/PhysRevD.87.113009}
}

@article{MiniBooNEpi02010,
  author = {Aguilar-Arevalo, A. A. and others},
  collaboration = {MiniBooNE Collaboration},
  title = {Measurement of neutral current $\pi^{0}$ production in neutrino interactions on mineral oil},
  journal = {Phys. Rev. D},
  volume = {81},
  pages = {013005},
  year = {2010},
  doi = {10.1103/PhysRevD.81.013005}
}

@article{MicroBooNEDetector2017,
  author = {Acciarri, R. and others},
  collaboration = {MicroBooNE Collaboration},
  title = {Design and construction of the {MicroBooNE} detector},
  journal = {J. Instrum.},
  volume = {12},
  pages = {P02017},
  year = {2017},
  doi = {10.1088/1748-0221/12/02/P02017}
}

@article{DUNECDR2020,
  author = {Abi, B. and others},
  collaboration = {DUNE Collaboration},
  title = {Deep Underground Neutrino Experiment ({DUNE}), far detector technical design report},
  journal = {J. Instrum.},
  volume = {15},
  pages = {T08008},
  year = {2020},
  doi = {10.1088/1748-0221/15/08/T08008}
}

@article{TKI,
  author = {Lu, X.-G. and Pickering, L. and Dolan, S. and Barr, G. and Coplowe, D. and Uchida, Y. and Wark, D. and Wascko, M. O. and Weber, A. and Yuan, T.},
  title = {Measurement of nuclear effects in neutrino interactions with minimal dependence on neutrino energy},
  journal = {Phys. Rev. C},
  volume = {94},
  pages = {015503},
  year = {2016},
  doi = {10.1103/PhysRevC.94.015503}
}

@article{TKI_Ar,
  author        = {Lars Bathe-Peters and Steven Gardiner and Roxanne Guenette},
  title         = {Comparing generator predictions of transverse kinematic imbalance in neutrino--argon scattering},
  eprint        = {2201.04664},
  archivePrefix = {arXiv},
  primaryClass  = {hep-ph},
  note          = {{FERMILAB-PUB-22-007-SCD}},
  journal       = {arXiv preprint},
  year          = {2022}
}

@article{uboone_ccpi0,
  author = {Abratenko, P. and others},
  collaboration = {MicroBooNE Collaboration},
  title = {Measurement of the differential cross section for neutral pion production in charged-current muon neutrino interactions on argon with the {MicroBooNE} detector},
  journal = {Phys. Rev. D},
  volume = {110},
  pages = {092014},
  year = {2024},
  doi = {10.1103/PhysRevD.110.092014}
}

@article{T2KTKI2018,
  author = {Abe, K. and others},
  collaboration = {T2K Collaboration},
  title = {Measurement of transverse kinematic imbalance in neutrino--nucleus interactions},
  journal = {Phys. Rev. D},
  volume = {98},
  pages = {032003},
  year = {2018},
  doi = {10.1103/PhysRevD.98.032003}
}

@article{T2K_pip_pTx,
  author = {Abe, K. and others},
  collaboration = {T2K Collaboration},
  title = {First {T2K} measurement of transverse kinematic imbalance in the muon neutrino charged-current single-$\pi^{+}$ production channel containing at least one proton},
  journal = {Phys. Rev. D},
  volume = {103},
  pages = {112009},
  year = {2021},
  doi = {10.1103/PhysRevD.103.112009}
}

@article{minerva_pi0,
  author = {Aliaga, L. and others},
  collaboration = {{MINERvA} Collaboration},
  title = {Probing nuclear effects with neutrino-induced charged-current neutral pion production},
  journal = {Phys. Rev. D},
  volume = {102},
  pages = {072007},
  year = {2020},
  doi = {10.1103/PhysRevD.102.072007}
}

@article{uboone_tki_ptxy,
  author = {Abratenko, P. and others},
  collaboration = {MicroBooNE Collaboration},
  title = {Measurement of nuclear effects in neutrino--argon interactions using generalized kinematic imbalance variables with the {MicroBooNE} detector},
  journal = {Phys. Rev. D},
  volume = {109},
  pages = {092007},
  year = {2024},
  doi = {10.1103/PhysRevD.109.092007}
}

@article{BNB2009,
  author = {Aguilar-Arevalo, A. A. and others},
  collaboration = {MiniBooNE Collaboration},
  title = {The neutrino flux prediction at {MiniBooNE}},
  journal = {Phys. Rev. D},
  volume = {79},
  pages = {072002},
  year = {2009},
  doi = {10.1103/PhysRevD.79.072002}
}

@article{MicroBooNECosmic2018,
  author = {Acciarri, R. and others},
  collaboration = {MicroBooNE Collaboration},
  title = {Cosmic ray background removal in the {MicroBooNE} {LArTPC}},
  journal = {J. Instrum.},
  volume = {12},
  pages = {P12030},
  year = {2017},
  doi = {10.1088/1748-0221/12/12/P12030}
}

@article{GEANT4,
  author = {Agostinelli, S. and others},
  collaboration = {GEANT4 Collaboration},
  title = {{GEANT4}--a simulation toolkit},
  journal = {Nucl. Instrum. Meth. A},
  volume = {506},
  pages = {250--303},
  year = {2003},
  doi = {10.1016/S0168-9002(03)01368-8}
}

@article{GENIEv3,
  author = {Andreopoulos, C. and others},
  collaboration = {GENIE Collaboration},
  title = {The {GENIE} neutrino {Monte} {Carlo} generator: physics and user manual},
  journal = {Nucl. Instrum. Meth. A},
  volume = {614},
  pages = {87--104},
  year = {2010},
  doi = {10.1016/j.nima.2009.12.009}
}

@article{hA2018,
  author = {Tena-Vidal, J. and others},
  title = {Neutrino--nucleus modeling in {GENIE} v3},
  journal = {Phys. Rev. D},
  volume = {104},
  pages = {072009},
  year = {2021},
  doi = {10.1103/PhysRevD.104.072009}
}

@article{LArSoft,
  author = {Church, E. D.},
  title = {LArSoft: A software package for liquid argon time projection drift chambers},
  journal = {arXiv preprint},
  eprint = {1311.6774},
  archivePrefix = {arXiv},
  primaryClass = {physics.ins-det},
  year = {2013}
}

@article{hA2018_issue,
  author = {Sharma, H. R. and Nagu, Srishti and Singh, Jyotsna and Singh, R. B.},
  title = {Study of pion production in {$\nu_{\mu}$} interactions on {$^{40}$Ar} in {DUNE} using {GENIE} and {NuWro} event generators},
  journal = {Phys. Part. Nucl. Lett.},
  volume = {19},
  number = {6},
  pages = {724--739},
  year = {2022},
  doi = {10.1134/S1547477122060267}
}

@article{EPJCDetectorSys,
  author = {Abratenko, P. and others},
  collaboration = {MicroBooNE Collaboration},
  title = {Novel approach for evaluating detector-related uncertainties in a {LArTPC} using {MicroBooNE} data},
  journal = {Eur. Phys. J. C},
  volume = {82},
  pages = {454},
  year = {2022},
  doi = {10.1140/epjc/s10052-022-10270-8}
}

@article{WCLee2018,
  author = {Abratenko, P. and others},
  collaboration = {MicroBooNE Collaboration},
  title = {Measurement of inclusive muon neutrino charged-current differential cross sections on argon with {Wire-Cell} reconstruction},
  journal = {Phys. Rev. Lett.},
  volume = {123},
  pages = {131801},
  year = {2019},
  doi = {10.1103/PhysRevLett.123.131801}
}

@article{WireCellImaging,
  author = {Wang, Y. and others},
  title = {{Wire-Cell} {3D} imaging reconstruction in liquid argon time projection chambers},
  journal = {J. Instrum.},
  volume = {13},
  pages = {P07016},
  year = {2018},
  doi = {10.1088/1748-0221/13/05/p05032}
}

@article{WCpi02023,
  author = {Abratenko, P. and others},
  collaboration = {MicroBooNE Collaboration},
  title = {Measurement of neutral pion production in $\nu_{\mu}$ charged-current interactions on argon with the {MicroBooNE} detector},
  journal = {Phys. Rev. D},
  volume = {107},
  pages = {012004},
  year = {2023},
  doi = {10.1103/PhysRevD.107.012004}
}

@article{WCNCpi0,
  author = {Abratenko, P. and others},
  collaboration = {MicroBooNE Collaboration},
  title = {First double-differential cross section measurement of neutral-current $\pi^{0}$ production in neutrino--argon scattering in the {MicroBooNE} detector},
  journal = {Phys. Rev. Lett.},
  volume = {134},
  pages = {161802},
  year = {2025},
  doi = {10.1103/PhysRevLett.134.161802}
}

@article{Ben_0pNp,
  author = {Abratenko, P. and others},
  collaboration = {MicroBooNE Collaboration},
  title = {Inclusive cross section measurements in final states with and without protons for charged-current $\nu_{\mu}$--Ar scattering in {MicroBooNE}},
  journal = {Phys. Rev. D},
  volume = {110},
  pages = {013006},
  year = {2024},
  doi = {10.1103/PhysRevD.110.013006}
}

@article{wiener_svd,
  author = {Tang, W. and Li, X. and Qian, X. and Wei, H. and Zhang, C.},
  title = {Data unfolding with {Wiener-SVD} method},
  journal = {J. Instrum.},
  volume = {12},
  pages = {P10002},
  year = {2017},
  doi = {10.1088/1748-0221/12/10/P10002},
  archivePrefix = {arXiv},
  primaryClass = {physics.data-an}
}

@article{blockwise,
  author = {Gardiner, Steven},
  title = {Mathematical methods for neutrino cross-section extraction},
  journal = {arXiv preprint},
  eprint = {2401.04065},
  archivePrefix = {arXiv},
  primaryClass = {hep-ex},
  note          = {{FERMILAB-PUB-23-692-CSAID}},
  year = {2024}
}

@article{wc_methods,
  author = {Abratenko, P. and others},
  collaboration = {MicroBooNE Collaboration},
  title = {Data-driven model validation for neutrino--nucleus cross section measurements},
  journal = {Phys. Rev. D},
  volume = {111},
  pages = {092010},
  year = {2025},
  doi = {10.1103/PhysRevD.111.092010}
}

@article{NuWro,
  author = {Prasad, Hemant and Sobczyk, Jan T. and Ankowski, Artur M. and Bonilla, J. Luis and Banerjee, Rwik Dharmapal and Graczyk, Krzysztof M. and Kowal, Beata E.},
  title = {Developments in {NuWro} {Monte} {Carlo} generator},
  journal = {arXiv preprint},
  eprint = {2501.11470},
  archivePrefix = {arXiv},
  primaryClass = {hep-ph},
  year = {2025}
}

@article{Neut,
  author = {Dolan, S. and McElwee, J. and Bolognesi, S. and Hayato, Y. and McFarland, K. and Megias, G. and Niewczas, K. and Pickering, L. and Sobczyk, J. and Thompson, L. and Wret, C.},
  title = {Electron--nucleus scattering in the {NEUT} event generator},
  journal = {arXiv preprint},
  eprint = {2301.09195},
  archivePrefix = {arXiv},
  primaryClass = {hep-ex},
  year = {2023}
}

@article{Hayato2021NEUT,
  author = {Hayato, Yoshinari and Pickering, Luke},
  title = {The {NEUT} neutrino interaction simulation program library},
  journal = {Eur. Phys. J. Spec. Top.},
  volume = {230},
  pages = {4469--4481},
  year = {2021},
  doi = {10.1140/epjs/s11734-021-00287-7}
}

@article{NUISANCE,
  author = {Stowell, P. and others},
  title = {{NUISANCE}: A neutrino cross-section generator tuning and comparison framework},
  journal = {J. Instrum.},
  volume = {12},
  pages = {P01016},
  year = {2017},
  doi = {10.1088/1748-0221/12/01/P01016}
}

@article{gibuu_pi0,
  author = {Yan, Qiyu and Wen, Kaile and Gallmeister, Kai and Lu, Xianguo and Mosel, Ulrich and Zheng, Yangheng},
  title = {Understanding neutrino pion production with the {GiBUU} model},
  journal = {Phys. Rev. D},
  volume = {112},
  pages = {093007},
  year = {2025},
  doi = {10.1103/PhysRevD.112.093007}
}

@misc{GENIEGitHub,
  author = {{GENIE Collaboration}},
  title = {{GENIE} neutrino {Monte} {Carlo} generator},
  howpublished = {\url{https://github.com/GENIE-MC/Generator}},
  note = {Accessed: 2026-02-16}
}

@misc{ar23,
  author = {Dolan, S. and others},
  title = {{DUNE} baseline model and uncertainties ({AR23} 20i {GENIE} tune documentation)},
  howpublished = {\url{https://indico.cern.ch/event/1607882/contributions/6775083/attachments/3166278/5627343/DUNE_model_documentation-3.pdf}},
  note = {{DUNE} Collaboration document, CERN Indico},
  year = {2025}
}

@article{Katori2012RMP,
  author = {Formaggio, J. A. and Zeller, G. P.},
  title = {From {eV} to {EeV}: Neutrino cross sections across energy scales},
  journal = {Rev. Mod. Phys.},
  volume = {84},
  pages = {1307--1341},
  year = {2012},
  doi = {10.1103/RevModPhys.84.1307}
}

@article{nuwro_updates,
  author = {Prasad, Hemant and Sobczyk, Jan T. and Ankowski, Artur M. and Bonilla, J. Luis and Banerjee, Rwik Dharmapal and Graczyk, Krzysztof M. and Kowal, Beata E.},
  title = {Developments in {NuWro} {Monte} {Carlo} generator},
  journal = {arXiv preprint},
  eprint = {2501.11470},
  archivePrefix = {arXiv},
  primaryClass = {hep-ph},
  year = {2025}
}

@article{ptx_sym,
  author = {Lu, X.-G. and Coplowe, D. and Shah, R. and Barr, G. and Wark, D. and Weber, A.},
  title = {Reconstruction of energy spectra of neutrino beams independent of nuclear effects},
  journal = {Phys. Rev. D},
  volume = {92},
  pages = {051302},
  year = {2015},
  doi = {10.1103/PhysRevD.92.051302}
}

@article{fsi_models_genie,
  author = {Dytman, S. and Hayato, Y. and Raboanary, R. and Sobczyk, J. T. and Tena-Vidal, J. and Vololoniaina, N.},
  title = {Comparison of validation methods of simulations for final-state interactions in hadron production experiments},
  journal = {Phys. Rev. D},
  volume = {104},
  pages = {053006},
  year = {2021},
  doi = {10.1103/PhysRevD.104.053006}
}

@article{cem_mashnik,
  author = {Mashnik, S. G. and Sierk, A. J. and Gudima, K. K. and Baznat, M. I.},
  title = {{CEM03} and {LAQGSM03}---new modeling tools for nuclear applications},
  journal = { J. Phys. Conf. Ser.},
  volume = {41},
  pages = {340--351},
  year = {2006},
  doi = {10.1088/1742-6596/41/1/037}
}

@article{charge_to_ionization,
  author = {Shibamura, Eido and Hitachi, Akira and Doke, Tadayoshi and Takahashi, Tan and Kubota, Shinzou and Miyajima, Mitsuhiro},
  title = {Drift velocities of electrons, saturation characteristics of ionization and W-values for conversion electrons in liquid argon, liquid argon--gas mixtures and liquid xenon},
  journal = { Nucl. Instrum. Meth.},
  volume = {131},
  pages = {249--258},
  year = {1975},
  doi = {10.1016/0029-554X(75)90327-4}
}

@article{uboone_tune,
  author = {Abratenko, P. and others},
  collaboration = {MicroBooNE Collaboration},
  title = {New {CC0$\pi$} {GENIE} model tune for {MicroBooNE}},
  journal = {Phys. Rev. D},
  volume = {105},
  pages = {072001},
  year = {2022},
  doi = {10.1103/PhysRevD.105.072001}
}

@article{uboone_recom,
  author = {Acciarri, R. and others},
  collaboration = {MicroBooNE Collaboration},
  title = {A study of electron recombination using highly ionizing particles in the {ArgoNeuT} liquid argon {TPC}},
  journal = {J. Instrum.},
  volume = {8},
  pages = {P08005},
  year = {2013},
  doi = {10.1088/1748-0221/8/08/P08005}
}

@article{uboone_sim,
  author = {Acciarri, R. and others},
  collaboration = {MicroBooNE Collaboration},
  title = {Ionization electron signal processing in single-phase {LArTPCs}. {Part I}. Algorithm description and quantitative evaluation with {MicroBooNE} simulation},
  journal = {J. Instrum.},
  volume = {13},
  pages = {P07006},
  year = {2018},
  doi = {10.1088/1748-0221/13/07/P07006}
}

@article{BergerSehgal2007,
  author = {Berger, C. and Sehgal, L. M.},
  title = {Lepton mass effects in single pion production by neutrinos},
  journal = {Phys. Rev. D},
  volume = {76},
  pages = {113004},
  year = {2007},
  doi = {10.1103/PhysRevD.76.113004}
}

@misc{supp,
  note = {See Supplemental Material at [URL will be inserted by publisher] for additional details}
}

@article{nuwro25_11,
  author = {Yan, Qiyu and Niewczas, Kajetan and Nikolakopoulos, Alexis and Gonz{\'a}lez-Jim{\'e}nez, Ra{\'u}l and Jachowicz, Natalie and Lu, Xianguo and Sobczyk, Jan and Zheng, Yangheng},
  title = {The {Ghent} hybrid model in {NuWro}: A new neutrino single-pion production model in the {GeV} regime},
  journal = {J. High Energy Phys.},
  volume = {2024},
  pages = {141},
  year = {2024},
  doi = {10.1007/JHEP12(2024)141}
}

@article{nieves,
  author = {Nieves, J. and Amaro, J. E. and Valverde, M.},
  title = {Inclusive quasielastic charged-current neutrino--nucleus reactions},
  journal = { Phys. Rev. C},
  volume = {70},
  pages = {055503},
  year = {2004},
  doi = {10.1103/PhysRevC.70.055503}
}

@article{nieves_2p2h,
  author = {Gran, R. and Nieves, J. and Sanchez, F. and Vicente Vacas, M. J.},
  title = {Neutrino--nucleus quasielastic and 2p2h interactions up to 10 {GeV}},
  journal = {Phys. Rev. D},
  volume = {88},
  pages = {113007},
  year = {2013},
  doi = {10.1103/PhysRevD.88.113007}
}

@article{Yang_2009,
  author = {Yang, T. and Andreopoulos, C. and Gallagher, H. and Hofmann, K. and Kehayias, P.},
  title = {A hadronization model for few-{GeV} neutrino interactions},
  journal = {Eur. Phys. J. C},
  volume = {63},
  pages = {1--10},
  year = {2009},
  doi = {10.1140/epjc/s10052-009-1094-z}
}

@article{dune_model_SP,
 author = {Chakrani, J. and others},
  title = {Parametrized uncertainties in the spectral function model of neutrino charged-current quasielastic interactions for oscillation analyses},
  journal = {Phys. Rev. D},
  volume = {109},
  pages = {072006},
  year = {2024},
  doi = {10.1103/PhysRevD.109.072006}
}

@article{SuSAv2,
  author = {Dolan, S. and Megias, G. D. and Bolognesi, S.},
  title = {Implementation of the {SuSAv2} meson-exchange current 1p1h and 2p2h models in {GENIE} and analysis of nuclear effects in {T2K} measurements},
  journal = {Phys. Rev. D},
  volume = {101},
  pages = {033003},
  year = {2020},
  doi = {10.1103/PhysRevD.101.033003}
}

@article{Salcedo,
  author = {Salcedo, L. L. and Oset, E. and Vicente Vacas, M. J. and Garcia-Recio, C.},
  title = {Computer simulation of inclusive pion nuclear reactions},
  journal = {Nucl. Phys. A},
  volume = {484},
  pages = {557--592},
  year = {1988},
  doi = {10.1016/0375-9474(88)90310-7}
}

@article{res_problems,
  author = {Freedman, R. A. and Miller, G. A. and Henley, E. M.},
  title = {Pion nucleus scattering and systematics of the {$\Delta$}-nucleus potential},
  journal = {Phys. Lett. B},
  volume = {103},
  pages = {397--400},
  year = {1981},
  doi = {10.1016/0370-2693(81)90068-X}
}

@article{london_3d,
  author = {Abratenko, P. and others},
  collaboration = {MicroBooNE Collaboration},
   title={Measurement of three-dimensional inclusive muon-neutrino charged-current cross sections on argon with the {MicroBooNE} detector},

   doi={10.1016/j.physletb.2025.139939},
   journal={Phys. Lett. B},
   volume={870},
   pages={139939},
   year={2025},
    }

@article{real_flux,
  title = {Treatment of flux shape uncertainties in unfolded, flux-averaged neutrino cross-section measurements},
  author = {Koch, Lukas and Dolan, Stephen},
  journal = {Phys. Rev. D},
  volume = {102},
  issue = {11},
  pages = {113012},
  numpages = {11},
  year = {2020},
  month = {Dec},
  publisher = {American Physical Society},
  doi = {10.1103/PhysRevD.102.113012},
}

@misc{genie_v2,
      title={The {GENIE} Neutrino {Monte} {Carlo} {Generator}: Physics and User Manual}, 
      author={Costas Andreopoulos and Christopher Barry and Steve Dytman and Hugh Gallagher and Tomasz Golan and Robert Hatcher and Gabriel Perdue and Julia Yarba},
      year={2015},
      eprint={1510.05494},
      archivePrefix={arXiv},
      primaryClass={hep-ph},
      note = {{FERMILAB-FN-1004-CD}}

}

@article{gibuu_fw,
   title={Transport-theoretical description of nuclear reactions},
   volume={512},
   ISSN={0370-1573},
   url={http://dx.doi.org/10.1016/j.physrep.2011.12.001},
   DOI={10.1016/j.physrep.2011.12.001},
   number={1-2},
   journal={Phys. Rep.},
   publisher={Elsevier BV},
   author={Buss, O. and Gaitanos, T. and Gallmeister, K. and van Hees, H. and Kaskulov, M. and Lalakulich, O. and Larionov, A.B. and Leitner, T. and Weil, J. and Mosel, U.},
   year={2012},
   month={mar},
   pages={1–124} 
   }

@article{mcs_paper,
  author        = {{MicroBooNE Collaboration}},
  title         = {Improved muon energy estimation using a detailed model of multiple Coulomb scattering in the {MicroBooNE} {LArTPC}},
  year          = {2026},
  eprint        = {2605.03048},
  archivePrefix = {arXiv},
  primaryClass  = {hep-ex},
  journal       = {arXiv preprint},
  url           = {https://arxiv.org/abs/2605.03048}
}

@article{geant4_rw,
	author = {Calcutt, J. and Thorpe, C. and Mahn, K. and Fields, L.},
	doi = {10.1088/1748-0221/16/08/P08042},
	journal = {J. Instrum.},
	month = {aug},
	number = {08},
	pages = {P08042},
	publisher = {IOP Publishing},
	title = {{Geant4Reweight}: a framework for evaluating and propagating hadronic interaction uncertainties in {Geant4}},
	url = {https://doi.org/10.1088/1748-0221/16/08/P08042},
	volume = {16},
	year = {2021}}

@article{light_yeild,
  author         = "{MicroBooNE Collaboration}",
  title          = "{Scintillation light calibrations, systematic uncertainties, and triggering efficiency in the {MicroBooNE} detector}",
  year           = "2026",
  eprint         = "2603.23691",
  archivePrefix  = "arXiv",
  primaryClass   = "physics.ins-det",
  journal        = "arXiv preprint"
}

\end{document}